\documentclass[12pt,preprint]{aastex}

\begin{document}

\title{Ionized Gas Towards Molecular Clumps: Physical Properties of Massive Star Forming Regions}
\shorttitle{Ionized Gas Towards Molecular Clumps}
\shortauthors{Johnston et al.}

\author{Katharine G. Johnston\altaffilmark{1}, Debra S. Shepherd\altaffilmark{2}, James E. Aguirre \altaffilmark{3}, Miranda K. Dunham \altaffilmark{4}, Erik Rosolowsky\altaffilmark{5}, and Kenneth Wood\altaffilmark{1}}

\altaffiltext{1}{SUPA, School of Physics and Astronomy, University of St Andrews, St Andrews, KY16 9SS, UK; kgj1@st-andrews.ac.uk}
\altaffiltext{2}{National Radio Astronomy Observatory, P.O. Box O, 1003
  Lopezville Rd, Socorro, NM 87801}
\altaffiltext{3}{Department of Physics and Astronomy, University of Pennsylvania, 209 South 33rd Street, Philadelphia, PA 19104}
\altaffiltext{4}{Department of Astronomy, University of Texas at Austin, 1 University Station, C1400, Austin, TX 78712-0259}
\altaffiltext{5}{University of British Columbia at Okanagan, 3333 University Way, Kelowna BC V1V 1V7, Canada}

\date{\today}

\newpage
\begin{abstract}

We have conducted a search for ionized gas at 3.6\,cm, using the Very Large Array,
towards 31 Galactic intermediate- and high-mass clumps
detected in previous millimeter continuum observations.
In the 10 observed fields, 35 HII regions are identified, of which 20 are newly discovered. 
Many of the HII regions are multiply peaked indicating the presence of a cluster
of massive stars. 
We find that the ionized gas tends to be associated towards the millimeter clumps; 
of the 31 millimeter clumps observed, 9 of these appear to be physically related
to ionized gas, and a further 6 have ionized gas emission within 1'.
For clumps with associated ionized gas, the combined mass of the ionizing massive stars is compared to
the clump masses to provide an estimate of the instantaneous star formation efficiency. 
These values range from a few percent to 25\%, and have an average of 7 $\pm$ 8\%. We also find a correlation between the clump mass
and the mass of the ionizing massive stars within it, which is consistent with a power law.
 This result is comparable to the prediction of star formation by competitive accretion
that a power law relationship exists between the mass of the most massive star in a cluster
 and the total mass of the remaining stars.

\end{abstract}


\maketitle

\section{INTRODUCTION}

The study of intermediate ($ 2 $M$_{\odot} <  $ M$_\star <8 $M$_{\odot}$) and high-mass (M$_\star> 8 $M$_{\odot}$) star formation has made significant progress in recent years, catching up dramatically with our understanding of low mass star formation \citep[e.g.,][]{testi03, beuther05, zinnecker07}. In fact, it appears that stars share many similar formation processes over a wide range of stellar masses. As in the case of low-mass star formation, outflows have been observed ubiquitously towards forming intermediate and high-mass stars \citep[e.g.,][]{fuente01,shepherd960,zhang05}, and there exist examples of stars as massive as early B-type surrounded by disks  \citep[e.g.,][]{cesaroni07}.

However, a key difference between these regimes is that the reduced Kelvin-Helmholtz contraction timescale of massive protostars results in massive stars reaching the main-sequence while they are still accreting \citep[][and references therein]{keto06}. Thus, we expect the circumstellar material of these accreting embedded stars to be progressively engulfed by an expanding HII region, and to also be affected by strong stellar winds and radiation pressure. This fact may provide an explanation for why outflows and disks are not commonly detected towards forming O stars \citep[e.g.,][]{shepherd03, cesaroni07}, as the above processes may be responsible for quickly photoevaporating or dispersing the circumstellar material. However, current observational biases may instead be responsible for the lack of disk detections for the most massive stars.

In this work we aim to provide preliminary studies of a selection of sites containing intermediate- and high-mass star formation, specifically to uncover the presence of ionized gas towards them. We also wish to study the relationship between the star forming gas, traced by millimeter continuum emission from dust, and the ionized gas created by massive stars, traced by radio continuum emission. We will select the most promising objects from this study for follow up with higher resolution observations, to map any outflows or disks towards these sources, and to study how the formation of an HII region affects the material within several hundreds of AU of the star. Therefore, we have selected sources that are within a declination range suitable for future study with Atacama Large Millimeter Array (ALMA) and the Expanded Very Large Array (EVLA).

To fulfill these aims, we have conducted radio continuum observations at 3.6~cm, using the Very Large Array (VLA) of the National Radio Astronomy Observatory \footnote{The National Radio Astronomy Observatory is a facility of the National Science Foundation operated under cooperative agreement by Associated Universities, Inc.}, of 31 clumps detected in previous millimeter continuum observations.   Ten sources were selected from preliminary images from the Bolocam Galactic Plane Survey (BGPS, Aguirre et al. 2009, in preparation), and 5 were selected from \citet[][hereafter B06]{beltran06}. The remaining 16 sources were observed serendipitously, as their positions lay within the observed VLA 3.6~cm fields.  Note that, in this work, we adopt the terminology that a molecular core produces a single star (or close binary system) while molecular clumps form clusters of stars.  In our study of massive star forming regions, our selected sources are several kiloparsecs away and, thus, we most likely detect clumps forming one or more massive stars along with many lower mass stars.   

Section \ref{sourceselect} presents our selection criteria for the observed millimeter sources. Section \ref{observation} provides details of our observations and data reduction. Section \ref{results} presents the observed 3.6~cm continuum images of each field; lists the measured positions, fluxes, and angular sizes of each detected VLA source; compares these results to existing millimeter and mid-IR observations; and presents derived properties for each source. Our discussion is presented in Section \ref{discussion}. Finally, our conclusions are given in Section \ref{conclusions}.

\section{SOURCE SELECTION \label{sourceselect}}

Clumps of molecular gas which are in the process of collapsing to form
stars contain dust at temperatures of several tens of Kelvin. This
dust is assumed to radiate as a gray body - a black body modified by a
frequency dependent opacity, producing emission peaking at
sub-millimeter and millimeter wavelengths. Thus observations in the
millimeter regime are able to detect possible sites of star formation.
The selected molecular clumps have masses large enough
to harbor a forming intermediate or high-mass star, ranging from 
approximately 50 to 2500\,M$_{\sun}$.  Sources were
chosen from BGPS preliminary images and the source list of B06.  
The BGPS (Aguirre et al. 2009, in preparation) is a 1.1 mm continuum survey of 150 square degrees 
of the Galactic Plane visible from the northern hemisphere, including a contiguous strip
from $l$ = -10.5 to 86.5, $b$ = $\pm$0.5, as well as selected regions beyond the solar circle.
The survey has a limiting non-uniform 1-$\sigma$ noise level in the range
20 and 50 mJy\,beam$^{-1}$ RMS at a resolution of 33".
The observations of B06 were taken with the 37-channel SEST Imaging Bolometer Array
(SIMBA) on the Swedish-ESO Submillimetre Telescope (SEST) to identify
1.2\,mm continuum emission within a 15' by 6.6' region centered on
selected IRAS sources. These observations have a resolution of 40".
Sources were chosen to have: 

\begin{enumerate}

\item A near distance of $\sim$1\,kpc, for clumps selected from the BGPS. 
A description of how the BGPS clump distances were calculated is given in
Section \ref{mmprops}. However, HI observations subsequently showed many of the 
selected BGPS clumps to be at the far distance (see Section \ref{mmprops}).
Clumps from B06 were selected to have distances given in their
Table 1 of less than 4\,kpc. Two of these sources however were also subsequently
found to be at the far distance.

\item A clump mass greater than 1\,M$_\odot$ at the near distance, with a preference for more
  massive clumps.  A description of how the BGPS source masses were
  calculated is given in Section \ref{mmprops}.  Masses of
  B06-selected sources are taken from their Table 2. 
  
\item Little or no associated emission in the NRAO VLA Sky Survey
  \citep[NVSS,][]{condon93} 1.4 GHz continuum images and MSX E band
  (21.34 $\mu$m) images \citep{price01}, suggesting that any massive
  protostellar objects are young, and/or deeply embedded. The RMS
  noise in the NVSS and MSX E band images is 0.45 mJy\,beam$^{-1}$ (in a
  synthesized beam size of 45"), and $\ge$13.3 MJy\,sr$^{-1}$ respectively.  

\item A declination between -20 and +40$^{\circ}$ (J2000). The lower
  declination limit of the VLA is -48$^{\circ}$, however, to ensure
  our observations had a reasonable beam shape, we chose a lower
  declination limit of -20$^{\circ}$. To avoid excessive shadowing in
  ALMA's compact configurations in follow up studies, we chose an
  upper declination limit of +40$^{\circ}$.  

\end{enumerate}

Preference was given to sources with no previous VLA observations at 3.6~cm that have sensitivities $\lesssim$30 $\mu$Jy\,beam$^{-1}$.
Fifteen sources were selected based on the above criteria while another sixteen were serendipitously present within the observed fields. Table~\ref{selectsrc} presents these sources, giving their positions; the velocity of associated $^{13}$CO Galactic Ring Survey \citep[GRS]{jackson06} emission, if available; their near and far distances; the assumed distance (near or far); the luminosity of associated IRAS sources; the millimeter flux (at 1.1 or 1.2 mm); and the estimated mass of the millimeter clumps. Millimeter clumps that were observed serendipitously are identified by ``Serend.'' in column 14 while clumps initially selected are identified by ``Select.'' The determination of the BGPS-selected source properties given in Table \ref{selectsrc} is described in Section \ref{mmprops}. B06-selected source 
properties in Table~\ref{selectsrc} are taken from their Tables 1 and 2, apart from the exceptions described below.
 
It was possible to determine the velocities and distances of the clumps towards IRAS\,18424-0329 and IRAS\,18571+0349, using the same methods as for the BGPS-selected sources (outlined in Section \ref{mmprops}). Clumps associated with IRAS\,18424-0329 were found to be at the far distance (see Section \ref{mmprops}), and IRAS\,18571+0349 was assumed to be at the far distance following the results of \citet{kuchar94}. Therefore the values of B06 are appropriately scaled to the far distance for these sources, and the velocity and distances found by the methods in Section \ref{mmprops} are instead quoted in columns 6, 7 and 8 of Table \ref{selectsrc}. 

\subsection{Properties of the BGPS Millimeter Sources \label{mmprops}}

The BGPS sources were identified by eye in preliminary BGPS images, prior to the final data release (Aguirre et al. 2009, in preparation).  Several of the sources are not included in the final BGPS catalog \citep{rosolowsky09}, which is designed to minimize false detections, however sources were confirmed by detection of morphologically similar emission in the  $^{13}$CO(J=1-0) channel maps of the Boston University-Five College Radio Astronomy Observatory Galactic Ring Survey \citep[GRS]{jackson06}. In order to extract flux densities for our objects, we reprocessed the images using the BGPS catalog pipeline, seeding the source identification portion of the catalog with the by-eye positions.  This enabled the catalog routine to extract source properties that directly correspond to the fields observed.  Source properties are determined as per the BGPS catalog; the integrated flux densities listed in Table \ref{selectsrc} are determined using the same method as the flux density $S$ in \citet{rosolowsky09}. The quoted uncertainties for the integrated flux densities shown in Table~\ref{selectsrc} are due to the photometry only; an additional 11\% error should be added to account for the uncertainty in the absolute flux calibration (10\% random and 5\% systematic, Aguirre et al. 2009, in preparation).
 
Distances to BGPS-selected molecular clumps were determined using
velocities of the corresponding $^{13}$CO emission. The velocity $v_R$ of
each source was measured from the GRS $^{13}$CO channel maps, taking
 the mean velocity of the associated $^{13}$CO cloud at the source position.
The distances of the clumps
were calculated assuming a flat Galactic rotation curve, and circular
orbits around the center of the Galaxy. IAU values were assumed for
the radius of the Sun's orbit around the galaxy $R_{\odot}$, and the
velocity of the Sun around the galaxy $v_{\odot}$
\citep[$R_{\odot}=8.5 \pm 1.1$ kpc, $v_{\odot}=222 \pm 20$ kms$^{-1}$,
][]{kerr86}. Uncertainties of the calculated
distances were estimated using a Monte Carlo simulation, assuming
Gaussian errors for $R_{\odot}$, $v_{\odot}$ and the measured velocity
$v_R$.   

Following a similar method to that outlined in \citet{roman-duval09}, distance ambiguities were resolved by examining HI channel maps from the VLA Galactic Plane Survey \citep[VGPS,][]{stil06} towards each source. Two methods of distance determination were used. Firstly, the HI channel maps were inspected for HI self-absorption at the velocity assigned to each millimeter clump. Cool HI in the outer layers of clouds can absorb emission from the warm diffuse galactic HI background; clouds at the near distance should show HI self-absorption of diffuse HI emission at the far distance, whereas clouds at the far distance will not since there is no background to absorb. It should be noted, however, that the presence of 21\,cm continuum emission from HII regions in the cloud in question can provide a background to absorb even at the far distance. Therefore, a second method of distance determination was employed for sources that displayed HI absorption at the same velocity as the cloud, but also significant 21\,cm emission with the same morphology. This continuum emission can also be absorbed by clouds in the line of sight. If the cloud is at the near distance, the HI channel maps and profile will exhibit absorption features at velocities only less than that of the cloud, and if the cloud is at the far distance, there will be absorption features present up to the tangent point velocity \citep[see Figure 2 of][]{roman-duval09}. 
Table\,\ref{selectsrc}, column\,9, indicates whether the near or far distance has been chosen based on the above methods, indicating whether presence or lack of HI self-absorption (HISA) or 21\,cm continuum absorption (CA) was the deciding factor. 

We were able to determine the velocities and distances to the clumps associated with  IRAS\,18424-0329 and IRAS\,18571+0349 using the above methods, as they have corresponding GRS and VGPS data. The clumps toward IRAS\,18424-0329 were found to be at the far distance. The distances for the clumps toward IRAS\,18571+0349 could not unambiguously be determined using the above methods, however HI observations by \citet{kuchar94} found the molecular cloud towards these sources to be at the far distance. Therefore the luminosities and masses quoted in B06 for these sources are accordingly scaled to the far distance.

Several of the selected clumps lie within the GRS clouds with distance determinations listed in \citet{roman-duval09}. These clouds are: GRSMC\,G029.14-00.16, GRSMC\,G048.59+00.04, GRSMC\,G048.84+00.24, and GRSMC\,G050.29-00.46, which were all determined to be at the far distance, in agreement with our results.

The luminosities of IRAS sources associated with BGPS-selected millimeter clumps 
(at the distances given in Table\,\ref{selectsrc}) were calculated by first 
integrating the flux under the four 12 to 100~${\mu}$m IRAS fluxes, and 
then adding the integrated flux of a blackbody peaking at 100~${\mu}$m
for wavelengths longer than 100~${\mu}$m.
When an upper limit was given for any of the four
IRAS fluxes, the upper limit to the integrated flux was found by
integrating under these values, and the corresponding lower limit was
then found by setting the fluxes in question to zero before
integration.   When two values
for the IRAS luminosity are given in Table~\ref{selectsrc}, separated
by a dash, they denote the lower and upper limits to the luminosity
respectively. Uncertainties in the IRAS luminosity due to the distance
uncertainty are not given in Table\,\ref{selectsrc}, however these can
be found by scaling the given luminosities to the upper and
lower distance limits.  

The masses of the observed BGPS clumps were found using the equation

\begin{equation} M_{gas} = \frac{g S_{\nu} d^2}{B(\nu,T) \kappa_{\nu}} \end{equation}

\noindent where $g$ is the gas-to-dust ratio, taken to be 100, $S_{\nu}$ is the
integrated flux density, $d$ is the clump distance, $B(\nu,T)$ is the
black body function, which is a function of frequency $\nu$ and
temperature $T$, and  $\kappa_{\nu}$ is the frequency-dependent
opacity. We have assumed a temperature of 30\,K for all sources, as a
compromise between the dust and NH$_3$ temperature results of \citet{sridharan02}, who
found an average dust temperature from IRAS emission towards their 
selection of high-mass stellar objects of 40\,K, and an average rotation
 temperature from (1, 1) and (2, 2) lines of NH$_3$ of 20\,K.
 Following \citet{enoch06}, we assumed an opacity of 1.14~cm$^{2}$g$^{-1}$ at
1.1~mm, derived from the results of \citet{ossenkopf94}.

The uncertainties in the 1.1\,mm flux density vary from $\sim$ 10-60\%. The temperatures of the observed clumps also may lie between 15 and 60\,K \citep[from both NH$_3$ and dust temperature measurements from][]{sridharan02}. Assuming the Rayleigh-Jeans approximation holds, mass is approximately inversely proportional to temperature. Therefore the maximum error in the temperature, which is assumed to be 30\,K, is 100\% when T=60\,K, and will give rise to the same error in the mass. Table~\ref{selectsrc} shows that the uncertainty in the clump distances lies in the range 10-20\%, therefore contributes an error of 40\%. The opacity is uncertain by a factor of two or less \citep{ossenkopf94}, and the gas-to-dust ratio has been derived, using extinction and gas-phase abundance measurements respectively, to be approximately 100 and 140 \citep{draine07}, giving an error of at least 40\%. Therefore the calculated masses are accurate to within a factor of approximately 2, as the uncertainty is dominated by the uncertainty in the opacity.

\section{OBSERVATIONS \label{observation}}

The selected millimeter sources were observed on two occasions with the D~configuration of the National Radio Astronomy Observatory VLA, in 3.6~cm (8.4~GHz) continuum mode on 2007 April 6 and  on 2007 May 1, under the program AS895. VLA continuum mode consists of four bands, each having 50~MHz bandwidth; two of these bands are centered on a frequency of 8.435~GHz, and the remaining two are centered on 8.485~GHz. The baseline lengths in D array range from 35~m to 1.03~km, resulting in the largest angular scale of observable structure being approximately 3' at 3.6~cm. However, note this value was lower in some instances due to flagging of the shortest baselines during data reduction. The average angular resolution of our final images is 8.35". 

The on-source integration time for each observation was approximately 30 minutes, which corresponds to a theoretical RMS noise of $\sim30~\mu$Jy\,beam$^{-1}$, and $\sim20~\mu$Jy\,beam$^{-1}$ in overlap regions of mosaiced fields. Twenty-six antennas were available for the first set of observations, and twenty-five were available during the second. The data were taken during the transition to the new VLA correlator 
controller, which resulted in there being a larger fraction of data than
expected with poor phases or amplitudes that had to be flagged, decreasing 
the final sensitivity in the images. As well as flagging to remove 
erroneous data, some of the shorter baselines were flagged to improve the 
final images, when bright, large-scale structure could not be adequately 
imaged. The percentage of data flagged for each field ranged from 33 to 
47\%. These values include $\sim$10\% of the data that was previously flagged by the on-line system.

The pointing centers for most target fields were shifted between the April 6 and May 1 observations (presented in Table \ref{obssum}). This was done primarily to move several bright sources on the edge of the VLA fields closer to the phase center of the map, to improve image deconvolution. For most of the observations, 1922+155 was used as a phase calibrator, with the exception of the source IRAS 18256-0742, which used 1822-096, and IRAS 18424-0329 and IRAS 18586+0106 which used 1832-105. The fluxes found during calibration for 1822-096, 1832-105 and 1922+155 were 1.34, 1.43 and 0.68~Jy, respectively. 

Data reduction was carried out using the Common Astronomy Software Applications (CASA)\footnote{http://casa.nrao.edu} package. The phase calibrator 1922+155 was extended, and therefore self-calibration was performed on 1922+155 before deriving and applying phase calibration to the target source data. The primary flux calibrator was 1331+305 (3C286), which has a flux of 5.23~Jy at a wavelength of 3.6~cm. We used a model of 1331+305 for flux calibration, as this source was slightly resolved, allowing data at all UV distances to be used for flux calibration. The error in absolute flux calibration is approximately 2\%. All of the data, with the exception of IRAS~18256-0742 and IRAS~18586+0106, were imaged with CASA's multi-scale mosaic deconvolution routine, which is based on the CLEAN algorithm, using either three or four scales. Imaging of the remaining two sources was performed using multi-scale CLEAN, as they share the same pointing center in both observations.

\section{RESULTS \label{results}}

Panels a) and b) of Figures \ref{G44587} to \ref{IRAS18586} present images of the VLA 3.6~cm continuum emission tracing ionized gas, and the millimeter continuum emission from the BGPS and B06 tracing warm dust, in the 10 observed fields. Panel a) of Figures \ref{G44587} to \ref{IRAS18586} presents the observed VLA 3.6~cm fields in grayscale and contours. Panel b) of Figures \ref{G44587} to \ref{IRAS18586} compares the 3.6\,cm images, in contours, to the corresponding BGPS 1.1~mm or SEST SIMBA 1.2~mm images (from B06), shown in grayscale. The BGPS 1.1~mm images have been Gaussian smoothed using $\sigma=$1.5 pixels. 

Hereafter, sources selected from the BGPS will be quoted in the format G44.587, as this is sufficient to discriminate between them. Several of the observed BGPS sources are close enough to one another to be shown in the same figure; these are G44.587 \& G44.661, G48.580 \& G48.616, G48.598 \& G48.656, and G50.271 \& G50.283. 

The noise in the 3.6~cm images was found not to be Gaussian, due to poor UV coverage of extended and very bright emission in the observed fields (causing stripes in the deconvolved images). Therefore an estimate of the noise in the image $\Delta S$ was taken to be above any residual artifacts. The 3.6~cm contour levels presented in Figures \ref{G44587} to \ref{IRAS18586} are given in multiples of $\Delta S$. 

\subsection{Positions, Fluxes and Angular Sizes of Detected 3.6~cm Sources \label{results1}} 

In the 10 observed fields, 35 HII regions are identified, of which 20 are newly discovered. Table \ref{vlaresult} lists the integrated flux density, peak position, peak flux density, angular size, position angle, and solid angle for each VLA source detected above 5$\times \Delta S$ in the observed fields. Also listed in the final column of Table \ref{vlaresult} are any pre-existing identifiers for each VLA source. For newly discovered radio continuum sources, this column contains the flag ``New". Sub-sources are denoted by ``A,B,C...", and components of sources which are fully or nearly unresolved (e.g., point-like), are denoted by ``-P."

With the exception of point-like sources (-P), measurements of the integrated and peak flux density were carried out using a custom-made irregular aperture photometry program. The integrated flux density was measured using a 1$\times \Delta S$ cut-off, within apertures placed so that they included all of the 1$\times \Delta S$ contour for each source. Errors in the aperture fluxes were calculated to be a combination of the error due to the image noise over the aperture, and the VLA absolute flux error, which is 2\% for our observations. Similarly, the peak flux error was found by combining the 1$\times \Delta S$ flux density with the VLA absolute flux error. 
 
The apertures for source sub-components were unavoidably arbitrary, as they cut across 1$\times \Delta S$ contours or higher. Therefore we have provided Figure \ref{polygons} to show the chosen apertures for sub-sources, denoted by ``A,B,C..." in Table \ref{vlaresult}. In addition to the noise and absolute flux errors mentioned above, which are accounted for and quoted for each sub-component in Table \ref{vlaresult}, an additional 10-20\% uncertainty in the integrated flux density should also be included for sub-sources, to account for the arbitrary placing of apertures.

Values for the solid angle quoted in Table \ref{vlaresult} are given for irregular or extended sources only, and are derived from the number of pixels above the 1$\times \Delta S$ within each photometry aperture.

For all extended sources, the angular size of the source at the 3$\times \Delta S$ level was measured by taking the major axis of the source to be along the direction which it is most extended, to within a position angle of 10 degrees. 

The peak and integrated flux densities, the angular size, and the position angle of fully or nearly unresolved sources (-P) were determined using the AIPS task IMFIT. 

The reported peak positions are measured from the peak pixel of the source, except for fully or nearly unresolved sources, whose peak positions are determined from the peak in the fitted Gaussian.

Due to the removal of some of the shortest baselines (c.f. \S \ref{observation}), and the fact that VLA D array observations are initially sensitive to size scales less than 3', some of the measured 3.6\,cm source flux densities may suffer from missing extended flux. This would have an impact on the derived properties of these sources by, for example, underestimating the luminosities and masses of their exciting stars (see Sections \ref{hII_discussion} and \ref{em_clump_mass}). 

\subsection{Analysis of Individual Fields \label{notes}}

Panel c) of Figures \ref{G44587} to \ref{IRAS18586} presents a three-color (Red: 8$\mu$m, Green: 4.5$\mu$m, Blue: 3.6$\mu$m) \textit{Spitzer} IRAC \citep{fazio04} GLIMPSE \citep{benjamin03} image of the observed fields, or an MSX \citep{price01} A Band (8.28$\mu$m) image for the two observed VLA fields not covered by the GLIMPSE survey, overlaid with contours of VLA 3.6~cm emission. 

Table~\ref{assoc_2} provides a summary of the evolutionary indicators associated with each millimeter clump. Column 2 of Table~\ref{assoc_2} lists whether there is 3.6\,cm emission detected within 60" of each clump peak position, and columns 3 to 7 list whether maser emission, dense gas tracers, GLIMPSE or MSX mid-IR emission, outflows, or IRAS sources have been detected towards each clump. The final column of Table~\ref{assoc_2} provides the IRAS luminosities of the associated IRAS sources, calculated using the method outlined in Section~\ref{mmprops}.

In the following subsections, Spitzer GLIMPSE point-source catalog names have been shortened to start with "GL" (e.g., SSTGLMC~G044.6598+00.3503 is shortened to be GL044.6598+00.3503).  A single source from the GLIMPSE Archive (indicating the photometry is less reliable) is indicated with the full name SSTGLMA G050.3179-00.4186.

\subsubsection{G44.587 \& G44.661}

The G44.587 \& G44.661 field (Figure \ref{G44587}) contains two 3.6~cm sources. VLA~1 is in the far S-W of the image, and is associated with the millimeter source G44.521. VLA~2 is found near the centre of the field, close to the millimeter source G44.617. There is no significant ionized emission found towards either G44.587 or G44.661.

The GLIMPSE image of the G44.587 \& G44.661 field, presented in Figure \ref{G44587}c), shows that there are clusters of mid-IR sources and diffuse 8.0\,$\mu$m emission associated with each of the three millimeter clumps. The inset panels show the areas marked by white boxes in the main panel, which surround the millimeter sources (corresponding from top to bottom) G44.661, G44.587, and G44.521.

The GLIMPSE source GL044.6598+00.3503, associated with the millimeter clump G44.661, and marked by a light blue circle in Figure \ref{G44587}c), has both IRAC colors [3.6] - [4.5] and [5.8] - [8.0] greater than 0.6 suggesting that the source is deeply embedded \citep{allen04}. There are also two extended mid-IR sources to the North-West and South of GL044.6598+00.3503. 

The millimeter clump G44.587 is associated with a cluster of GLIMPSE sources which have positive IRAC colors or rising mid-IR SEDs: GL044.5823+00.3689, GL044.5901+00.3697, GL044.5914+00.3689, and GL044.5802+00.3660 (marked by a yellow, light green, red, and turquoise circle respectively). This is also the case for G44.521, which is associated with GL044.5215+00.3902, GL044.5223+00.3858, GL044.5206+0.3866, and GL044.5231+00.3828 (marked by dark green, orange, dark blue, and magenta circles respectively). Both the reddened and clustered nature of these sources suggest that the millimeter clumps G44.521, G44.587, and G44.661 are associated with star forming clusters. However, the lack of ionized emission towards these clusters suggests that their members are either low-mass or very young.

\subsubsection{G48.580 \& G48.616}

Figure \ref{G48580} shows that there are six sources detected in the G48.580 \& G48.616 field with 3.6~cm emission greater than 5$\times \Delta S$. The emission is dominated by the HII region VLA~5, the most extended source in the field; the integrated flux of VLA~5 is 3.23~Jy, constituting $\sim 3/4$ of the total emission from sources in the field brighter than 5$\times \Delta S$. Both VLA~4 and VLA~5 consist of several components. VLA~4 has a compact component to the South, (VLA~4A) and an extended component to the North (VLA~4B). VLA~5 consists of an extended Western component, approximately extended along the NE-SW direction (VLA~5A), two compact components which also include some fainter diffuse emission (VLA~5C and VLA~5D being the NE and SE components respectively), and a component which extends from VLA~5D to the South-East: VLA~5B. The edge of VLA~5A is associated with the millimeter source G48.580. The sources VLA~5B,  VLA~5C, and VLA~5D are associated with the millimeter clump G48.605.

VLA~5 has been observed previously by many authors; it is known most commonly as G48.6+0.0 or IRAS~19181+1349. HI, OH and H$_2$CO absorption have been detected towards VLA 5 \citep{caswell75, silverglate78, wilson78, downes80, kuchar94}, placing this source at the far kinematic distance, between 10.3 and 12.0 kpc.  This is consistent with our determination that the millimeter sources G48.580 and G48.605, are at the far distance ($d\sim10$\,kpc, see Section \ref{mmprops}).  

VLA 5A, 5B, 5C, and 5D have been observed in radio continuum observations by \citet{matthews78}, \citet{zoonematkermani90}, and \citet[][using the VLA D array at 3.6~cm, see their Figure 7]{kurtz99}. Although the integration time of our original D array observations of this field was four times longer than those of \citet{kurtz99}, we obtained the same sensitivity in our final image (1.2 mJy\,beam$^{-1}$). Nevertheless, we show our image as it presents a larger field of view which covers both millimeter sources G48.580 and G48.616. 

In addition, 5C and 5D were observed at higher resolution by \citet{kurtz94}, at 2 and 3.6~cm with the VLA in B array. We calculated the spectral index of VLA 5C and VLA 5D between 2 and 3.6~cm, using only the B array flux densities given in \citet{kurtz94}, as they result from the most similar UV coverages. We take $S_{\nu} \propto \nu^{\alpha}$, where $\alpha$ is the spectral index. The spectral index of VLA 5C, named G48.609+0.027 by \citet{kurtz94}, was found to be -0.4$\pm$0.2, which is consistent with mixed thermal and non-thermal (synchrotron) emission from ionized gas. The spectral indices of the two components of VLA~5D resolved by the VLA B array observations, G48.606+0.023 and G48.606+0.024, were found to be 0.2$\pm$0.1 and 0.3$\pm$0.5 respectively, which are consistent with an ionized wind or optically thick free-free emission. However, in the case of G48.606+0.024, the larger error also allows the spectral index to be explained by optically thin free-free emission. 

\citet{de-buizer05} have performed a high resolution ($\sim$1-2") mid-infrared survey towards sites of water maser emission, in which mid-IR emission was detected towards VLA~5C and VLA~5D. Both water and hydroxyl masers have been detected toward VLA~5D \citep[e.g.,][]{forster89}, and a wealth of molecular tracers have also been observed  \citep[e.g. CS(J=7-6) and CO(J=3-2)][]{plume92}. \citet{mueller02} have also detected the source in the far-IR, at 350$\mu$m. These detections, in combination with the coincidence of VLA~5D with the millimeter source G48.605, suggest this source is at a very early stage of evolution.

Figure \ref{G48580}c) presents a GLIMPSE image of the G48.580 \& G48.616 field. The crosses mark the positions of the millimeter clumps (in increasing R.A.: G48.616, G48.540, G48.580, and G48.605). The inner panel shows the sources VLA~5B, 5C and 5D, and covers the area shown by the black box in the main panel. Comparison between the GLIMPSE and VLA 3.6~cm images reveals that VLA~5B is associated with the GLIMPSE source GL048.6021+00.0257, and VLA~5C is associated with the GLIMPSE source GL048.6093+00.0270 (whose positions are shown by a yellow and blue circle respectively in Figure \ref{G48580}c). There are no mid-IR IRAC sources directly associated with the peak of the compact HII region VLA~5D, in fact the source appears to lie within a dark filament. However, VLA~5D appears to have associated 4.5 $\mu$m emission (green) extending in the N-S direction. Emission in the 4.5 $\mu$m band is thought to be produced by shocked H$_2$ or CO gas in outflows \citep[see][and references within]{cyganowski08}. If we are seeing shocked gas from the outflow of this source, this provides further evidence towards its youth, and suggests it may still be in the process of outflow and accretion.

In summary, there are several newly discovered HII regions in the field, for example VLA~1 and VLA~3. However these HII regions are not coincident with any BGPS clumps. In contrast, the extended bright HII region VLA~5 contains several compact components: VLA~5B, C and D, of which two have spectral indices that can not be explained by optically thin free-free emission expected from classical HII regions. There are also many indicators of (massive) star formation towards these components, in particular VLA~5D, including mid-IR, mazer, and 350$\mu$m emission. Therefore we conclude that VLA~5 is a current site of massive star formation.

\subsubsection{G48.598 \& G48.656}

Figure \ref{G48598} shows there are six 3.6~cm sources in the G48.598 \& G48.656 field. VLA~1 is a diffuse extended source, lying along the NW-SE direction. VLA~2 is an unresolved source, found to the South of VLA~1. VLA~3 is again an unresolved source, and VLA~4 is a roughly circular source to the West of VLA~5. VLA~5 is a complex extended source, with 6 components (A to F in increasing R.A., which are labeled in Figure \ref{polygons}). Component VLA~5C is composed of a brighter unresolved source surrounded by fainter extended emission. VLA~6 is also unresolved, with the brightest peak flux in the field (25.5 mJy\,beam$^{-1}$).

VLA~4 is coincident with the millimeter source G48.634, and  VLA~5D is coincident with G48.656. The millimeter source G48.610 is also associated with slightly resolved 3.6~cm emission at the 4$\times \Delta S$ level.

 \citet{sridharan02} have carried out VLA B array observations of this field at 3.6~cm, with a sensitivity of 0.1 mJy\,beam$^{-1}$. As the 3.6~cm images were not presented in \citet{sridharan02}, we downloaded the data from the NRAO archive\footnote{http://archive.cv.nrao.edu/} (observation date: 1998 July 2, project code: AS643) and imaged the field around the observed source 19175+1357, which is coincident with G48.634. Sources VLA~5C and VLA~6 were detected in their B array observations; Figure \ref{a21a25_sridharan} presents both B and D array observations of these sources in its right hand panels, with the B array image shown in contours, and the D array image in grayscale. The subcomponent VLA~5C, which is observed as an unresolved source in the D array observations at 3.6~cm, is resolved into a double-peaked source in the B array observations, the North-West lobe being its brightest component. VLA~6 is also detected as faint ($\sim$5-$\sigma$) emission in the B array observations, which contains two 5-$\sigma$ peaks.

The distance toward G48.598 and G48.656 has been determined, using observations of radio recombination lines (H110$\alpha$ and H138$\beta$), and H$_2$CO and HI absorption, to be between 10.5 and 12.5 kpc \citep{watson03, kuchar94, planesas91}. This is consistent with our determination that these sources are at the far distance, with  distances of $10.6$ and $10.3$\,kpc respectively (see Section \ref{mmprops}).  

\citet{beuther020} has observed the field at 1.2~mm using the MAMBO bolometer array at the IRAM 30~m telescope. Comparing their Figure 1, panel 19175+1357, to the Bolocam observations of the field shown in Figure \ref{G48598} of this work, one can see that very similar dust structures are traced by the two sets of observations. The MAMBO resolution of 11" at 1.2~mm resolves both millimeter sources G48.634 and G48.610 into two components; these are identified by \citeauthor{beuther020} as sources 1 and 2, and 4 and 5 respectively. Source 1 is coincident with the HII region VLA~4, and G48.598 and G48.656 correspond respectively to sources 6 and 7. \citet{williams04} have observed G48.634 at 450 and 850\,$\mu$m, detecting two sources: WFS70 and WFS71, whose positions are proximate to the \citeauthor{beuther020} sources 19175+1357:\,2 and 1 respectively (with offsets of 2.9" and 6.2"). 

Figure \ref{G48598}c) shows a three-color \textit{Spitzer} IRAC GLIMPSE image of the G48.598 \& G48.656 field, overlaid with contours of the VLA 3.6~cm emission. The white crosses mark the positions of the millimeter clumps (in increasing R.A.: G48.598, G48.610, G48.634, and G48.656). The two right panels show the areas of the \textit{Spitzer} IRAC image marked by white boxes in the left panel, and are also overlaid with contours of 3.6~cm emission. The bottom right panel shows the IRAC emission towards G48.634. There are two extended sources shown in this panel, one to the North which is associated with VLA~4, and a smaller extended source to the south, which is coincident with 19175+1357:\,2, and WFS70. By studying the separate IRAC band images, the Northern source was found to contain three point sources superimposed upon the extended emission, at approximately 19$^h$19$^m$49$^s.$2 +14$^{\circ}$02'49", 19$^h$19$^m$49$^s.$0 +14$^{\circ}$02'46" and 19$^h$19$^m$48$^s.$6 +14$^{\circ}$02'52" (J2000). The top right panel shows a close up of the IRAC emission towards VLA~6. The mid-IR emission associated with this source is mostly extended, and is therefore not in the GLIMPSE catalog. However there is a central point source seen in IRAC band 1, at 19$^h$19$^m$55$^s.$5 +14$^{\circ}$04'58" (J2000). 

The GLIMPSE source GL048.6557+00.2285 appears to be associated with VLA 5D, and the source GL048.6024+00.2394 is close to VLA 2. The position of these two GLIMPSE sources are shown in Figure \ref{G48598}c) by a blue and yellow circle respectively. In addition, two extended GLIMPSE sources are associated with the millimeter source G48.598, near VLA~2, with positions 19$^h$19$^m$41$^s.$6 +14$^{\circ}$01'19" and 19$^h$19$^m$42$^s.$1 +14$^{\circ}$01'24".

By comparing the 3.6\,cm, 1.1\,mm and GLIMPSE images, we have shown there are several sites of massive star formation in the G48.598 \& G48.656 field. Two examples are VLA~4 and VLA~5D, which are coincident with several signposts of massive star formation (e.g. dust continuum, mid-IR and ionized gas emission).

\subsubsection{G48.751}

There is one faint (F$_{peak}$=0.431 mJy\,beam$^{-1}$) VLA 3.6~cm source detected above 5$\times \Delta S$ in the G48.751 field shown in Figure \ref{G48751}. This source is not associated with any of the detected Bolocam millimeter sources in the field. There is a nearby IRAS source, IRAS 19191+1352, however it is not related to any millimeter or 3.6~cm continuum emission. The extended GLIMPSE source GL048.7750-00.1507 (whose position is shown by a yellow circle in Figure \ref{G48751}c) is  $\sim 20"$ from the peak of G48.771, but still appears to be associated with the extended millimeter dust emission towards this source. There is also extended 8$\mu$m emission covering a large proportion of the bottom half of Figure \ref{G48751}c), which extends to the South of G48.751 and VLA~1. Therefore it appears that the G48.751 field contains two mostly quiescent clumps, one of which may now contain a protostar shown in the mid-IR by GL048.7750-00.1507.

\subsubsection{G49.912}

The 3.6~cm image of the G49.912 field shown in Figure \ref{G49912} uncovers no significant emission from ionized gas towards G49.912. However there is a double-lobed 3.6~cm source, VLA~1, in the S-W of the field, which is associated with the millimeter source G49.830.

Figure \ref{G49912}c) shows a three-colour GLIMPSE image of the region. The white crosses mark the positions of the millimeter clumps (in increasing R.A.: G49.830 and G49.912). Figure \ref{G49912}c) shows that there is a cluster of mid-IR sources associated with G49.912 (see inset panel, which covers the area marked by the top white box in the main panel). The most conspicuous source in the cluster at these wavelengths is an extended mid-IR source at 19$^h$21$^m$48$^s.$1 +15$^{\circ}$14'32" (J2000). There are several other mid-IR sources in the cluster having IRAC colors which suggest they are young objects. The sources GL049.9113+00.3719 and GL049.9134+00.3723 (indicated by a yellow and green circle respectively in Figure \ref{G49912}c) both have IRAC [3.6] - [4.5] and [5.8] - [8.0] colors greater than 0.6 showing that they are deeply embedded. This, in combination with the clustering of these sources and the presence of an associated millimeter clump, provides evidence that these stars constitute a young star forming cluster. However, as we observe no significant ionized gas emission towards G49.912, it appears that there are either no stars in the cluster massive enough to produce significant free-free emission, or that any forming massive stars have not yet reached the main sequence.

Emission at 8\,$\mu$m can originate from Polycyclic Aromatic Hydrocarbons (PAHs) which are thought to fluoresce at the boundaries of HII regions \citep{churchwell04}, and, indeed, the 8\,$\mu$m emission associated with the HII region VLA\,1A \& 1B traces the edges of the two observed lobes of ionized gas (Figure \ref{G49912}c, right panel).  The presence of PAH emission near the boundaries of this bipolar, ionized gas structure suggests that VLA\,1A \& 1B represent the bipolar, ionized outflow lobes from a massive star or star cluster rather than an extragalactic radio galaxy. The morphology of IRAC emission is also similar to the bipolar bubble (S97 and S98) observed by \citet[][see their Figure 2f]{churchwell06}. 

There is an extended mid-IR source lying between the two lobes of the VLA~1, slightly offset to the West of the axis of this bipolar source, which the shape of the 3.6~cm emission can be seen to follow. It may be that this extended mid-IR source marks the current site of star formation towards VLA~1. 

Therefore the presented observations reveal that there is star forming cluster embedded in the clump G49.912 which may only be forming low-mass stars, or has not yet formed a massive star detectable in our 3.6\,cm observations. In contrast, the clump G49.830 is coincident with an evolved bipolar HII region, however there is a mid-IR source which may be the site of current star formation in this clump.

\subsubsection{G50.271 \& G50.283}

There are four 3.6~cm continuum sources detected above 5$\times \Delta S$ in the G50.271 \& G50.283 field (Figure \ref{G50271}). VLA~1 dominates the integrated flux in the region; the source has an extended peak which is elongated in roughly the E-W direction, and two fainter lobes. The main lobe extends to the North, and the second to the S-E.   The source VLA~2 is to the S-E of VLA~1, detected at the 5$\times \Delta S$ level. VLA~3 is a double-lobed source, and VLA~4 is an unresolved source in the far S-E of the field.

VLA~1 is coincident with the millimeter source G50.283, however the peak of the ionized gas emission is  slightly offset (14") from the millimeter peak. There is no significant 3.6~cm emission associated with the other millimeter source in the field, G50.271.

Using H110$\alpha$ recombination line and H$_2$CO absorption observations, \citet{watson03} found the distance to VLA~1 to be 10.0 kpc. This is consistent with our determination that the source is at a far distance of $\sim 9.6$\,kpc (see Section \ref{mmprops}).

Figure \ref{G50271}c) shows a  three-color GLIMPSE image of the G50.271 \& G50.283 field, overlaid with contours of the VLA 3.6~cm emission. The white crosses mark the positions of the millimeter clumps (in increasing R.A.: G50.283, and G50.271). The inset panel shows a close up of the IRAC emission associated with G50.283 (also VLA~1), the area of which is marked by the white box in the main panel. From Figure \ref{G50271}c) we see that the 8\,$\mu$m PAH emission associated with VLA~1 traces out the edges of the Northern lobe of ionized gas, which is similar to the IRAC emission associated with VLA~1 in Figure \ref{G49912}c). 

The GLIMPSE source GL050.2830-00.3904 (indicated by a blue circle in Figure \ref{G50271}c), which is located within the HII region VLA~1, is coincident with the peak of G50.283. There is also a diffuse 4.5 $\mu$m extension of GL050.2830-00.3904 that points into the main Northern lobe, which may be either tracing a less luminous source, or instead is shocked emission from an outflow associated with GL050.2830-00.3904. Thus, the mid-IR emission from G50.283, in conjunction with the existence of warm dust and ionized gas shown by the $\sim$1~mm and 3.6~cm images, points to this being a site of current massive star formation. There is a fainter point source, not listed in the GLIMPSE catalogs, at 19$^h$25$^m$18$^s.$49 +15$^{\circ}$12'28.5" (J2000), which is closer to the peak position of VLA~1 than GL050.2830-00.3904. It is possible that this instead is the source creating the HII region VLA~1. A saturated GLIMPSE source, SSTGLMA G050.3179-00.4186 (indicated by a red circle in Figure \ref{G50271}c), is associated with VLA~3B. There is also diffuse IRAC emission near the millimeter source G50.271; however, as there are no associated IRAC point sources and no coincident ionized gas emission, this clump appears to be quiescent, or at an early stage of star formation.

Therefore we conclude that VLA~1, which is coincident with the millimeter clump G50.283, is a site of current massive star formation, and that the clump G50.271 is quiescent or in a very early stage of star formation.

\subsubsection{IRAS 18256-0742}

The VLA 3.6~cm image of the IRAS~18256-0742 region, shown in Figure \ref{IRAS18256}, contains five sources. VLA~1,  3 and 5 are compact sources, whilst VLA~2 and 4 are extended. VLA~2 is associated with the millimeter Clump 1 in the IRAS~18256-0742 field observed by B06. \citet{molinari98} have previously detected the source VLA~5 at 6~cm using B configuration VLA observations. 

As there is no GLIMPSE coverage of this field, an MSX A band image of the field at 8.28$\mu$m is shown in Figure \ref{IRAS18256}c). There is 8$\mu$m emission associated with VLA~2 and millimeter Clump 1. Therefore this clump is likely to be a current site of massive star formation. There is also a ``shell" of 8$\mu$m emission surrounding VLA~4, tracing PAH emission at the boundary of the HII region. As the HII region VLA~4 is approximately circular, has a diameter of almost a parsec, and is not associated with any dense gas traced by millimeter emission, we conclude that it is more evolved than VLA~2.

\subsubsection{IRAS 18424-0329}

Figure \ref{IRAS18424} shows that there are two 3.6~cm continuum sources detected in the IRAS~18424-0329 field. VLA~1 is a compact unresolved source in the South of the field, and VLA~2 is cometary in shape and lies close to the field center. VLA~2 is also associated with millimeter emission detected in the field (shown in the panel b of Figure \ref{IRAS18424}).  B06 find three millimeter clumps listed to be within $\sim$ 1' of VLA~2: Clumps 2, 4, and 6. The given positions of these three clumps are shown by small circles in the panel b) of Figure \ref{IRAS18424}. The peak of VLA~2 lies closest to Clump 2. \citet{becker94} have previously observed both VLA~1 and 2 at 1.4 and 5 GHz, and \citet{molinari98} have also detected VLA~1 at 6~cm. 

As can be seen from the GLIMPSE image of the region (Figure \ref{IRAS18424}c) there is diffuse 8.0$\mu$m PAH emission which follows the edges of VLA~2. VLA~2 has a calculated physical size, $\Delta s$, of 3.33 pc and is therefore a classical HII region. It may be that the massive star which created VLA~2 formed from part of the same cloud of molecular gas as the three existing clumps, which do not yet appear to have formed stars. It is also possible that VLA~2 is not physically associated with Clumps 2, 4, and 6 from B06, and that these clumps constitute a quiescent complex of molecular gas. There is no IRAC emission associated with the HII region VLA~1. 

\subsubsection{IRAS~18571+0349}

The 3.6~cm image of the field surrounding IRAS 18571+0349 (Figure \ref{IRAS18571}) shows that there are eight sources in the field detected at greater than 5$\times \Delta S$. All of the sources bar one, VLA~3, are extended. VLA~3 and VLA~5 are associated with the most massive millimeter clump in this field, Clump 1. In addition, VLA~7 is associated with the millimeter Clump 4. 

\citet{molinari98} have previously detected VLA~5 at 6~cm using the VLA in B array. HI absorption observations taken by \citet{kuchar94} indicate that the sources in this field are at the far distance ($d$=10.4 kpc). Therefore, we have assumed the far distances calculated from our measured GRS $^{13}$CO velocities in Section \ref{sourceselect} for all clumps in the field, and have scaled the clump properties given by B06 accordingly. 

The IRAC \textit{Spitzer} image of the region (Figure \ref{IRAS18571}c) shows a ridge of diffuse mid-IR emission, with condensations containing bright objects or clusters situated along it. The VLA 3.6~cm sources VLA~3, 5 and 7 are associated with GLIMPSE sources in this ridge. The compact source VLA~3 is associated with extended mid-IR emission at approximately 18$^h$59$^m$41$^s.$6 +3$^{\circ}$53'30" (J2000), and the peak of VLA~5 is associated with another mid-IR source with the position 18$^h$59$^m$43$^s.$0 +3$^{\circ}$53'37" (J2000). There is also a bright GLIMPSE source associated with VLA~5, GL037.3418-00.0591 (identified by a blue circle in Figure \ref{IRAS18571}c), which has IRAC colors [3.6] - [4.5]=2.31 and [5.8] - [8.0]=1.07, showing that this is a highly embedded object. The clustering of these sources, along with the associated ionized and molecular gas, suggest that this is a current site of clustered massive star formation.

Further to the North, there is also a dust condensation (Clump 4 in B06, associated with VLA~7) in a ``pillar" which extends from the ridge to the S-E. The ridge contains a GLIMPSE source (GL037.3815-00.0820, indicated by a yellow circle in Figure \ref{IRAS18571}c) that has a rising SED between 2.17$\mu$m (2MASS K band) and 8.0$\mu$m. Methanol masers have also been detected towards Clump 4 (VLA~7) by \citet{pandian07} within a 40" beam, suggesting star formation is currently occurring within it. It is interesting to note however that the peak of VLA~7 is not coincident with the GLIMPSE source, but instead is displaced to the East by 7.5". It may be that the object associated with GL037.3815-00.0820 is not creating an HII region, but instead a nearby star, such as the one directly West of the tip of the pillar, is ionizing molecular material in its proximity. The peak of the source VLA~8 is coincident with the GLIMPSE source GL037.3816-00.0922 (shown by a green circle in Figure \ref{IRAS18571}c).

Although unrelated to any diffuse mid-IR emission, VLA~2 and 4 appear to be related to the GLIMPSE sources GL037.3199-00.0641 and GL037.3241-00.0690 respectively (indicated by a red and orange circle in Figure \ref{IRAS18571}c). 

In summary, the field surrounding IRAS~18571+0349 contains several sites of massive star formation; in particular, those surrounding Clumps 1 and 4 observed by B06.

\subsubsection{IRAS 18586+0106}
There were no detections above 3$\times \Delta S$ (=0.48 mJy\,beam$^{-1}$) in the IRAS 18586+0106 field (Figure \ref{IRAS18586}). Deconvolution of the image was hampered, and therefore the RMS noise in the final image was increased, by a bright source (G35.20-1.74) on the edge of the field, which caused significant stripes through the image. \citet{molinari98} detected a faint source in this field at 6\,cm, denoted Mol~87, lying between Clumps 1 and 5 (19$^h$01$^m$15$^s.$271 +01$^{\circ}$11'00.04", J2000) with a peak flux density of 6.89 mJy\,beam$^{-1}$. 

There are no \textit{Spitzer} IRAC images which cover this field, however we provide an MSX A band image of the field shown in Figure \ref{IRAS18586}. The 8$\mu$m emission is coincident with the IRAS source, however these are both displaced from the peak of the nearest millimeter clump, Clump 5.

\subsection{Derived Properties of HII Regions\label{hII_discussion}}

Table \ref{vlaproperties} lists the derived physical properties of the detected VLA 3.6~cm sources given in Table \ref{vlaresult}. Properties of unresolved and partially resolved sources were derived assuming they are spherically symmetric, optically thin, homogeneous, and ionization-bounded HII regions. In the case of irregularly shaped sources, we assume they are optically thin, homogeneous, and ionization-bounded HII regions, and that the observed area in the plane of the sky is projected along a depth $\Delta s$, estimated from the geometrical mean of the size of the source on the sky. We do not take the effect of dust into our calculations, which means our calculated spectral types and luminosties are in fact lower limits. We also assume shocks in outflows do not contribute a significant amount to the ionized gas emission.

In Table \ref{vlaproperties}, the assumed distance $d$ in kpc for each 3.6~cm source was taken to be the distance of the millimeter source closest in projection on the sky within the observed fields. The effective angular diameter of the source $\Delta\theta$, given in degrees, was found by calculating the geometrical mean of the angular sizes given in Table \ref{vlaresult}. For unresolved and partially unresolved sources, it was necessary to correct the Gaussian angular diameter $\Delta\theta$ to the angular size of a sphere, using the multiplicative correction factors given in \citet{panagia78}.

The physical size $\Delta s$ in pc was calculated from $\Delta\theta$ and $d$. In the case of unresolved and partially unresolved sources, $\Delta s$ was taken to be the physical diameter of the spherical HII region. For irregular sources, by assuming that the size of the sources in the plane of the sky is similar to that along the line of sight, $\Delta s$ was taken to be the depth of the emitting region.

The brightness temperature $T_b$ was found using the Rayleigh-Jeans approximation:

\begin{equation} T_b = \frac{S_{\nu} 10^{-29} c^2}{2\nu^2 k \Omega_s}~\rm{[K]}, \end{equation}

\noindent where $S_{\nu}$ is the integrated flux density of the source in mJy, $c$ is the speed of light in ms$^{-1}$, $\nu$ is the frequency of the radiation in Hz, $k$ is the Boltzmann constant, and $\Omega_s$ is the solid angle covered by the source. For unresolved and partially unresolved sources, $\Omega_s = \pi \Delta\theta^2 /4$. For irregular sources, $\Omega_s$ was found by summing the area of all pixels associated with the sources with a flux density per beam above $\Delta S$, the significance level in the image.

The optical depth, $\tau$, was then calculated by solving the equation of radiative transfer $T_b=T_e(1-e^{-\tau})$ for $\tau$, assuming the emitting region is uniformly filled with $T_e = 8200$\,K ionized gas. The assumed value for the electron temperature was calculated from the results of \citet{quireza06}; the mean electron temperature of their observed sources was $T_e=8200 \pm 2330$ K. The calculated values for the optical depth given in Table \ref{vlaproperties} are all less than 0.1. Therefore, with the exception of the sources G48.587 \& G48.661 VLA~2-P, G48.580 \& G48.616 VLA~4A-P, IRAS~18256-0742 VLA~1-P, and IRAS~18256-0742 VLA~3-P, for which the angular size is an upper limit, and hence their properties should be viewed with caution, the assumption holds that the observed sources are optically thin.

The emission measure, EM, is given by:

\begin{equation} \rm{EM} = \frac{\tau}{8.235 \times 10^{-2}~\alpha(\nu,T_e)~T_{e}^{-1.35}~\nu^{-2.1}}~\rm{[cm^{-6}~pc]}, \label{EM} \end{equation}

\noindent where $T_e=8200$ K and $\nu$ is in GHz. The correction factor $\alpha(\nu,T_e)$, which is of order unity, rectifies the small discrepancy between the approximation shown in equation \ref{EM}, given by \citet{altenhoff60}, and the original derivation by \citet{oster61}. We take $\alpha(\nu,T_e) = 0.9828$ from Table 6 in \cite{mezger67}, for $T_e=8000$ K and $\nu=8$ GHz. 

For sources modeled as a spherical HII region, the number density of electrons, $n_e$, was estimated using

\begin{equation} n_e = \sqrt{\frac{3 ~ \rm{EM}}{2 \Delta s}}~\rm{[cm^{-3}]} \label{sphne} \end{equation}

\noindent where the factor of 2/3 is the ratio of the volume of a sphere to that of a cylinder. For irregular sources, the number density is given by:

\begin{equation} n_e = \sqrt{\frac{EM}{\Delta s}}~\rm{[cm^{-3}]}. \end{equation}

For a spherical HII region, the excitation parameter U can be calculated using the equation 

\begin{equation} U = rn_e^{2/3} = 4.553 {\left[ \frac{ \nu^{0.1} T_e^{0.35} S_{\nu} d^2}{\alpha(\nu,T_e)} \right]}^{1/3}~\rm{[pc~cm^{-2}]}, \end{equation}

\noindent which can be derived using equations \ref{EM} and \ref{sphne}, and by also assuming that the mean free path of photons within the HII region $r$ is equal to $\Delta s/2$. The frequency $\nu$ is in GHz, $T_e$ is in kelvin, $S_\nu$ is in Jy, and $d$ is in kpc.  An equivalent expression can be derived for the excitation parameter $U$ of irregular sources:

\begin{equation} U = 3.669 \times 10^{-2} {\left[ \frac{ \nu^{0.1} T_e^{0.35} S_{\nu} {\Delta s}^{2}}{\alpha(\nu,T_e) \Omega_s} \right]}^{1/3}~ \rm{[pc~cm^{-2}]}. \end{equation}

The Lyman photon flux required to sustain the HII region, $N_{Ly}$, can be estimated via the equation $N_{Ly} = V n_e^2 \alpha_{B}$, where $V$ is the volume of the nebula and $\alpha_{B}$ is the case B hydrogen recombination coefficient, which does not include recombinations to the ground level, as these are balanced by the diffuse radiation field due to recombinations of the ionized gas. The recombination coefficient $\alpha_{B}$ was taken to be 2.59$\times$10$^{-13}$\,s$^{-1}$cm$^3$ \citep[Table 2.1 of][for $T=10^4$\,K]{osterbrock06}.

Therefore, for both spherical and irregular cases, the ionization rate $N_{Ly}$ can be expressed as:

\begin{equation} N_{Ly} = 3.020 \times 10^{45}~\frac{T_e^{0.35} \nu^{0.1} d^2 S_{\nu}}{\alpha(\nu,T_e)}~\rm{[s^{-1}]}. \end{equation}
 
Assuming the observed HII regions are created by single OB stars, the corresponding spectral type and luminosity, L$_{\rm{cm}}$, of the exciting star required to sustain each observed HII region are listed in Table \ref{vlaproperties}. These were found using the results of \citet{panagia73}. 

Table \ref{vlaproperties} shows that there is a large range in the properties of the observed HII regions. For instance, their physical sizes extend from $<$0.05\,pc to 7.88\,pc, and their spectral types cover B2 to O5.

\subsubsection{Uncertainties on Derived HII region Properties \label{unc_vla_prop}}

Uncertainties on the properties quoted in Table \ref{vlaproperties} were calculated using a Monte Carlo error propagation code, assuming Gaussian errors for the 3.6\,cm integrated flux density, the 3.6\,cm source size, the assumed distance for each 3.6\,cm source, and the electron temperature $T_e$. The error in the source size was estimated to be 5" in the case of extended sources, and in the case of unresolved sources the errors given by IMFIT were assumed, which was usually $\lesssim5$\%. However the sources G50.271 \& G50.283 VLA~4-P; IRAS 18256-0742 VLA~3-P; IRAS 18256-0742 VLA~5-P; and IRAS 18424-0329:VLA~1-P have uncertainties in their sizes ranging from 10 to 40\%, and the size of IRAS 18256-0742: VLA~1-P has a 110\% uncertainty. The additional uncertainty on the integrated fluxes due to aperture placement (see Section \ref{results1}) was not included in the error analysis. 

The approximate uncertainties in the calculated HII region properties are as follows: uncertainty in $\Delta \theta \sim 9\%$, uncertainty in $\Delta s \sim 20\%$, uncertainty in $T_b\sim 15\%$, uncertainty in $\tau_{\nu}\sim 45\%$, uncertainty in EM $\sim 20\%$, uncertainty in $n_e$ and $U \sim 15\%$, and uncertainty in $\log_{10}{N_{Ly}} \sim 15\%$. Therefore the quoted spectral types are at least correct to within a spectral type, which is also the case for the corresponding luminosities. 

\section{Discussion\label{discussion}}

\subsection{Association of Molecular and Ionized Gas}
Table \ref{assoc} lists the nearest 3.6~cm source in projection to each millimeter clump, noting whether the peak of the associated ionized gas is observed within a radius of 60 arcseconds. Columns 4 and 5 of Table \ref{assoc} list the projected distances, in arcseconds and parsecs respectively, to the peak of the nearest 3.6~cm emission, and the final column provides the stellar luminosity required to create the nearest HII region within 60'', taken from the results in Table~\ref{vlaproperties}.

Figure \ref{ment} shows the distribution of millimeter clumps as a function of their projected distance from the nearest ionized gas in parsecs. Similarly, Figure \ref{mass_displ} shows the relationship between the mass of the millimeter clumps and their projected distance in parsecs from the nearest ionized gas. It can be seen from Figure \ref{ment} that the observed millimeter clumps tend to be associated with the ionized gas. This is not a artifact of our selection criteria, which selected millimeter clumps with little or no 21\,cm continuum emission toward them in the NVSS. In Figure \ref{mass_displ}, there appears to be no obvious correlation between the mass of a clump and the projected distance to the peak of the nearest ionized gas emission.

It can be seen from Figures \ref{G44587} to \ref{IRAS18586} that although the ionized gas tends to be associated with the millimeter clumps, there exist examples of HII regions which, if at the same distance as the observed clumps, appear to have formed at the edges of the dense gas. For instance, in the field G48.580 \& G48.616, the sources VLA~1, 2 and 3 do not appear to be associated with the millimeter emission in the field. This may be evidence that massive stars can also form at the edges of their parent molecular clouds. Therefore it would be interesting to carry out follow up observations of these sources to determine whether they do lie at the same velocities as the millimeter clumps in each field.

\subsection{Inferred Ages of the Observed Sites of Star Formation}

Single early B stars (M$_{\star} \sim$ 10 M$_{\sun}$) reach the Zero Age Main Sequence (ZAMS) in about $10^5$\,years whereas mid-O stars with M$_{\star} \gtrsim 25 M_{\sun}$ reach the ZAMS in $\sim 5 \times 10^4$\,years \citep[e.g.,][and references therein]{yorke04}.  In comparison, the time scale for the formation of O star clusters for which the most massive star in the cluster has a mass M$_{\star} > 25 M_{\sun}$ (roughly a mid-O star) appears to be $<$3\,Myrs \citep{massey95}.  Early B stars in these clusters continue to form at least 1 Myrs after the formation of the O stars.  

Several fields show examples in which HII regions are interspersed throughout the gas and dust traced by 1\,mm emission: the HII region G48.598 \& G48.656 VLA~5 (Figure \ref{G48598}) is produced by a cluster of early B to late O stars, and G48.580 \& G48.616 VLA~5 (Figure \ref{G48580}) by a late O star cluster.  The presence of such a well-defined cluster suggests that these regions are likely to be more evolved (since multiple ZAMS stars have formed already) and, indeed, the HII regions tend to be extended and/or complex.  Yet there is the potential for significantly more star formation to occur, as there are still large amounts of gas ($\sim$1000-5000 M$_{\sun}$) traced by warm dust emission that are able to continue to form massive stars in the cluster. Given the presence of ongoing massive star formation in these two clusters, their age is likely to be greater than $\sim 10^6$ years but less than 3\,Myrs.  

Several regions also have massive millimeter clumps that show only a singly peaked HII region forming at or near the center of a millimeter clump. Examples include G44.587 \& G44.661 VLA1 (Figure \ref{G44587}), that has the required luminosity of a early B star, and G50.271 \& G50.283 VLA1 (Figure \ref{G50271}) with the luminosity of a late O star.  The presence of a single HII region suggests that these sites have just started the massive star formation process and only a single star or a compact star cluster producing the HII region has reached the ZAMS so far. These regions are likely to be roughly $5 \times 10^4$ to $10^5$\,years old.  

A number of regions have millimeter clumps that show no ionized gas emission near the warm dust peak. Examples include the clump G48.616 with a mass of 484 M$_{\sun}$; G48.540 with a mass of 333 M$_{\sun}$ (both Figure \ref{G48580}), and G50.271 with a mass of 157 M$_{\sun}$ (Figure \ref{G50271}). Thus, these sources are candidate sites for the earliest stages of star formation in which any massive forming star has not yet reached the ZAMS.  If this is the case, they are likely to be less than $\sim 5 \times 10^4$\,years old. Alternatively, these millimeter sources could represent sites of more evolved low mass star formation. The sources G44.587 and G44.661 are good examples of this, where the GLIMPSE image of this field (Figure \ref{G44587}c) shows that there are clusters associated with these two millimeter clumps.

\subsection{Comparison of 1~mm and $^{13}$CO Images \label{1mm_13co_sect}} 

Figure \ref{1mm_13co} presents a comparison of the millimeter continuum emission (from BGPS or B06) tracing warm dust, the $^{13}$CO emission from the GRS survey \citep{jackson06} tracing gas at comparatively lower densities and temperatures than the 1\,mm images, and the locations of the detected ionized gas peaks, for several of the observed fields. Only those sources with GRS data (8 of the 10 observed fields) are presented. Integrated $^{13}$CO emission is shown in greyscale, warm dust emission in contours, and cross symbols mark the peak positions of the HII regions listed in Table \ref{vlaresult}. The integrated $^{13}$CO images were made by integrating the channels containing $^{13}$CO emission from the clouds associated with each millimeter clump. In many cases, the same cloud contained several of the clumps in the field. From Figure \ref{1mm_13co}, we can see that in all cases, the warm, 1\,mm dust emission follows the morphology of the $^{13}$CO gas emission. This demonstrates how it was possible to match the morphologies of the millimeter and $^{13}$CO emission, and therefore assign velocities and distances to our observed sources with confidence.

Masses of the material associated with the observed millimeter clumps were also derived from the $^{13}$CO and $\sim$1~mm emission over the area of sky covering the $^{13}$CO cloud, using the same photometry aperture for both images. The $^{13}$CO masses were derived using the equations of \citet{scoville86}, assuming the emitting material is optically thin. We also assumed a temperature of 10K, in agreement with the results of \cite{rathborne09}, and found the required [$^{12}$CO/$^{13}$CO] abundance ratios from the results of \citet{wilson94}. The millimeter dust masses were calculated using the same method as outlined in Section \ref{mmprops}. For dust masses calculated from the B06 1.2\,mm images, the dust opacity was assumed to be 1\,cm$^2$\,g$^{-1}$, consistent with their assumed value.

Table \ref{m13co} presents the calculated masses of the associated cloud material in $^{13}$CO and 1\,mm dust emission, M$_{\rm{cloud}}(^{13}$CO) and M$_{\rm{cloud}}(1\,\rm{mm})$ respectively, along with the ratios M$_{\rm{cloud}}(1\,\rm{mm})$/M$_{\rm{cloud}}(^{13}$CO) and $\Sigma$M$_{\rm{clump}}(1\,\rm{mm})$/M$_{\rm{cloud}}(^{13}$CO). The value $\Sigma$M$_{\rm{clump}}(1\,\rm{mm})$/M$_{\rm{cloud}}(^{13}$CO) is the sum of the masses of the associated millimeter clumps selected from BGPS or B06 over the total cloud mass traced by $^{13}$CO.  The cloud masses given are highly uncertain, due to the arbitrary apertures used, and therefore the ratio $\Sigma$M$_{\rm{clump}}(1\,\rm{mm})$/M$_{\rm{cloud}}(^{13}$CO) is correspondingly uncertain. However, as the cloud masses have been calculated from the fluxes measured in the same photometry apertures, their ratio, M$_{\rm{cloud}}(1\,\rm{mm})$/M$_{\rm{cloud}}(^{13}$CO), does not suffer from this uncertainty.

The ratio M$_{\rm{cloud}}(1\,\rm{mm})$/M$_{\rm{cloud}}(^{13}$CO) ranges from 0.08 to 1.0 with a mean and median value of 0.33$\pm$0.25 and 0.26 respectively. This ratio gives an estimate of the fraction of mass warm, more compact, dust structures traced by 1\,mm emission take up in the cloud compared to the cooler, lower density material traced by $^{13}$CO.

The ratio $\Sigma$M$_{\rm{clump}}(1\,\rm{mm})$/M$_{\rm{cloud}}(^{13}$CO) gives an estimate of the amount of mass the most compact structures, traced by the detected clumps in the 1\,mm images, contribute to the mass of the cloud. The values for this ratio range from 0.04 to 0.43 with a mean and median of 0.13$\pm$0.12 and 0.07 respectively. 

\subsection{Comparison of the Mass of Embedded Stars to Clump Masses \label{em_clump_mass}}

Table \ref{mclump_mstar} compares the mass of each millimeter clump, M$_{\rm{clump}}$, with the combined mass of its embedded ionizing stars, M$_{\star}$, derived from the luminosities given in Table~\ref{vlaproperties}.  Only sources where the millimeter and centimeter emission are coincident, suggesting that the massive star producing the ionized gas is associated with the millimeter clump, are included in Table \ref{mclump_mstar}. The stellar masses were calculated using the relation $\rm{L}_{\rm{cm}}\propto \rm{M}_{\star}^{\phantom{\star}3.6}$.  
The final column of Table \ref{mclump_mstar} gives the ratio M$_{\star}$/(M$_{\rm{clump}}$+M$_{\star}$).  This quantity represents the instantaneous massive star formation efficiency (MSFE) in the dense gas traced by warm dust emission.  The MSFEs in these millimeter clumps range from a few percent to 25\%, while the mean MSFE is 7 $\pm$ 8\%.

Figure \ref{mclump_mstar_plot} shows the relationship between M$_{\rm{clump}}$ and M$_{\star}$. We fit a polynomial function to the results of \citet{panagia73}, to find a function describing the relationship between $\log_{10}{N_{Ly}}$ and $\log_{10}{\rm{L}_{\rm{cm}}}$. Assuming $\rm{L}_{\rm{cm}}\propto \rm{M}_{\star}^{\phantom{\star}3.6}$, we then found the relationship between $\log_{10}{N_{Ly}}$ and $\log_{10}{\rm{M}_{\star}}$, from which we could calculate values of $\log_{10}{\rm{M}_{\star}}$ and derive uncertainties. Due to the different calculation method, the values of {M}$_{\star}$ plotted in Figure \ref{mclump_mstar_plot} differ slightly from those given in Table \ref{mclump_mstar}. 

Error bars are also shown in Figure \ref{mclump_mstar_plot}. The error in the clump mass M$_{\rm{clump}}$ is approximately a factor of two (see Section \ref{mmprops}). The main source of error in the combined stellar mass M$_{\star}$ originates from the uncertainties in the distance to the HII region and the electron temperature $T_e$, which are the largest errors involved in the calculation of $N_{Ly}$. However, it is possible that our assumptions, such as that each HII region is powered by a single star, are incorrect, which may explain the large scatter in the data around the trend line.

Figure \ref{mclump_mstar_plot} shows a possible power law relationship between M$_{\rm{clump}}$ and M$_{\star}$. Without taking in to account the errors on M$_{\rm{clump}}$ and M$_{\star}$, these data can be fit by the following power law: M$_{\star} = 1.0 \pm 0.9 \times \rm{M}_{\rm{clump}}^{\phantom{clump} 0.5 \pm 0.1}$, which is drawn upon the data in Figure \ref{mclump_mstar_plot}. The correlation coefficient between $\log_{10}{\rm{M}_{\rm{clump}}}$ and $\log_{10}{\rm{M}_{\star}}$ was found to be 0.74. This result is consistent with the the idea that the mass of the clump should determine the mass of the massive stars forming within it. It also bears resemblance to the prediction of competitive accretion simulations that the most massive star in a forming cluster is related to the mass of the remaining stars in the cluster \citep{bonnell04}. These simulations predict a power law relationship of the form M$_{\rm{max}} \propto $\,M$_{\rm{stars}}^{\phantom{stars}2/3}$, similar to the power law we observe. We would expect these two relationships to resemble one another if the mass of a clump is linearly related to the total mass of all the stars that form from it. As we are only able to show a small number of data points in Figure \ref{mclump_mstar_plot}, it would be useful for future studies to confirm whether the same result is found for a larger number of clumps with embedded massive stars.

\subsection{Follow-up of the Observed Fields with the EVLA and ALMA}

By comparing 3.6\,cm, $\sim$1\,mm, $^{13}$CO, and mid-IR observations, we have been able to gain insight into several regions of massive star formation. However future studies with the EVLA and ALMA will be able to further enhance our understanding. The EVLA will have 5-20 times the continuum sensitivity of the current VLA, allowing the detection of fainter centimeter continuum emission from ionized gas. In addition, just over a third of our detected 3.6\,cm sources are unresolved: 17 of the 47 detected sources (including sub-components). Therefore follow-up EVLA continuum observations in A, B, or C array of these fields will allow us to further resolve the ionized gas towards these star formation regions. 

ALMA is set to revolutionize our understanding of star formation. This is primarily because it will provide spatial resolutions on the order of 0.01" at 1\,mm, which is a vast improvement in resolution over the $\sim$1\,mm observations presented in this paper (33" and 44"). Even at distances of 10\,kpc this will still provide a physical resolution of 100\,AU, which will be sufficient to map any outflows and disks around the youngest massive stars. Examples of some of the transitions which will be observable at these wavelengths are $^{13}$CO(J=2-1), $^{13}$CO(J=3-2), HCN(J=3-2), CS(J=7-6), and CH$_{3}$CN(J=12-11), all suitable for tracing outflows or disks around forming stars.

One example of where EVLA and ALMA observations would improve our understanding of our observed 3.6\,cm and 1.1\,mm sources is the millimeter clump G48.605, which is coincident with the 3.6\,cm HII region VLA~5D. High sensitivity A array 3.6\,cm observations of this region may further resolve the two centimeter continuum sources G48.606+0.023 and G48.606+0.024  towards VLA~5D detected by \cite{kurtz94}, and also be able to pick up fainter compact emission towards them. In addition, ALMA observations would be able to determine whether there is an outflow toward VLA~5D which could produce shocked H$_{2}$ emission at 4.5$\mu$m, and also be able to study the molecular lines in this region to detect any disk or torus-like structures.

\section{Summary and Conclusions \label{conclusions}}

Using 3.6\,cm continuum observations with the VLA, we have surveyed the ionized
gas towards 31 intermediate and massive molecular clumps
previously observed at millimeter wavelengths. These millimeter clumps were selected 
from preliminary 1.1\,mm Bolocam Galactic Plane Survey images, 
and the 1.2\,mm observations of \citet{beltran06}. 

In the 10 observed fields, 35 HII regions were identified, 20 of them being newly discovered.
The observed HII regions display a wide range of morphologies, and many
are multiply peaked, indicating the presence of a cluster of massive stars.

Images comparing the warm dust and ionized gas emission in the 10 observed fields are presented,
as well as GLIMPSE survey images of \textit{Spizter} IRAC emission in these fields. 
The properties of the millimeter continuum clumps and the HII regions were calculated, 
and are listed in Tables \ref{selectsrc} and \ref{vlaproperties}. 
There is a large range in the properties of the observed HII regions; their physical sizes extend 
from $<$0.05\,pc to 7.88\,pc, and their spectral types cover B2 to O5.
The wealth of information we have provided about the observed millimeter clumps allows them to 
be followed up in future with more powerful instruments such as the EVLA and ALMA. 

By comparing the positions of the millimeter clumps and ionized gas, we have shown that
the ionized gas tends to be associated with the millimeter clumps, 
however this association does not depend on the mass of the clump. 
Of the 31 millimeter clumps observed, 9 appear to be physically related
to ionized gas, and a further 6 have ionized gas emission within 1'. 

We also infer ages for several ``types" of millimeter clump: those with multiply peaked 3.6\,cm continuum emission,
 indicating they are clusters, are most likely to pinpoint the most evolved sites of star formation. These are followed 
 by millimeter clumps associated with singly peaked HII regions, which are likely to be a single star or
 compact cluster, and finally millimeter clumps with no associated ionized gas emission may provide examples of the youngest sources.

Comparing the 1\,mm Bolocam or SEST images with $^{13}$CO images from 
the Galactic Ring Survey (GPS), we have shown that the emission from these two tracers is correlated, 
confirming that distance determinations using $^{13}$CO data can be made for the observed 1\,mm clumps.

We have compared the masses of massive stars or star clusters to
the actual amount of gas available traced by millimeter emission to provide an estimate of the
instantaneous star formation efficiency for a clump, giving values ranging from a few percent to 25\%, with an average of 7 $\pm$ 8\%.

We find that the mass of a clump is correlated with the mass of the massive stars which form within it,
and that this relationship is consistent with a power law. This is comparable to the results of \citet{bonnell04}, 
who find that there is a power-law relationship between the most massive star in a cluster and the combined mass of the 
remaining stars.

\begin{figure}
\center{
\includegraphics[]{fig1ab.eps}
\\
\vspace{1cm}
\includegraphics[]{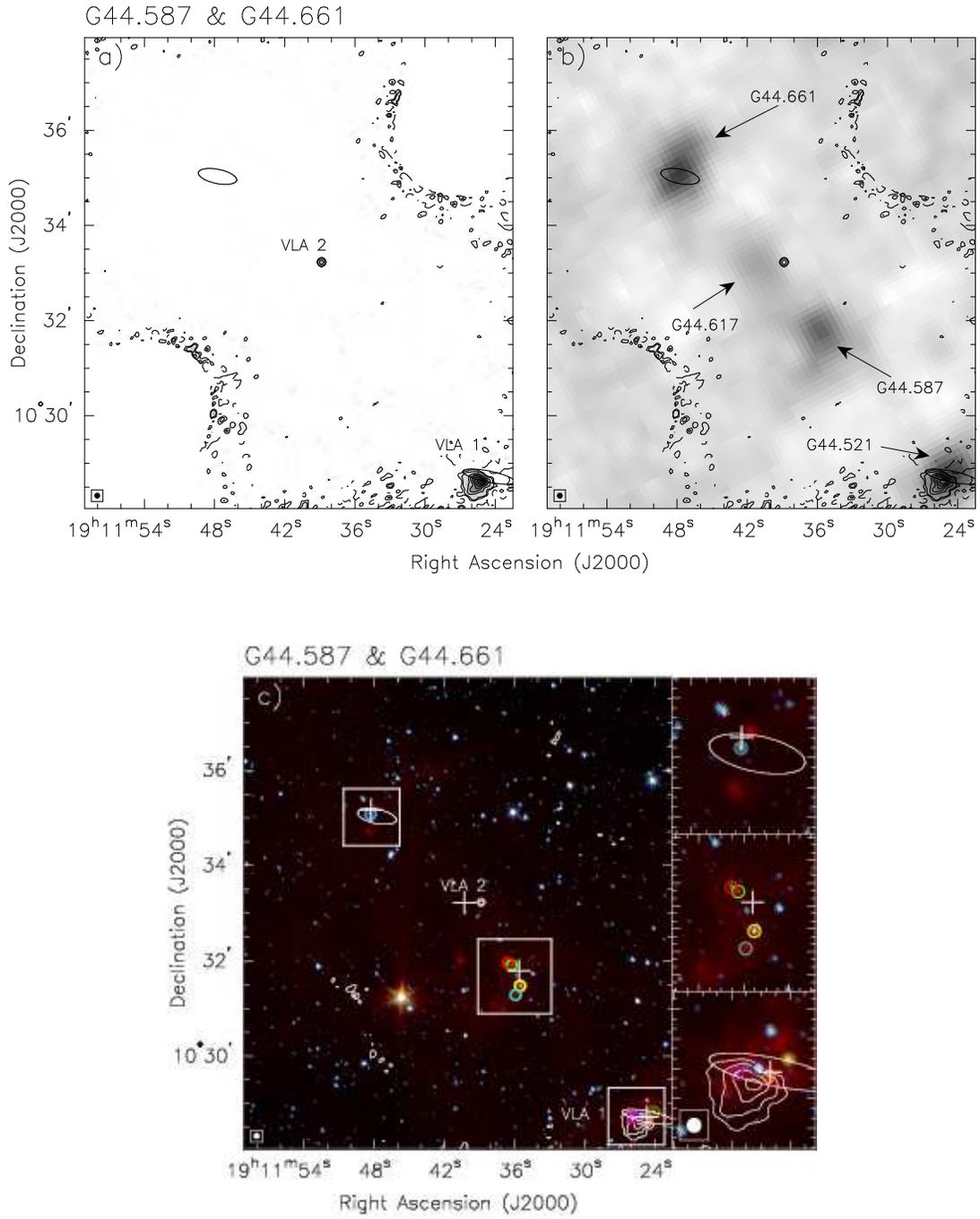}
\caption{\small 3.6\,cm, 1.1\,mm, and Mid-IR GLIMPSE images of the G44.587 \& G44.661 field. \label{G44587}}
}
\end{figure}

\begin{figure}
\center{
\includegraphics[]{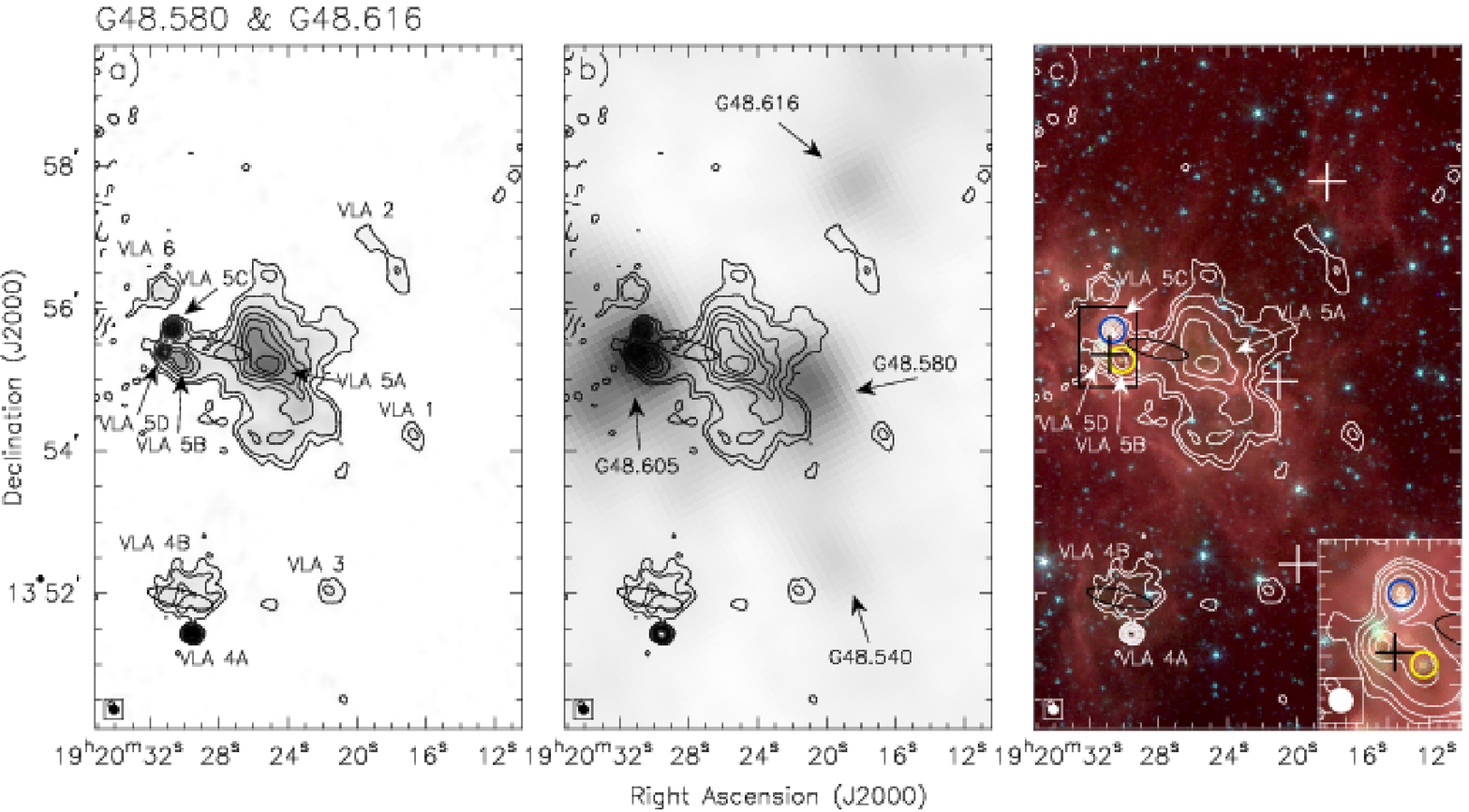}
\caption{\small 3.6\,cm, 1.1\,mm, and Mid-IR GLIMPSE images of the G48.580 \& G48.616 field. \label{G48580}}
}
\end{figure}

\begin{figure}
\center{
\includegraphics[]{fig3ab.eps}
\\
\vspace{1cm}
\includegraphics[]{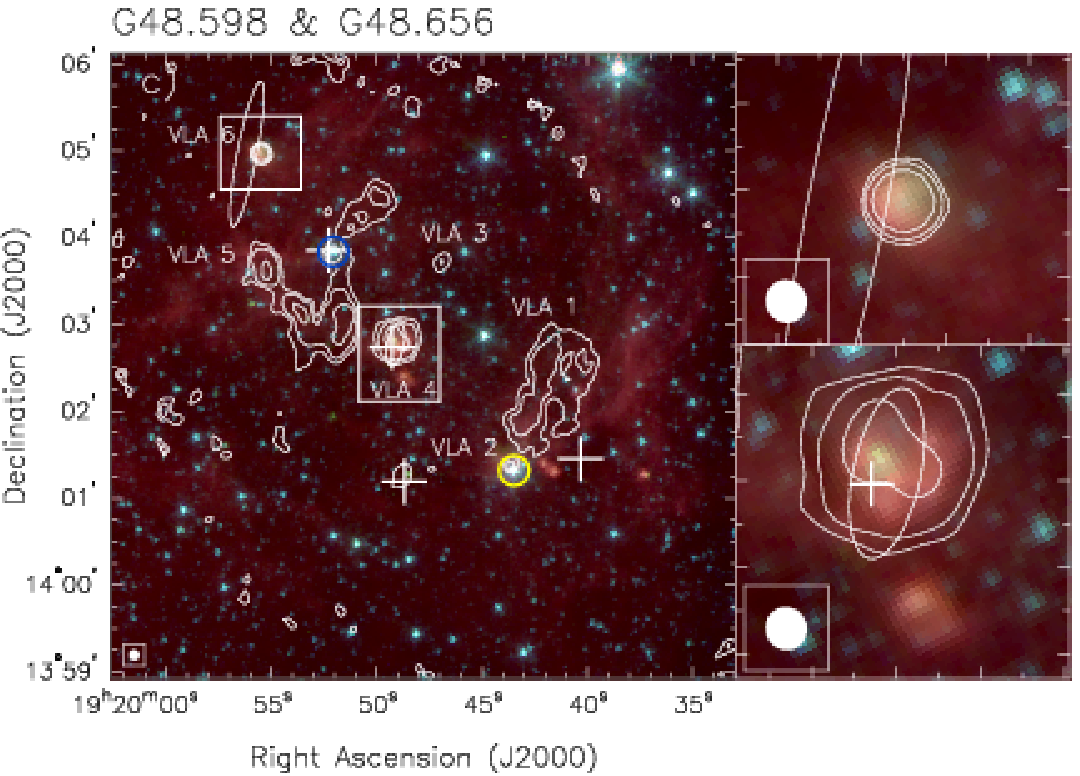}
\caption{\small 3.6\,cm, 1.1\,mm, and Mid-IR GLIMPSE images of the G48.598 \& G48.656 field. \label{G48598}}
}
\end{figure}

\begin{figure}
\center{
\includegraphics[]{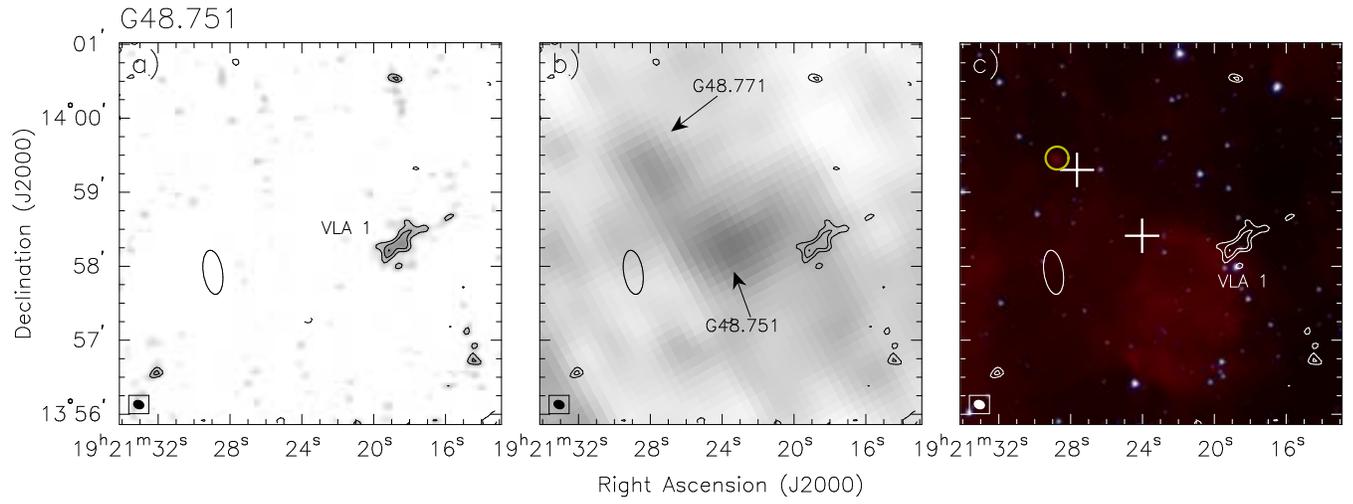}
\caption{\small 3.6\,cm, 1.1\,mm, and Mid-IR GLIMPSE images of the G48.751 field. \label{G48751}}
}
\end{figure}

\begin{figure}
\center{
\includegraphics[]{fig5ab.eps}
\\
\vspace{1cm}
\includegraphics[]{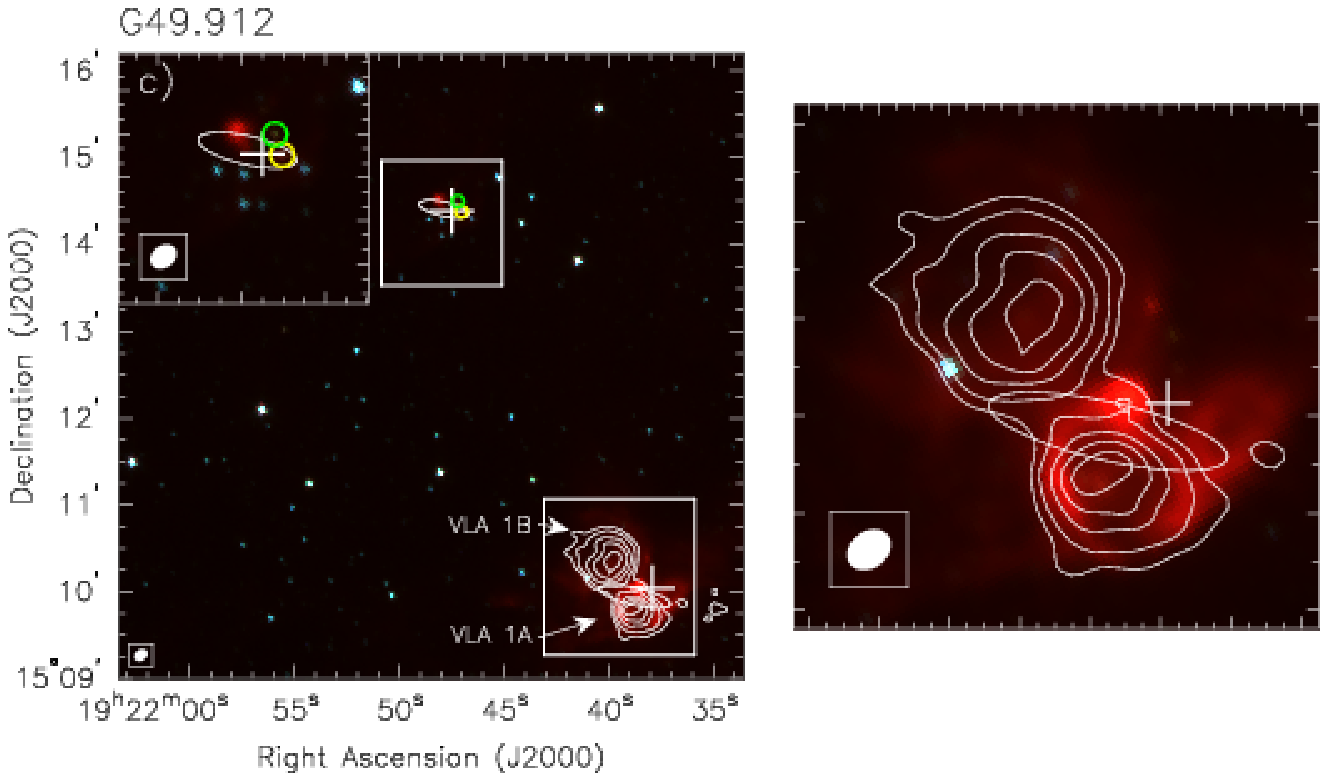}
\caption{\small 3.6\,cm, 1.1\,mm, and Mid-IR GLIMPSE images of the G49.912 field. \label{G49912}}
}
\end{figure}

\begin{figure}
\center{
\includegraphics[]{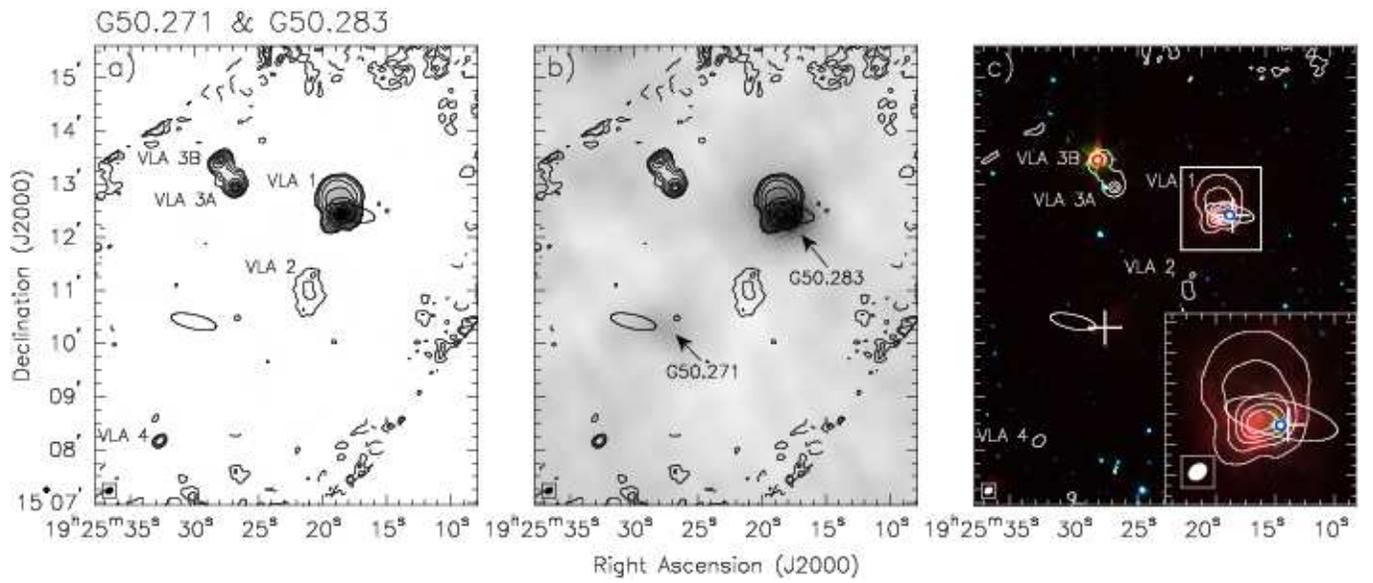}
\caption{\small 3.6\,cm, 1.1\,mm, and Mid-IR GLIMPSE images of the G50.271 \& G50.283 field. \label{G50271}}
}
\end{figure}

\begin{figure}
\center{
\includegraphics[]{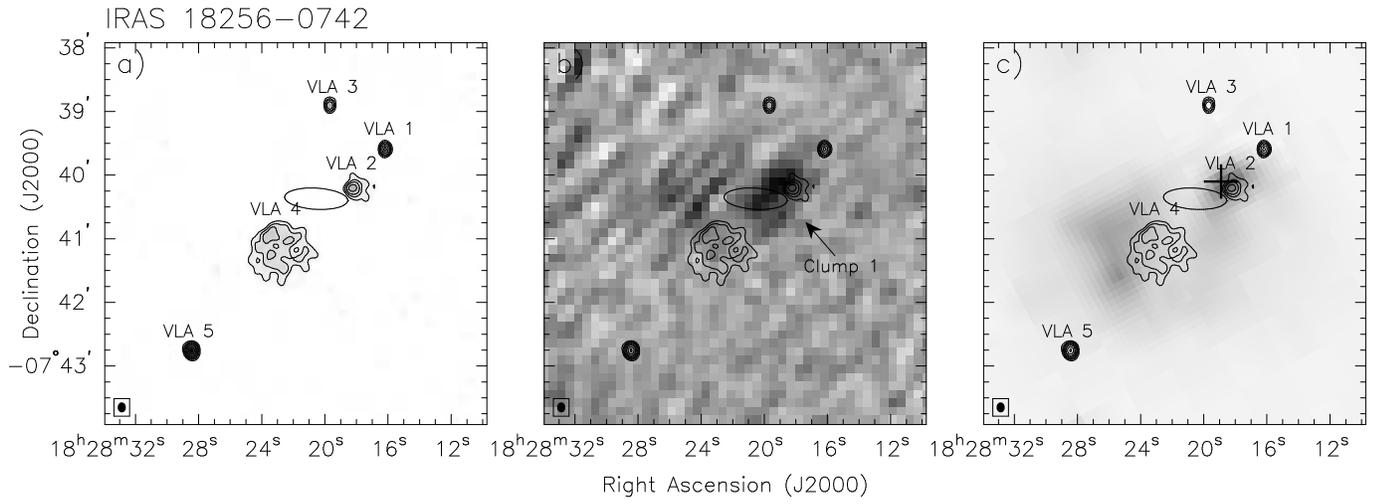}
\caption{\small 3.6\,cm, 1.2\,mm, and Mid-IR MSX 8.28$\mu$m images of the IRAS 18256-0742 field. \label{IRAS18256}}
}
\end{figure}

\begin{figure}
\center{
\includegraphics[]{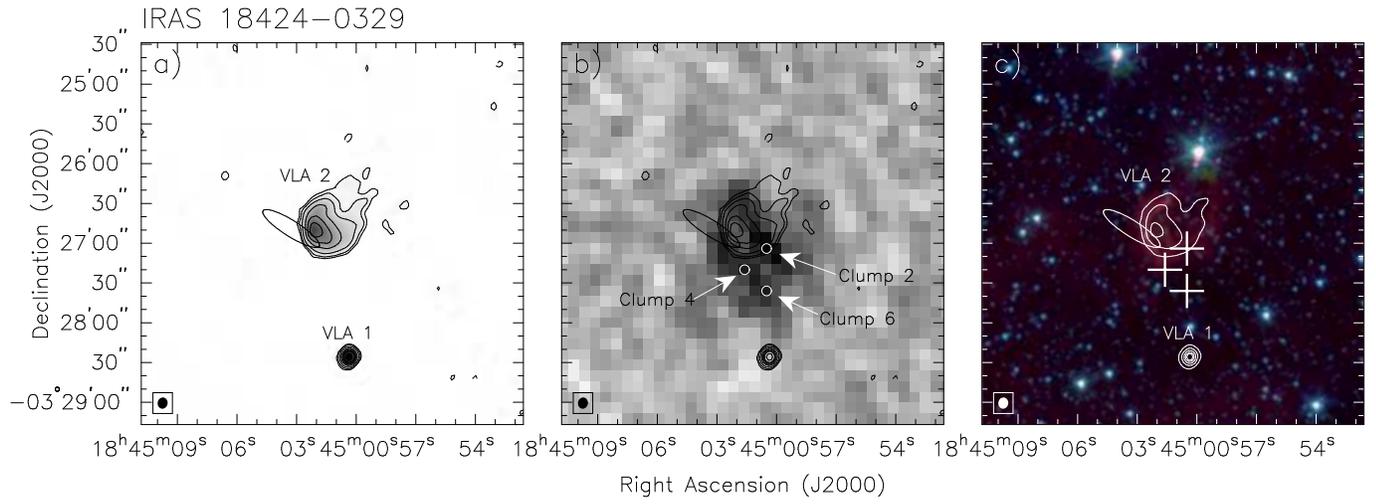}
\caption{\small 3.6\,cm, 1.2\,mm, and Mid-IR GLIMPSE images of the IRAS 18424-0329 field. \label{IRAS18424}}
}
\end{figure}

\begin{figure}
\center{
\includegraphics[]{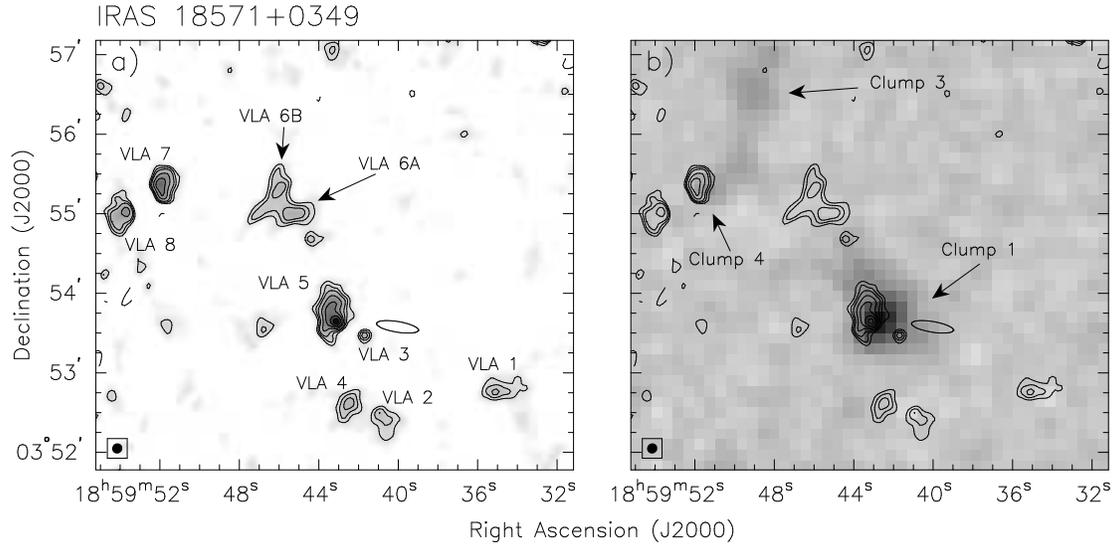}
\\
\vspace{1cm}
\includegraphics[]{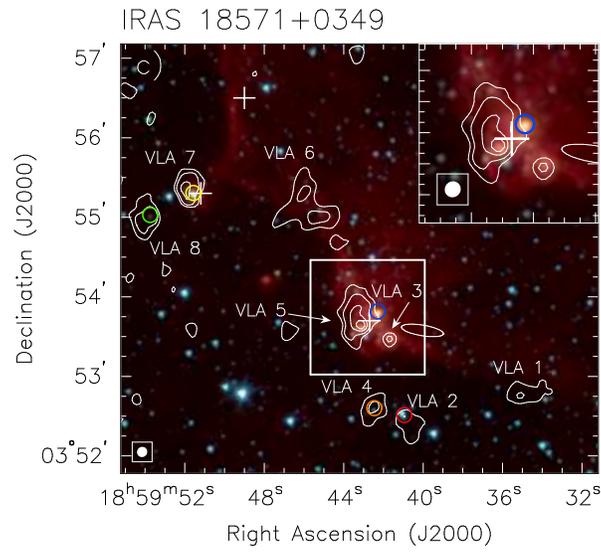}
\caption{\small 3.6\,cm, 1.2\,mm, and Mid-IR GLIMPSE images of the IRAS 18571+0349 field. \label{IRAS18571}}
}
\end{figure}

\begin{figure}
\center{
\includegraphics[]{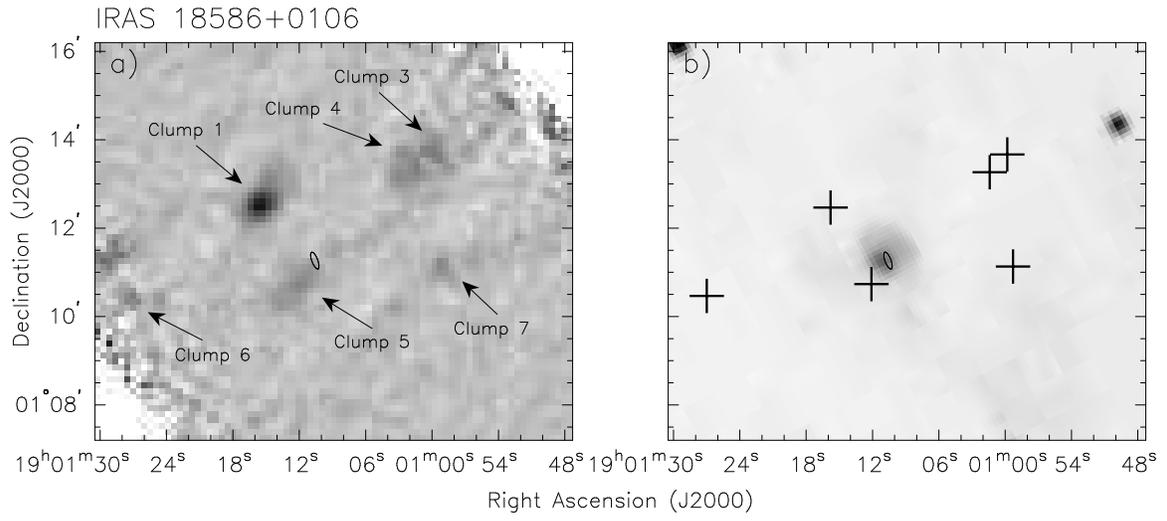}
\caption{\small 3.6\,cm, and Mid-IR MSX 8.28$\mu$m images of the IRAS 18586+0106 field. \label{IRAS18586}}
}
\end{figure}

\acknowledgments
We would like to thank the anonymous referee for providing helpful comments which allowed us to improve this paper. We also thank M. T. Beltran for providing digital versions of four maps from B06; T. Robitaille for providing the background matched \textit{Spitzer} GLIMPSE images used in panel c) of Figures \ref{G44587} to \ref{IRAS18586}, and for many helpful discussions; J. Roman-Duval for providing results from \citet{roman-duval09} before publication and giving invaluable advice concerning distance determinations; S. Kurtz for advice on the calculation of HII region properties; and D. Johnstone for helpful comments. We are also very grateful for the help and support provided by the Bolocam Galactic Plane Survey Team.

This publication makes use of molecular line data from the Boston University-FCRAO Galactic Ring Survey (GRS). The GRS is a joint project of the Boston University and the Five College Radio Astronomy Observatory, funded by the National Science Foundation under grants AST-9800334, AST-0098562, AST-0100793, AST-0228993, and AST-0507657. This research has made use of data products from the Midcourse Space Experiment. Processing of the MSX data was funded by the Ballistic Missile Defense Organization with additional support from NASA Office of Space Science. This work relies on observations made using the Spitzer Space Telescope, and uses the NASA/ IPAC Infrared Science Archive, which are operated by the Jet Propulsion Laboratory, California Institute of Technology, under contract with the National Aeronautics and Space Administration. The VGPS is supported by a grant from the Natural Sciences and Engineering Research Council of Canada and from the U.S. National Science Foundation. 

K.J. acknowledges support from STFC and the NRAO Graduate Student Internship Program. 

\bibliography{}

\begin{thebibliography}{90}
\expandafter\ifx\csname natexlab\endcsname\relax\def\natexlab#1{#1}\fi

\bibitem[{{Allen} {et~al.}(2004){Allen}, {Calvet}, {D'Alessio}, {Merin},
  {Hartmann}, {Megeath}, {Gutermuth}, {Muzerolle}, {Pipher}, {Myers}, \&
  {Fazio}}]{allen04}
{Allen}, L.~E. {et~al.} 2004, \apjs, 154, 363

\bibitem[{{Altenhoff} {et~al.}(1979){Altenhoff}, {Downes}, {Pauls}, \&
  {Schraml}}]{altenhoff79}
{Altenhoff}, W.~J., {Downes}, D., {Pauls}, T., \& {Schraml}, J. 1979, \aaps,
  35, 23

\bibitem[{{Altenhoff} {et~al.}(1960){Altenhoff}, {Mezger}, {Wendker}, \&
  {Westerhout}}]{altenhoff60}
{Altenhoff}, W.~J., {Mezger}, P.~G., {Wendker}, H., \& {Westerhout}, G. 1960,
  Veroff. Sternwarte, Bonn, No. 59, 48

\bibitem[{{Baudry} {et~al.}(1997){Baudry}, {Desmurs}, {Wilson}, \&
  {Cohen}}]{baudry97}
{Baudry}, A., {Desmurs}, J.~F., {Wilson}, T.~L., \& {Cohen}, R.~J. 1997, \aap,
  325, 255

\bibitem[{{Becker} {et~al.}(1994){Becker}, {White}, {Helfand}, \&
  {Zoonematkermani}}]{becker94}
{Becker}, R.~H., {White}, R.~L., {Helfand}, D.~J., \& {Zoonematkermani}, S.
  1994, \apjs, 91, 347

\bibitem[{{Beltr{\'a}n} {et~al.}(2006){Beltr{\'a}n}, {Brand}, {Cesaroni},
  {Fontani}, {Pezzuto}, {Testi}, \& {Molinari}}]{beltran06}
{Beltr{\'a}n}, M.~T., {Brand}, J., {Cesaroni}, R., {Fontani}, F., {Pezzuto},
  S., {Testi}, L., \& {Molinari}, S. 2006, \aap, 447, 221,
  arXiv:astro-ph/0510422

\bibitem[{{Benjamin} {et~al.}(2003){Benjamin}, {Churchwell}, {Babler}, {Bania},
  {Clemens}, {Cohen}, {Dickey}, {Indebetouw}, {Jackson}, {Kobulnicky},
  {Lazarian}, {Marston}, {Mathis}, {Meade}, {Seager}, {Stolovy}, {Watson},
  {Whitney}, {Wolff}, \& {Wolfire}}]{benjamin03}
{Benjamin}, R.~A. {et~al.} 2003, \pasp, 115, 953, arXiv:astro-ph/0306274

\bibitem[{{Beuther} {et~al.}(2002){Beuther}, {Schilke}, {Menten}, {Motte},
  {Sridharan}, \& {Wyrowski}}]{beuther020}
{Beuther}, H., {Schilke}, P., {Menten}, K.~M., {Motte}, F., {Sridharan}, T.~K.,
  \& {Wyrowski}, F. 2002, \apj, 566, 945, arXiv:astro-ph/0110370

\bibitem[{{Beuther} \& {Shepherd}(2005)}]{beuther05}
{Beuther}, H., \& {Shepherd}, D. 2005, in Cores to Clusters: Star Formation
  with Next Generation Telescopes, ed. M.~S.~N. {Kumar}, M.~{Tafalla}, \&
  P.~{Caselli}, 105--119

\bibitem[{{Bonnell} {et~al.}(2004){Bonnell}, {Vine}, \& {Bate}}]{bonnell04}
{Bonnell}, I.~A., {Vine}, S.~G., \& {Bate}, M.~R. 2004, \mnras, 349, 735,
  arXiv:astro-ph/0401059

\bibitem[{{Bronfman} {et~al.}(1996){Bronfman}, {Nyman}, \& {May}}]{bronfman96}
{Bronfman}, L., {Nyman}, L.-A., \& {May}, J. 1996, \aaps, 115, 81

\bibitem[{{Caswell} {et~al.}(1975){Caswell}, {Murray}, {Roger}, {Cole}, \&
  {Cooke}}]{caswell75}
{Caswell}, J.~L., {Murray}, J.~D., {Roger}, R.~S., {Cole}, D.~J., \& {Cooke},
  D.~J. 1975, \aap, 45, 239

\bibitem[{{Cesaroni} {et~al.}(2007){Cesaroni}, {Galli}, {Lodato}, {Walmsley},
  \& {Zhang}}]{cesaroni07}
{Cesaroni}, R., {Galli}, D., {Lodato}, G., {Walmsley}, C.~M., \& {Zhang}, Q.
  2007, in Protostars and Planets V, ed. B.~{Reipurth}, D.~{Jewitt}, \&
  K.~{Keil}, 197--212

\bibitem[{{Churchwell} {et~al.}(2006){Churchwell}, {Povich}, {Allen}, {Taylor},
  {Meade}, {Babler}, {Indebetouw}, {Watson}, {Whitney}, {Wolfire}, {Bania},
  {Benjamin}, {Clemens}, {Cohen}, {Cyganowski}, {Jackson}, {Kobulnicky},
  {Mathis}, {Mercer}, {Stolovy}, {Uzpen}, {Watson}, \& {Wolff}}]{churchwell06}
{Churchwell}, E. {et~al.} 2006, \apj, 649, 759

\bibitem[{{Churchwell} {et~al.}(2004){Churchwell}, {Whitney}, {Babler},
  {Indebetouw}, {Meade}, {Watson}, {Wolff}, {Wolfire}, {Bania}, {Benjamin},
  {Clemens}, {Cohen}, {Devine}, {Dickey}, {Heitsch}, {Jackson}, {Kobulnicky},
  {Marston}, {Mathis}, {Mercer}, {Stauffer}, \& {Stolovy}}]{churchwell04}
------. 2004, \apjs, 154, 322

\bibitem[{{Condon} {et~al.}(1993){Condon}, {Griffith}, \& {Wright}}]{condon93}
{Condon}, J.~J., {Griffith}, M.~R., \& {Wright}, A.~E. 1993, \aj, 106, 1095

\bibitem[{{Copetti} \& {Schmidt}(1991)}]{copetti91}
{Copetti}, M.~V.~F., \& {Schmidt}, A.~A. 1991, \mnras, 250, 127

\bibitem[{{Crowther} \& {Conti}(2003)}]{crowther03}
{Crowther}, P.~A., \& {Conti}, P.~S. 2003, \mnras, 343, 143,
  arXiv:astro-ph/0302481

\bibitem[{{Cyganowski} {et~al.}(2008){Cyganowski}, {Whitney}, {Holden},
  {Braden}, {Brogan}, {Churchwell}, {Indebetouw}, {Watson}, {Babler},
  {Benjamin}, {Gomez}, {Meade}, {Povich}, {Robitaille}, \&
  {Watson}}]{cyganowski08}
{Cyganowski}, C.~J. {et~al.} 2008, \aj, 136, 2391, 0810.0530

\bibitem[{{De Buizer} {et~al.}(2005){De Buizer}, {Radomski}, {Telesco}, \&
  {Pi{\~n}a}}]{de-buizer05}
{De Buizer}, J.~M., {Radomski}, J.~T., {Telesco}, C.~M., \& {Pi{\~n}a}, R.~K.
  2005, \apjs, 156, 179, arXiv:astro-ph/0410630

\bibitem[{{Douglas} {et~al.}(1996){Douglas}, {Bash}, {Bozyan}, {Torrence}, \&
  {Wolfe}}]{douglas96}
{Douglas}, J.~N., {Bash}, F.~N., {Bozyan}, F.~A., {Torrence}, G.~W., \&
  {Wolfe}, C. 1996, \aj, 111, 1945

\bibitem[{{Downes} {et~al.}(1980){Downes}, {Wilson}, {Bieging}, \&
  {Wink}}]{downes80}
{Downes}, D., {Wilson}, T.~L., {Bieging}, J., \& {Wink}, J. 1980, \aaps, 40,
  379

\bibitem[{{Draine} {et~al.}(2007){Draine}, {Dale}, {Bendo}, {Gordon}, {Smith},
  {Armus}, {Engelbracht}, {Helou}, {Kennicutt}, {Li}, {Roussel}, {Walter},
  {Calzetti}, {Moustakas}, {Murphy}, {Rieke}, {Bot}, {Hollenbach}, {Sheth}, \&
  {Teplitz}}]{draine07}
{Draine}, B.~T. {et~al.} 2007, \apj, 663, 866, arXiv:astro-ph/0703213

\bibitem[{{Edris} {et~al.}(2007){Edris}, {Fuller}, \& {Cohen}}]{edris07}
{Edris}, K.~A., {Fuller}, G.~A., \& {Cohen}, R.~J. 2007, \aap, 465, 865,
  arXiv:astro-ph/0701652

\bibitem[{{Enoch} {et~al.}(2006){Enoch}, {Young}, {Glenn}, {Evans}, {Golwala},
  {Sargent}, {Harvey}, {Aguirre}, {Goldin}, {Haig}, {Huard}, {Lange},
  {Laurent}, {Maloney}, {Mauskopf}, {Rossinot}, \& {Sayers}}]{enoch06}
{Enoch}, M.~L. {et~al.} 2006, \apj, 638, 293, arXiv:astro-ph/0510202

\bibitem[{{Fazio} {et~al.}(2004){Fazio}, {Hora}, {Allen}, {Ashby}, {Barmby},
  {Deutsch}, {Huang}, {Kleiner}, {Marengo}, {Megeath}, {Melnick}, {Pahre},
  {Patten}, {Polizotti}, {Smith}, {Taylor}, {Wang}, {Willner}, {Hoffmann},
  {Pipher}, {Forrest}, {McMurty}, {McCreight}, {McKelvey}, {McMurray}, {Koch},
  {Moseley}, {Arendt}, {Mentzell}, {Marx}, {Losch}, {Mayman}, {Eichhorn},
  {Krebs}, {Jhabvala}, {Gezari}, {Fixsen}, {Flores}, {Shakoorzadeh}, {Jungo},
  {Hakun}, {Workman}, {Karpati}, {Kichak}, {Whitley}, {Mann}, {Tollestrup},
  {Eisenhardt}, {Stern}, {Gorjian}, {Bhattacharya}, {Carey}, {Nelson},
  {Glaccum}, {Lacy}, {Lowrance}, {Laine}, {Reach}, {Stauffer}, {Surace},
  {Wilson}, {Wright}, {Hoffman}, {Domingo}, \& {Cohen}}]{fazio04}
{Fazio}, G.~G. {et~al.} 2004, \apjs, 154, 10, arXiv:astro-ph/0405616

\bibitem[{{Forster} \& {Caswell}(1989)}]{forster89}
{Forster}, J.~R., \& {Caswell}, J.~L. 1989, \aap, 213, 339

\bibitem[{{Fuente} {et~al.}(2001){Fuente}, {Neri}, {Mart{\'{\i}}n-Pintado},
  {Bachiller}, {Rodr{\'{\i}}guez-Franco}, \& {Palla}}]{fuente01}
{Fuente}, A., {Neri}, R., {Mart{\'{\i}}n-Pintado}, J., {Bachiller}, R.,
  {Rodr{\'{\i}}guez-Franco}, A., \& {Palla}, F. 2001, \aap, 366, 873

\bibitem[{{Fuller} {et~al.}(2005){Fuller}, {Williams}, \&
  {Sridharan}}]{fuller05}
{Fuller}, G.~A., {Williams}, S.~J., \& {Sridharan}, T.~K. 2005, \aap, 442, 949,
  arXiv:astro-ph/0508098

\bibitem[{{Gregory} {et~al.}(1996){Gregory}, {Scott}, {Douglas}, \&
  {Condon}}]{gregory96}
{Gregory}, P.~C., {Scott}, W.~K., {Douglas}, K., \& {Condon}, J.~J. 1996,
  \apjs, 103, 427

\bibitem[{{Handa} {et~al.}(1987){Handa}, {Sofue}, {Nakai}, {Hirabayashi}, \&
  {Inoue}}]{handa87}
{Handa}, T., {Sofue}, Y., {Nakai}, N., {Hirabayashi}, H., \& {Inoue}, M. 1987,
  \pasj, 39, 709

\bibitem[{{Jackson} {et~al.}(2006){Jackson}, {Rathborne}, {Shah}, {Simon},
  {Bania}, {Clemens}, {Chambers}, {Johnson}, {Dormody}, {Lavoie}, \&
  {Heyer}}]{jackson06}
{Jackson}, J.~M. {et~al.} 2006, \apjs, 163, 145, arXiv:astro-ph/0602160

\bibitem[{{Jourdain de Muizon} {et~al.}(1990){Jourdain de Muizon}, {Cox}, \&
  {Lequeux}}]{Jourdain-de-Muizon90}
{Jourdain de Muizon}, M., {Cox}, P., \& {Lequeux}, J. 1990, \aaps, 83, 337

\bibitem[{{Kerr} \& {Lynden-Bell}(1986)}]{kerr86}
{Kerr}, F.~J., \& {Lynden-Bell}, D. 1986, \mnras, 221, 1023

\bibitem[{{Keto} \& {Wood}(2006)}]{keto06}
{Keto}, E., \& {Wood}, K. 2006, \apj, 637, 850, astro-ph/0510176

\bibitem[{{Kuchar} \& {Bania}(1994)}]{kuchar94}
{Kuchar}, T.~A., \& {Bania}, T.~M. 1994, \apj, 436, 117

\bibitem[{{Kuchar} \& {Clark}(1997)}]{kuchar97}
{Kuchar}, T.~A., \& {Clark}, F.~O. 1997, \apj, 488, 224

\bibitem[{{Kundu} \& {Velusamy}(1967)}]{kundu67}
{Kundu}, M.~R., \& {Velusamy}, T. 1967, Annales d'Astrophysique, 30, 59

\bibitem[{{Kurtz} {et~al.}(1994){Kurtz}, {Churchwell}, \& {Wood}}]{kurtz94}
{Kurtz}, S., {Churchwell}, E., \& {Wood}, D.~O.~S. 1994, \apjs, 91, 659

\bibitem[{{Kurtz} \& {Hofner}(2005)}]{kurtz050}
{Kurtz}, S., \& {Hofner}, P. 2005, \aj, 130, 711, arXiv:astro-ph/0507039

\bibitem[{{Kurtz} {et~al.}(1999){Kurtz}, {Watson}, {Hofner}, \&
  {Otte}}]{kurtz99}
{Kurtz}, S.~E., {Watson}, A.~M., {Hofner}, P., \& {Otte}, B. 1999, \apj, 514,
  232

\bibitem[{{Langston} {et~al.}(2000){Langston}, {Minter}, {D'Addario},
  {Eberhardt}, {Koski}, \& {Zuber}}]{langston00}
{Langston}, G., {Minter}, A., {D'Addario}, L., {Eberhardt}, K., {Koski}, K., \&
  {Zuber}, J. 2000, \aj, 119, 2801

\bibitem[{{Lockman}(1989)}]{lockman89}
{Lockman}, F.~J. 1989, \apjs, 71, 469

\bibitem[{{Massey} {et~al.}(1995){Massey}, {Johnson}, \&
  {Degioia-Eastwood}}]{massey95}
{Massey}, P., {Johnson}, K.~E., \& {Degioia-Eastwood}, K. 1995, \apj, 454, 151

\bibitem[{{Matthews} {et~al.}(1978){Matthews}, {Shaver}, {Goss}, \&
  {Habing}}]{matthews78}
{Matthews}, H.~E., {Shaver}, P.~A., {Goss}, W.~M., \& {Habing}, H.~J. 1978,
  \aap, 63, 307

\bibitem[{{Mezger} \& {Henderson}(1967)}]{mezger67}
{Mezger}, P.~G., \& {Henderson}, A.~P. 1967, \apj, 147, 471

\bibitem[{{Molinari} {et~al.}(1996){Molinari}, {Brand}, {Cesaroni}, \&
  {Palla}}]{molinari96}
{Molinari}, S., {Brand}, J., {Cesaroni}, R., \& {Palla}, F. 1996, \aap, 308,
  573

\bibitem[{{Molinari} {et~al.}(1998){Molinari}, {Brand}, {Cesaroni}, {Palla}, \&
  {Palumbo}}]{molinari98}
{Molinari}, S., {Brand}, J., {Cesaroni}, R., {Palla}, F., \& {Palumbo},
  G.~G.~C. 1998, \aap, 336, 339

\bibitem[{{Mueller} {et~al.}(2002){Mueller}, {Shirley}, {Evans}, \&
  {Jacobson}}]{mueller02}
{Mueller}, K.~E., {Shirley}, Y.~L., {Evans}, II, N.~J., \& {Jacobson}, H.~R.
  2002, in Astronomical Society of the Pacific Conference Series, Vol. 267, Hot
  Star Workshop III: The Earliest Phases of Massive Star Birth, ed.
  P.~{Crowther}, 395--+

\bibitem[{{Olnon} {et~al.}(1986){Olnon}, {Raimond}, {Neugebauer}, {van Duinen},
  {Habing}, {Aumann}, {Beintema}, {Boggess}, {Borgman}, {Clegg}, {Gillett},
  {Hauser}, {Houck}, {Jennings}, {de Jong}, {Low}, {Marsden}, {Pottasch},
  {Soifer}, {Walker}, {Emerson}, {Rowan-Robinson}, {Wesselius}, {Baud},
  {Beichman}, {Gautier}, {Harris}, {Miley}, \& {Young}}]{olnon86}
{Olnon}, F.~M. {et~al.} 1986, \aaps, 65, 607

\bibitem[{{Onello} \& {Phillips}(1995)}]{onello95}
{Onello}, J.~S., \& {Phillips}, J.~A. 1995, \apj, 448, 727

\bibitem[{{Ossenkopf} \& {Henning}(1994)}]{ossenkopf94}
{Ossenkopf}, V., \& {Henning}, T. 1994, \aap, 291, 943

\bibitem[{{Oster}(1961)}]{oster61}
{Oster}, L. 1961, Reviews of Modern Physics, 33, 525

\bibitem[{{Osterbrock} \& {Ferland}(2006)}]{osterbrock06}
{Osterbrock}, D.~E., \& {Ferland}, G.~J. 2006, {Astrophysics of gaseous nebulae
  and active galactic nuclei} (Astrophysics of gaseous nebulae and active
  galactic nuclei, 2nd.~ed.~by D.E.~Osterbrock and G.J.~Ferland.~Sausalito, CA:
  University Science Books, 2006)

\bibitem[{{Panagia}(1973)}]{panagia73}
{Panagia}, N. 1973, \aj, 78, 929

\bibitem[{{Panagia} \& {Walmsley}(1978)}]{panagia78}
{Panagia}, N., \& {Walmsley}, C.~M. 1978, \aap, 70, 411

\bibitem[{{Pandian} {et~al.}(2007){Pandian}, {Goldsmith}, \&
  {Deshpande}}]{pandian07}
{Pandian}, J.~D., {Goldsmith}, P.~F., \& {Deshpande}, A.~A. 2007, \apj, 656,
  255, arXiv:astro-ph/0702147

\bibitem[{{Planesas} {et~al.}(1991){Planesas}, {G{\'o}mez-Gonz{\'a}lez},
  {Rodr{\'{\i}}guez}, \& {Cant{\'o}}}]{planesas91}
{Planesas}, P., {G{\'o}mez-Gonz{\'a}lez}, J., {Rodr{\'{\i}}guez}, L.~F., \&
  {Cant{\'o}}, J. 1991, Revista Mexicana de Astronomia y Astrofisica, 22, 19

\bibitem[{{Plume} {et~al.}(1992){Plume}, {Jaffe}, \& {Evans}}]{plume92}
{Plume}, R., {Jaffe}, D.~T., \& {Evans}, II, N.~J. 1992, \apjs, 78, 505

\bibitem[{{Price} {et~al.}(2001){Price}, {Egan}, {Carey}, {Mizuno}, \&
  {Kuchar}}]{price01}
{Price}, S.~D., {Egan}, M.~P., {Carey}, S.~J., {Mizuno}, D.~R., \& {Kuchar},
  T.~A. 2001, \aj, 121, 2819

\bibitem[{{Quireza} {et~al.}(2006){Quireza}, {Rood}, {Bania}, {Balser}, \&
  {Maciel}}]{quireza06}
{Quireza}, C., {Rood}, R.~T., {Bania}, T.~M., {Balser}, D.~S., \& {Maciel},
  W.~J. 2006, \apj, 653, 1226, arXiv:astro-ph/0609006

\bibitem[{{Rathborne} {et~al.}(2009){Rathborne}, {Johnson}, {Jackson}, {Shah},
  \& {Simon}}]{rathborne09}
{Rathborne}, J.~M., {Johnson}, A.~M., {Jackson}, J.~M., {Shah}, R.~Y., \&
  {Simon}, R. 2009, \apjs, 182, 131, 0904.1217

\bibitem[{{Reich} {et~al.}(1984){Reich}, {Fuerst}, {Haslam}, {Steffen}, \&
  {Reif}}]{reich84}
{Reich}, W., {Fuerst}, E., {Haslam}, C.~G.~T., {Steffen}, P., \& {Reif}, K.
  1984, \aaps, 58, 197

\bibitem[{{Reich} {et~al.}(1990){Reich}, {Reich}, \& {Fuerst}}]{reich90}
{Reich}, W., {Reich}, P., \& {Fuerst}, E. 1990, \aaps, 83, 539

\bibitem[{{Roman-Duval} {et~al.}(2009){Roman-Duval},{Jackson},{Heyer},{Johnson},{Rathborne},{Shah}, \& {Simon}}]{roman-duval09} 
{Roman-Duval}, J., {Jackson}, J.~M., {Heyer}, M., {Johnson}, A.,{Rathborne}, J., {Shah}, R., \& {Simon}, R. 2009, \apj, 699, 1153, 0905.0723

\bibitem[Rosolowsky et al.(2009)]{rosolowsky09} Rosolowsky, E., et 
al.\ 2009, arXiv:0909.2871 

\bibitem[{{Scoville} {et~al.}(1986){Scoville}, {Sargent}, {Sanders},
  {Claussen}, {Masson}, {Lo}, \& {Phillips}}]{scoville86}
{Scoville}, N.~Z., {Sargent}, A.~I., {Sanders}, D.~B., {Claussen}, M.~J.,
  {Masson}, C.~R., {Lo}, K.~Y., \& {Phillips}, T.~G. 1986, \apj, 303, 416

\bibitem[{{Shepherd}(2003)}]{shepherd03}
{Shepherd}, D. 2003, in Astronomical Society of the Pacific Conference Series,
  Vol. 287, Galactic Star Formation Across the Stellar Mass Spectrum, ed. J.~M.
  {De Buizer} \& N.~S. {van der Bliek}, 333--344

\bibitem[{{Shepherd} \& {Churchwell}(1996{\natexlab{a}})}]{shepherd960}
{Shepherd}, D.~S., \& {Churchwell}, E. 1996{\natexlab{a}}, \apj, 472, 225

\bibitem[{{Shepherd} \& {Churchwell}(1996{\natexlab{b}})}]{shepherd96}
------. 1996{\natexlab{b}}, \apj, 457, 267

\bibitem[{{Shirley} {et~al.}(2003){Shirley}, {Evans}, {Young}, {Knez}, \&
  {Jaffe}}]{shirley03}
{Shirley}, Y.~L., {Evans}, II, N.~J., {Young}, K.~E., {Knez}, C., \& {Jaffe},
  D.~T. 2003, \apjs, 149, 375, arXiv:astro-ph/0308310

\bibitem[{{Silverglate} \& {Terzian}(1978)}]{silverglate78}
{Silverglate}, P., \& {Terzian}, Y. 1978, \aj, 83, 1412

\bibitem[{{Solomon} {et~al.}(1987){Solomon}, {Rivolo}, {Barrett}, \&
  {Yahil}}]{solomon87}
{Solomon}, P.~M., {Rivolo}, A.~R., {Barrett}, J., \& {Yahil}, A. 1987, \apj,
  319, 730

\bibitem[{{Sridharan} {et~al.}(2002){Sridharan}, {Beuther}, {Schilke},
  {Menten}, \& {Wyrowski}}]{sridharan02}
{Sridharan}, T.~K., {Beuther}, H., {Schilke}, P., {Menten}, K.~M., \&
  {Wyrowski}, F. 2002, \apj, 566, 931, arXiv:astro-ph/0110363

\bibitem[{{Stil} {et~al.}(2006){Stil}, {Taylor}, {Dickey}, {Kavars}, {Martin},
  {Rothwell}, {Boothroyd}, {Lockman}, \& {McClure-Griffiths}}]{stil06}
{Stil}, J.~M. {et~al.} 2006, \aj, 132, 1158, arXiv:astro-ph/0605422

\bibitem[{{Taylor} {et~al.}(1996){Taylor}, {Goss}, {Coleman}, {van Leeuwen}, \&
  {Wallace}}]{taylor96}
{Taylor}, A.~R., {Goss}, W.~M., {Coleman}, P.~H., {van Leeuwen}, J., \&
  {Wallace}, B.~J. 1996, \apjs, 107, 239

\bibitem[{{Te Lintel Hekkert} \& {Chapman}(1996)}]{te-lintel-hekkert96}
{Te Lintel Hekkert}, P., \& {Chapman}, J.~M. 1996, \aaps, 119, 459

\bibitem[{{Testi}(2003)}]{testi03}
{Testi}, L. 2003, in Astronomical Society of the Pacific Conference Series,
  Vol. 287, Galactic Star Formation Across the Stellar Mass Spectrum, ed. J.~M.
  {De Buizer} \& N.~S. {van der Bliek}, 163--173

\bibitem[{{Thomas} \& {Fuller}(2008)}]{thomas08}
{Thomas}, H.~S., \& {Fuller}, G.~A. 2008, \aap, 479, 751, arXiv:0712.1512

\bibitem[{{Volk} \& {Kwok}(1987)}]{volk87}
{Volk}, K., \& {Kwok}, S. 1987, \apj, 315, 654

\bibitem[{{Watson} {et~al.}(2003){Watson}, {Araya}, {Sewilo}, {Churchwell},
  {Hofner}, \& {Kurtz}}]{watson03}
{Watson}, C., {Araya}, E., {Sewilo}, M., {Churchwell}, E., {Hofner}, P., \&
  {Kurtz}, S. 2003, \apj, 587, 714

\bibitem[{{Williams} {et~al.}(2004){Williams}, {Fuller}, \&
  {Sridharan}}]{williams04}
{Williams}, S.~J., {Fuller}, G.~A., \& {Sridharan}, T.~K. 2004, \aap, 417, 115,
  arXiv:astro-ph/0401633

\bibitem[{{Wilson} {et~al.}(1970){Wilson}, {Mezger}, {Gardner}, \&
  {Milne}}]{wilson70}
{Wilson}, T.~L., {Mezger}, P.~G., {Gardner}, F.~F., \& {Milne}, D.~K. 1970,
  \aap, 6, 364

\bibitem[{{Wilson} {et~al.}(1978){Wilson}, {Pankonin}, \& {Dickey}}]{wilson78}
{Wilson}, T.~L., {Pankonin}, V., \& {Dickey}, J. 1978, \aap, 68, 303

\bibitem[{{Wilson} \& {Rood}(1994)}]{wilson94}
{Wilson}, T.~L., \& {Rood}, R. 1994, \araa, 32, 191

\bibitem[{{Wink} {et~al.}(1983){Wink}, {Wilson}, \& {Bieging}}]{wink83}
{Wink}, J.~E., {Wilson}, T.~L., \& {Bieging}, J.~H. 1983, \aap, 127, 211

\bibitem[{{Wu} \& {Evans}(2003)}]{wu03}
{Wu}, J., \& {Evans}, II, N.~J. 2003, \apjl, 592, L79, arXiv:astro-ph/0306543

\bibitem[{{Yorke}(2004)}]{yorke04}
{Yorke}, H.~W. 2004, in IAU Symposium, Vol. 221, Star Formation at High Angular
  Resolution, ed. M.~{Burton}, R.~{Jayawardhana}, \& T.~{Bourke}, 141--+

\bibitem[{{Zhang} {et~al.}(2005){Zhang}, {Hunter}, {Brand}, {Sridharan},
  {Cesaroni}, {Molinari}, {Wang}, \& {Kramer}}]{zhang05}
{Zhang}, Q., {Hunter}, T.~R., {Brand}, J., {Sridharan}, T.~K., {Cesaroni}, R.,
  {Molinari}, S., {Wang}, J., \& {Kramer}, M. 2005, \apj, 625, 864

\bibitem[{{Zinnecker} \& {Yorke}(2007)}]{zinnecker07}
{Zinnecker}, H., \& {Yorke}, H.~W. 2007, \araa, 45, 481, 0707.1279

\bibitem[{{Zoonematkermani} {et~al.}(1990){Zoonematkermani}, {Helfand},
  {Becker}, {White}, \& {Perley}}]{zoonematkermani90}
{Zoonematkermani}, S., {Helfand}, D.~J., {Becker}, R.~H., {White}, R.~L., \&
  {Perley}, R.~A. 1990, \apjs, 74, 181

\end{thebibliography}

\clearpage
\begin{center} Figure Captions \end{center}

\noindent 
Figure 1. - a) VLA 3.6\,cm D-array continuum, b) Bolocam Galactic Plane Survey 1.1\,mm, and c) three-color mid-IR (Red: 8$\mu$m, Green: 4.5$\mu$m, Blue: 3.6$\mu$m) GLIMPSE images of the G44.587 \& G44.661 field. Each image is overlaid with contours of 3.6\,cm continuum emission. The detected 3.6\,cm and 1.1\,mm sources are labeled in panels a) and b) respectively. In all three panels, ellipses mark the positions of any associated IRAS sources. Panel a) Contour levels: -3, 3, 5, 10, 15, 20, 25, 30 $\times \Delta S =$ 0.07 mJy\,beam$^{-1}$. Synthesized beam: 7.5 $\times$ 7.5" PA=38 degrees. Range of grayscale: 0.07 - 2.7 mJy\,beam$^{-1}$. Panel b) Contour levels and beam as in a). Range of grayscale: -0.03 - 0.3 Jy\,beam$^{-1}$. Panel c) Contour levels: 5, 10, 20, 30 $\times \Delta S =$ 0.07 mJy\,beam$^{-1}$. Synthesized beam: 7.5 $\times$ 7.5" PA=38 degrees. GLIMPSE image stretch: logarithmic, R: 5-500, G: 5-500, B: 20-500 MJy/Sr. Inset or smaller panels cover the area shown by the boxes displayed in the main panel. Crosses show the peak positions of detected millimeter sources in the field, and colored circles show the positions of GLIMPSE sources mentioned in Section \ref{notes}.

\noindent 
Figure 2. - a) 3.6\,cm continuum, b) 1.1\,mm, and c) GLIMPSE images of the G48.580 \& G48.616 field. Each image is overlaid with contours of 3.6\,cm continuum emission. The detected 3.6\,cm and 1.1\,mm sources are labeled in panels a) and b) respectively. Panel a) Contour levels: -3, 3, 5, 10, 15, 20, 25, 30, 35, 40, 60 $\times \Delta S =$ 1.2 mJy\,beam$^{-1}$. Synthesized beam: 9.1 $\times$ 8.7" PA=56 degrees. Range of grayscale: 1.2 - 68 mJy\,beam$^{-1}$. Panel b) Contour levels and beam as in a). Range of grayscale: -0.06 - 1.3 Jy\,beam$^{-1}$. Panel c) Contour levels: 3, 5, 10, 20, 30, 40, 60 $\times \Delta S =$ 1.2 mJy\,beam$^{-1}$. Synthesized beam: 9.1 $\times$ 8.7" PA=56 degrees. GLIMPSE image stretch: logarithmic, R: 20-1300, G: 2-600, B: 2-1000 MJy/Sr.

\begin{figure}
\epsscale{1.0}
\plotone{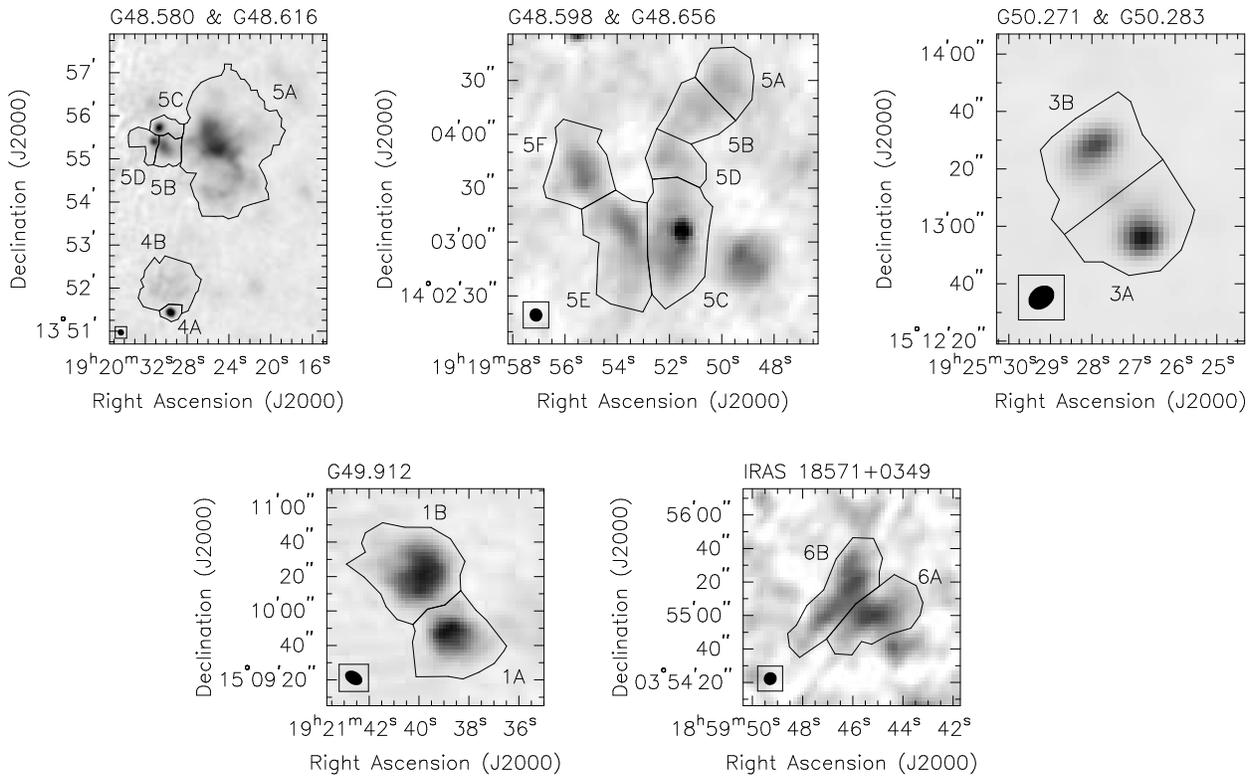}
\caption{Photometry apertures.  \label{polygons}}
\end{figure}

\begin{figure}
\epsscale{0.8}
\plotone{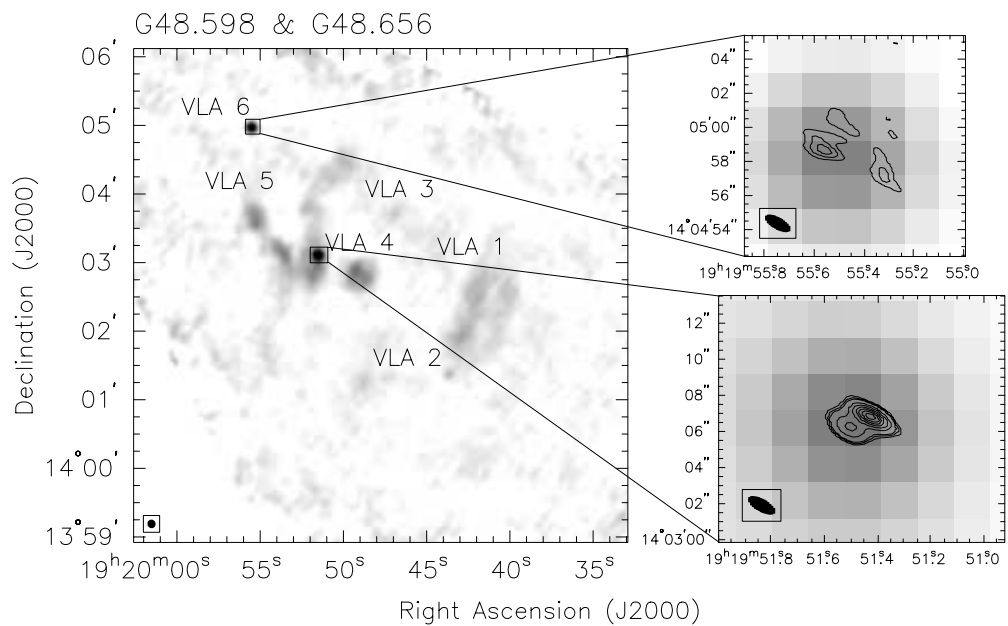}
\caption{VLA 3.6~cm B and D array images of G48.598 \& G48.656.  \label{a21a25_sridharan}}
\end{figure}

\begin{figure}
\center{
\includegraphics[angle=-90, width=4.in]{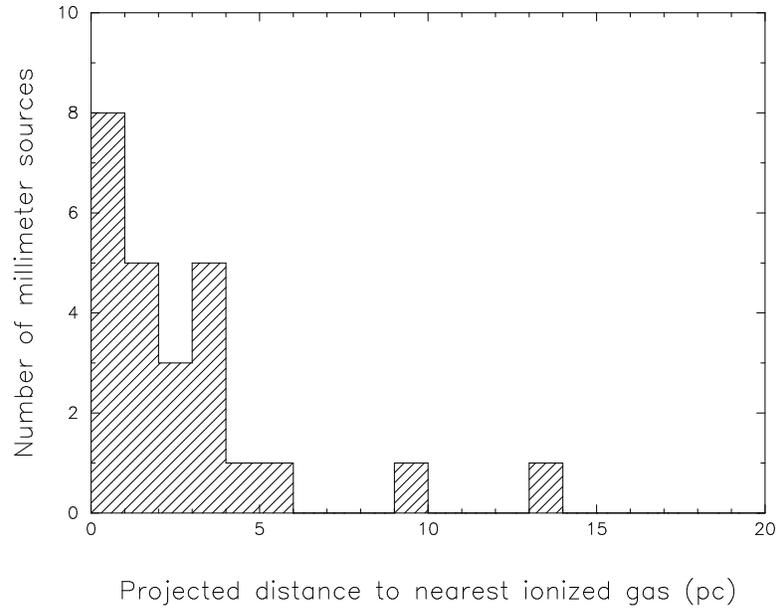}
\caption{Projected distance, in parsecs, between millimeter clump peaks and the nearest ionized gas peak. \label{ment}}
}
\end{figure}

\begin{figure}
\center{
\includegraphics[angle=-90, width=4.in]{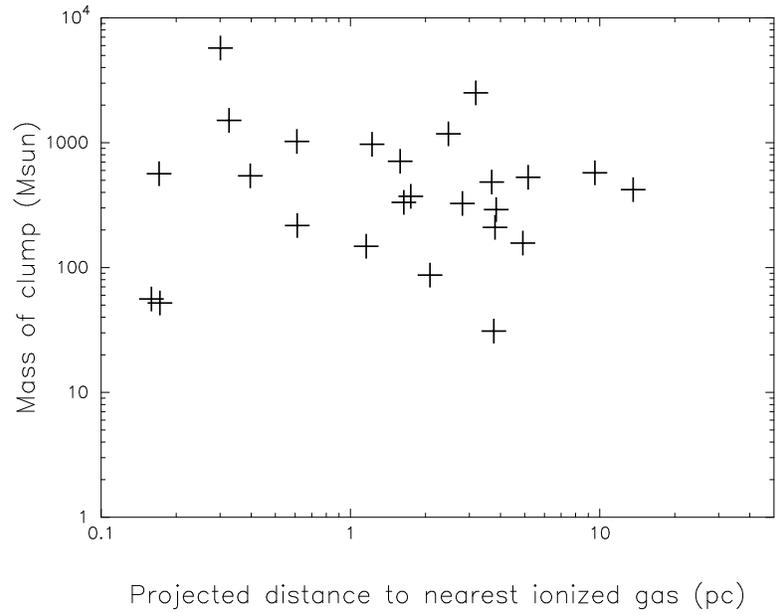}
\caption{The mass of the molecular clumps plotted against the projected distance between their peak positions and the nearest ionized gas peak. \label{mass_displ}}
}
\end{figure}

\noindent 
Figure 3. - a) 3.6\,cm continuum, b) 1.1\,mm, and c) GLIMPSE images of the G48.598 \& G48.656 field. Each image is overlaid with contours of 3.6\,cm continuum emission. The detected 3.6\,cm and 1.1\,mm sources are labeled in panels a) and b) respectively. Panel a) Contour levels: -3, 3, 4, 5, 7, 9, 12, 15, 20, 30 $\times \Delta S =$ 0.21 mJy\,beam$^{-1}$. Synthesized beam: 7.5 $\times$ 7.3" PA=36 degrees. Range of grayscale: 0.21 - 6.8 mJy\,beam$^{-1}$. The subcomponents of VLA 5 in the G48.598 \& G48.656 field are not labeled, but are shown instead in Figure \ref{polygons}. Panel b) Contour levels and beam as in a). Range of grayscale: -0.04 - 0.4 Jy\,beam$^{-1}$. Panel c) Contour levels: 3, 5, 10 $\times \Delta S =$ 0.21 mJy\,beam$^{-1}$. Synthesized beam: 7.5 $\times$ 7.3" PA=36 degrees. GLIMPSE image stretch: logarithmic, R: 20-1300, G: 2-600, B: 2-1000 MJy/Sr.

\noindent 
Figure 4. - a) 3.6\,cm continuum, b) 1.1\,mm, and c) GLIMPSE images of the G48.751 field. Each image is overlaid with contours of 3.6\,cm continuum emission. The detected 3.6\,cm and 1.1\,mm sources are labeled in panels a) and b) respectively. Panel a) Contour levels: -3, 3, 4, 5 $\times \Delta S =$ 0.07 mJy\,beam$^{-1}$. Synthesized beam: 9.6 $\times$ 7.5" PA=71 degrees. Range of grayscale: 0.07 - 0.43 mJy\,beam$^{-1}$. Panel b) Contour levels and beam as in a). Range of grayscale: -0.04 - 0.4 Jy\,beam$^{-1}$. Panel c) Contour levels: 3, 4, 5 $\times \Delta S =$ 0.07 mJy\,beam$^{-1}$. Synthesized beam: 9.6 $\times$ 7.5" PA=71 degrees. GLIMPSE image stretch: logarithmic, R: 30-500, G: 10-700, B: 5-1000 MJy/Sr.

\noindent 
Figure 5. - a) 3.6\,cm continuum, b) 1.1\,mm, and c) GLIMPSE images of the G49.912 field. Each image is overlaid with contours of 3.6\,cm continuum emission. The detected 3.6\,cm and 1.1\,mm sources are labeled in panels a) and b) respectively. Panel a) Contour levels:  -5, 5, 10, 20, 30, 40 $\times \Delta S =$ 0.04 mJy\,beam$^{-1}$. Synthesized beam: 11.1 $\times$ 7.5" PA=61 degrees. Range of grayscale:  0.04 - 2.1 mJy\,beam$^{-1}$. Panel b) Contour levels and beam as in a). Range of grayscale: -0.04 - 0.4 Jy\,beam$^{-1}$. Panel c) Contour levels: 5, 10, 20, 30, 40 $\times \Delta S =$ 0.04 mJy\,beam$^{-1}$. Synthesized beam: 11.1 $\times$ 7.5" PA=61 degrees. GLIMPSE image stretch: linear, R: 0-500, G: 0-400, B: 10-600 MJy/Sr.

\noindent 
Figure 6. - a) 3.6\,cm continuum, b) 1.1\,mm, and c) GLIMPSE images of the G50.271 \& G50.283 field. Each image is overlaid with contours of 3.6\,cm continuum emission. The detected 3.6\,cm and 1.1\,mm sources are labeled in panels a) and b) respectively. Panel a) Contour levels:  -3, 3, 5, 10, 25, 50, 75, 100, 150, 200, 225 $\times \Delta S =$ 0.15 mJy\,beam$^{-1}$. Synthesized beam: 10.0 $\times$ 7.6" PA=61 degrees. Range of grayscale:  0.15 - 34 mJy\,beam$^{-1}$. Panel b) Contour levels and beam as in a). Range of grayscale: -0.07 - 0.7 Jy\,beam$^{-1}$. Panel c) Contour levels: 5, 50, 100, 150, 200 $\times \Delta S =$ 0.15 mJy\,beam$^{-1}$. Synthesized beam: 10.0 $\times$ 7.6" PA=57 degrees. GLIMPSE image stretch: linear, R: 0-800, G: 0-150, B: 0-150 MJy/Sr.

\noindent 
Figure 7. - a) 3.6\,cm continuum, b) SEST 1.2\,mm, and c) MSX A Band (8.28$\mu$m) images of the IRAS\,18256-0742 field. Each image is overlaid with contours of 3.6\,cm continuum emission. The detected 3.6\,cm and 1.2\,mm sources are labeled in panels a) and b) respectively. Panel a) Contour levels:  -3, 3, 5, 7, 10, 15, 20, 30, 40 $\times \Delta S =$ 0.05 mJy\,beam$^{-1}$. Synthesized beam: 9.5 $\times$ 7.8" PA=-179 degrees. Range of grayscale:  0.05 - 2.0 mJy\,beam$^{-1}$. Panel b) Contour levels and beam as in a). Range of grayscale: -0.15 - 0.25 Jy\,beam$^{-1}$. Panel c) Contour levels and beam as in a). MSX 8.28$\mu$m image stretch: linear, Range: 0.0-5.0 $\times$ 10$^{-5}$ Wm$^{-2}$sr$^{-1}$.

\noindent 
Figure 8. - a) 3.6\,cm continuum, b) SEST 1.2\,mm, and c) GLIMPSE images of the IRAS\,18424-0329 field. Each image is overlaid with contours of 3.6\,cm continuum emission. The detected 3.6\,cm and 1.2\,mm sources are labeled in panels a) and b) respectively. Panel a) Contour levels:  -3, 3, 5, 10, 20, 30, 40, 50, 75, 100 $\times \Delta S =$ 0.09 mJy\,beam$^{-1}$. Synthesized beam: 8.5 $\times$ 7.5" PA=-10 degrees. Range of grayscale: 0.09 - 6.0 mJy\,beam$^{-1}$. Panel b) Contour levels and beam as in a). Range of grayscale: -0.1 - 0.2 Jy\,beam$^{-1}$. The white circles mark the positions of Clumps 2, 4 and 6 listed in B06. Panel c) Contour levels: 5, 20, 40, 60, 75, 100 $\times \Delta S =$ 0.09 mJy\,beam$^{-1}$. Synthesized beam: 8.5 $\times$ 7.5" PA=-10 degrees. GLIMPSE image stretch: logarithmic, R: 2-2000, G: 5-2000, B: 60-2000 MJy/Sr.

\noindent 
Figure 9. - a) 3.6\,cm continuum, b) SEST 1.2\,mm, and c) GLIMPSE images of the IRAS\,18571+0349 field. Each image is overlaid with contours of 3.6\,cm continuum emission. The detected 3.6\,cm and 1.2\,mm sources are labeled in panels a) and b) respectively. Panel a) Contour levels:  -3, 3, 4, 5, 7, 9, 12, 15 $\times \Delta S =$ 0.24 mJy\,beam$^{-1}$. Synthesized beam: 7.9 $\times$ 7.5" PA=-48 degrees. Range of grayscale: 0.24 - 4.1 mJy\,beam$^{-1}$. Panel b) Contour levels and beam as in a). Range of grayscale: -0.2 - 0.68 Jy\,beam$^{-1}$. Panel c) Contour levels: 3, 5, 9, 13, 17 $\times \Delta S =$ 0.24 mJy\,beam$^{-1}$. Synthesized beam: 7.9 $\times$ 7.5" PA=-48 degrees. GLIMPSE image stretch: logarithmic, R: 50-1000, G: 5-500, B: 5-700 MJy/Sr.

\noindent 
Figure 10. - b) SEST 1.2\,mm, and c) MSX A Band (8.28$\mu$m) images of the IRAS\,18586+0106 field. No significant 3.6~cm emission above 3 $\times \Delta S$ was detected in this field, $\Delta S =$ 0.16 mJy\,beam$^{-1}$, therefore the 3.6\,cm image is not shown. Synthesized beam: 9.6 $\times$ 8.0" PA=-39 degrees. Panel b) Range of grayscale: -0.2 - 0.6 Jy\,beam$^{-1}$. The detected 1.2\,mm sources are labeled. Panel c) MSX 8.28$\mu$m image stretch: linear, Range: 0.0-3.0 $\times$ 10$^{-5}$ Wm$^{-2}$sr$^{-1}$.

\noindent
Figure 11. - Photometry apertures used for the six multiply-peaked sources detected in our 3.6~cm VLA observations. VLA sub-sources are labeled in each panel. Grayscale in each panel: VLA 3.6~cm emission. Grayscale ranges: G48.580 \& G48.616 field: -$5 - 50$ mJy\,beam$^{-1}$, G48.598 \& G48.656 field: -$0.5 - 5$ mJy\,beam$^{-1}$, G50.271 \& G50.283 field: -$2 - 20$ mJy\,beam$^{-1}$, G49.912 field: -$0.2 - 2$ mJy\,beam$^{-1}$, and IRAS~18571+0106 field: -$0.2 - 2$ mJy\,beam$^{-1}$. When calculating the integrated fluxes, a cut-off above the 1$\times \Delta S$ level was applied within the aperture. 
An additional 10-20\% error should be added to the measured integrated fluxes for these sub-sources, due to the arbitrary nature of the positions of the aperture boundaries which have been chosen to separate the source components.

\noindent
Figure 12. - VLA 3.6~cm B and D array continuum images of the G48.598 \& G48.656 field. \textit{Left panel:} VLA 3.6~cm D array grayscale image of entire region. Grayscale range: 0.0 - 5.0 mJy\,beam$^{-1}$. \textit{Top right panel:} VLA~6. Contours: VLA 3.6~cm B array image, levels: -3, 3, 4, 5 $\times \sigma =$ 0.1 mJy\,beam$^{-1}$. Grayscale: VLA 3.6~cm D array image, range: 0.2 - 14 mJy\,beam$^{-1}$. \textit{Bottom right panel:} VLA~5C. Contours: VLA 3.6~cm B array image, levels: -3, 3, 4, 5, 10, 15, 20, 25, 30, 35 $\times \sigma =$0.1 mJy\,beam$^{-1}$. Grayscale: VLA 3.6~cm D array image, range: 0.2 - 14 mJy\,beam$^{-1}$.

\noindent
Figure 13. - Distribution of projected distances (in parsecs) between the position of the millimeter clump peak and the peak of the nearest ionized gas.  

\noindent
Figure 14. - The mass of the molecular clump taken from Table \ref{selectsrc} plotted against the projected distance, in parsecs, between the millimeter clump peak positions and the nearest ionized gas peak.  

\noindent
Figure 15. - Comparison of the dense gas traced by $^{13}$CO(J=1-0), molecular clumps traced by millimeter continuum, and ionized gas peaks for each field. $^{13}$CO(J=1-0) is shown in grey scale ranging from 0 to 1.1 times the peak flux density. 
Contours trace the millimeter continuum emission with contour levels plotted at 10, 20, 30, 40, 50, 60, 70, 80, and 90\% of the peak emission. The peak $^{13}$CO(J=1-0) temperature and $\sim$1.1\,mm flux density are given below each sub-Figure. The locations of the ionized gas peaks from HII regions shown in Figure 1 are shown by crosses.

\noindent
Figure 16. - The mass of the associated ionizing stars, derived from the HII region properties given in Table \ref{vlaproperties}, plotted against the clump mass. A line of best fit is plotted upon the data: M$_{\star}=1.0 \times \rm{M}_{\rm{clump}}^{\phantom{clump}0.5}$.  

\clearpage

\clearpage

\clearpage

\clearpage

\clearpage

\clearpage

\clearpage

\clearpage

\clearpage

\clearpage

\clearpage

\clearpage

\clearpage

\clearpage

\begin{figure}
\epsscale{0.5}
\plotone{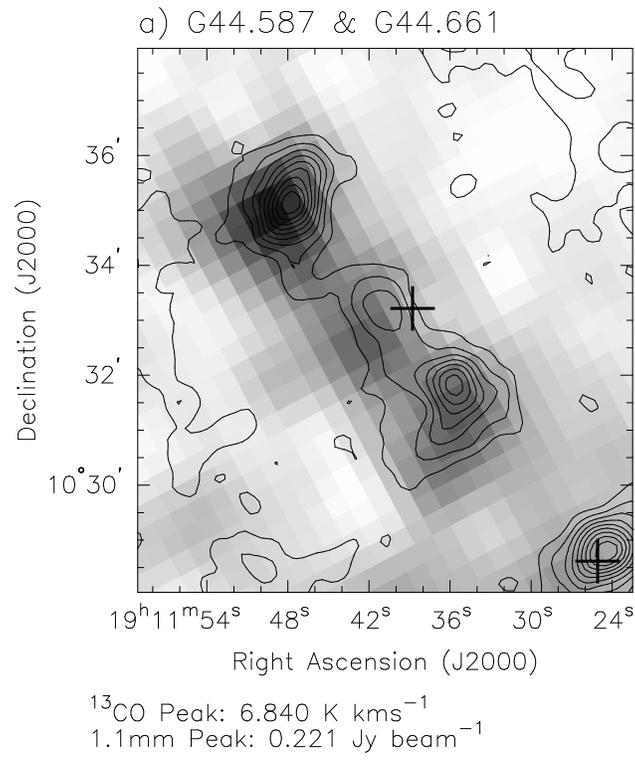}
\caption{Relationship between the dense gas traced by $^{13}$CO and millimeter clumps.\label{1mm_13co}}
\end{figure}

\clearpage

\begin{figure}
\epsscale{0.5}
\figurenum{15}
\plotone{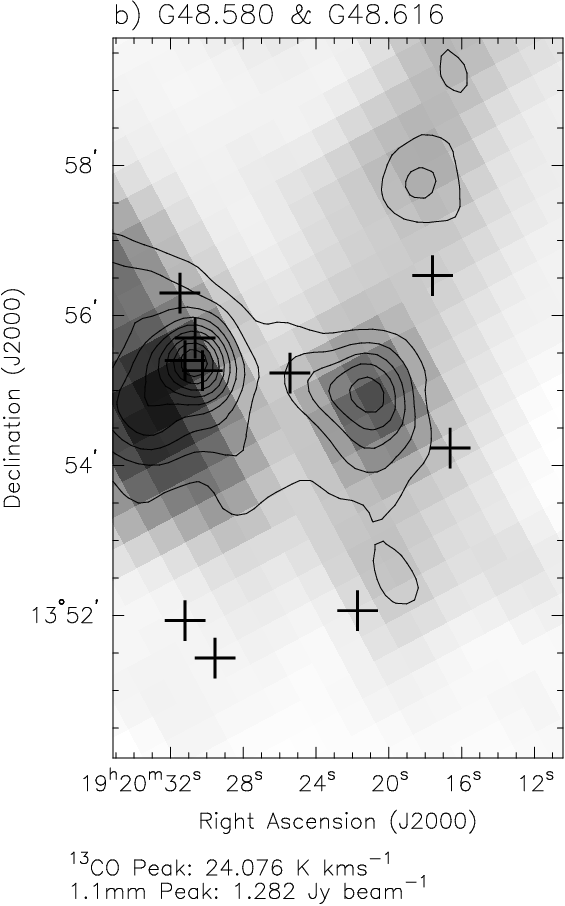}
\caption{continued}
\end{figure}

\clearpage

\begin{figure}
\epsscale{0.5}
\figurenum{15}
\plotone{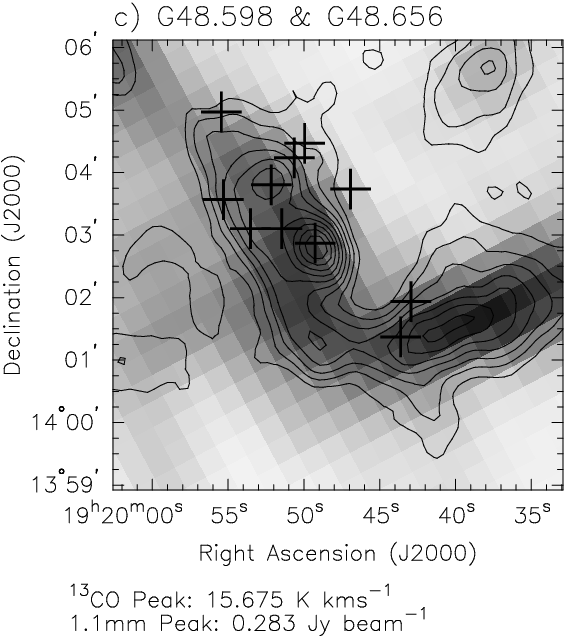}\\
\vspace{1cm}
\plotone{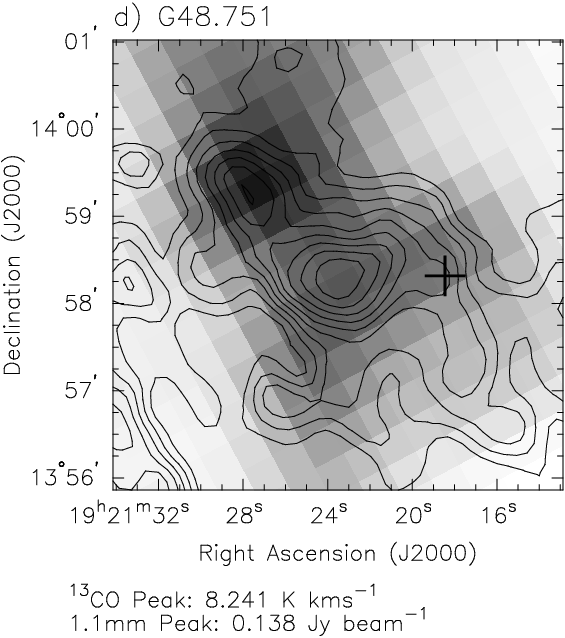}
\caption{continued}
\end{figure}

\clearpage

\begin{figure}
\epsscale{0.5}
\figurenum{15}
\plotone{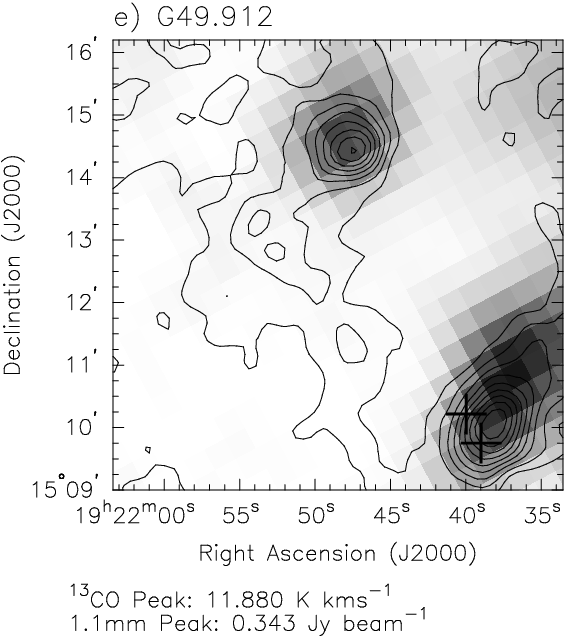}\\
\vspace{1cm}
\plotone{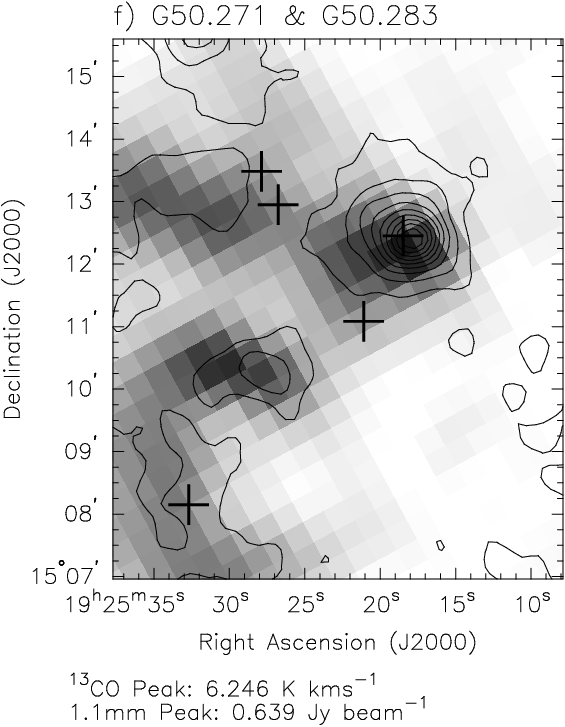}
\caption{continued}
\end{figure}

\clearpage

\begin{figure}
\epsscale{0.5}
\figurenum{15}
\plotone{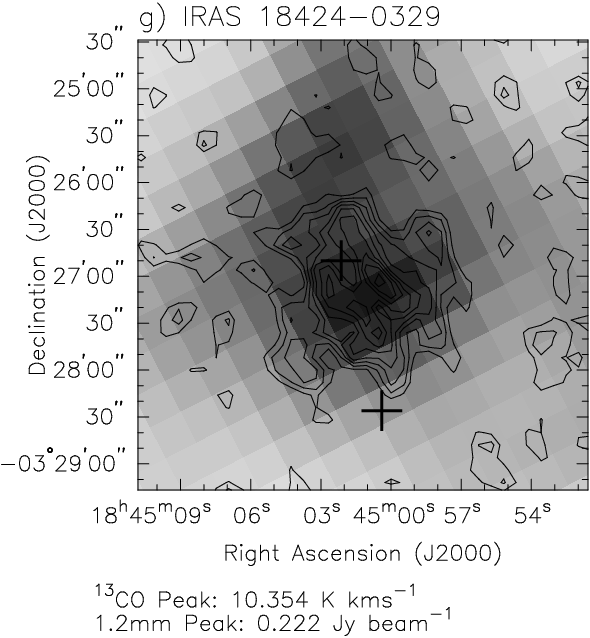}\\
\vspace{1cm}
\plotone{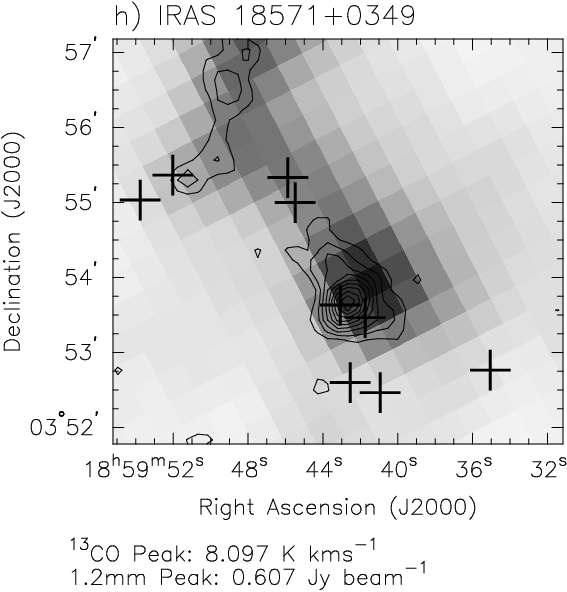}
\caption{continued}
\end{figure}

\clearpage

\begin{figure}
\epsscale{0.5}
\center{
\includegraphics[angle=-90, width=4.in]{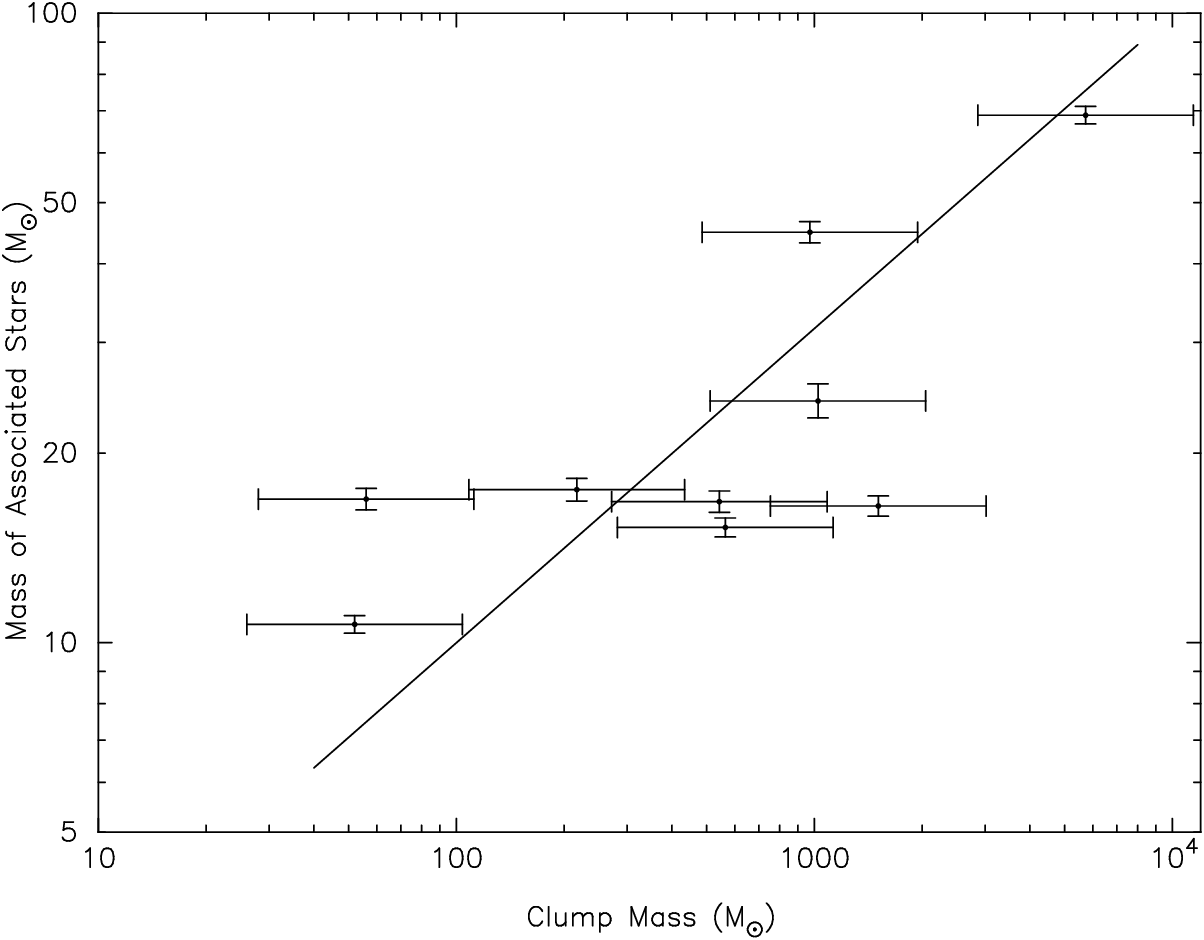}
\caption{The relationship between clump mass and the combined mass of the associated stars. \label{mclump_mstar_plot}}
}
\end{figure}

\clearpage

\begin{deluxetable}{lccccccclccccl}
\rotate
\tabletypesize{\tiny}
\tablecolumns{14}
\tablewidth{0pt}
\tablecaption{Observed Millimeter Clumps \label{selectsrc}}
\tablehead{ (1)&(2)&(3)&(4)&(5)&(6)&(7)&(8)&(9)&(10)&(11)&(12) & (13) & (14) \\
\colhead{Source Name} & \colhead{R.A.~(2000)}&\colhead{Dec.~(2000)} & \colhead{Gal. $\ell$}&\colhead{Gal. b} & \colhead{v$_{\rm{13CO}}$} &  \colhead{d$_{\rm near}$} &  \colhead{d$_{\rm far}$} & Assumed & \colhead{L$_{\rm{IRAS}}$} & \colhead{S$_{1.0/1.2\rm{mm}}$} & \colhead{M}& \colhead{Ref.} & \colhead{Source} \\
  & (h~m~s) & ($^{\circ}$~'~")& (deg.) & (deg.) & (kms$^{-1}$) & (kpc) & (kpc) & Distance & (10$^{3}~$L$_\sun$) & (Jy) & (M$_\sun$) & &Type\\}
\startdata
G044.521+00.387 & 19 11 24.7 & +10 28 43 & 44.5211 & 0.3871 & 51.3 $\pm$ 0.7 &  3.8 $\pm ^{0.8}_{0.6}$ & 8.3 $\pm {1.2}$ & near, HISA & 0.932 - 2.13 & 0.49 $\pm$ 0.11 & 56 & BGPS & Serend.\\ 
G044.587+00.371 & 19 11 35.6 & +10 31 47 & 44.5871 & 0.3711 & 16.3 $\pm$ 1.5 & 1.2 $\pm ^{0.3}_{0.2}$ & 10.9 $\pm {1.4}$ & far, HISA & \nodata & 0.56 $\pm$ 0.14 & 528 & BGPS &Select.\\
G044.617+00.365 & 19 11 40.3 & +10 33 13 & 44.6171 & 0.3652 & 17.7 $\pm$ 1.4 & 1.3 $\pm ^{0.3}_{0.2}$ & 10.8 $\pm {1.4}$ & far, HISA &  \nodata & 0.16 $\pm$ 0.08 & 148 & BGPS & Serend.\\ 
G044.661+00.351 & 19 11 48.3 & +10 35 10 & 44.6611 & 0.3512 & 17.6 $\pm$ 1.8 & 1.3 $\pm ^{0.3}_{0.2}$ & 10.8 $\pm {1.4}$ & far, HISA & 24.3 - 24.6 & 0.62 $\pm$ 0.12 & 574 & BGPS &Select.\\
G048.540+00.040  & 19 20 19.9 & +13 52 25 & 48.5405 & 0.0398 & 15.6 $\pm$ 1.5 &  1.2 $\pm ^{0.3}_{0.2}$ & 10.0 $\pm {1.3}$  & far, HISA & \nodata & 0.42 $\pm$ 0.12 & 333 & BGPS & Serend.\\
G048.580+00.056  & 19 20 21.0 & +13 54 59 & 48.5805 & 0.0558 & 16.2 $\pm$ 1.8 & 1.2 $\pm ^{0.3}_{0.2}$ & 10.0 $\pm {1.3}$ & far, CA & \nodata & 3.16 $\pm$ 0.29 & 2508 & BGPS &Select. \\
G048.598+00.252  & 19 19 40.3 & +14 01 27 & 48.5984 & 0.2518 & 8.4 $\pm$ 2.4 & 0.6 $\pm {0.2}$ & 10.6 $\pm {1.4}$ & far, CA & \nodata & 1.32 $\pm$ 0.23 & 1177 & BGPS &Select.\\
G048.605+00.024 & 19 20 30.8 & +13 55 21 & 48.6045 & 0.0238 & 18.0 $\pm$ 1.8 & 1.4 $\pm ^{0.3}_{0.2}$ & 9.9 $\pm {1.3}$ & far, CA & 927 - 932 & 7.36 $\pm$ 0.55 & 5725 & BGPS & Serend.\\ 
G048.610+00.220 & 19 19 48.7 & +14 01 11 & 48.6104 & 0.2198 & 9.2 $\pm$ 2.8 & 0.7 $\pm ^{0.3}_{0.2}$ & 10.5 $\pm {1.4}$ & far, CA & \nodata & 0.24 $\pm$ 0.10 & 210 & BGPS & Serend.\\ 
G048.616+00.088  & 19 20 18.2 & +13 57 48 & 48.6165 & 0.0878 & 16.9 $\pm$ 1.3 & 1.3 $\pm ^{0.3}_{0.2}$ & 10.0 $\pm {1.3}$ & far, CA & \nodata & 0.61 $\pm$ 0.13 & 484 & BGPS &Select. \\
G048.634+00.230 & 19 19 49.3 & +14 02 44 & 48.6344 & 0.2298 & 9.4 $\pm$ 3.6 & 0.7 $\pm {0.3}$ & 10.5 $\pm {1.4}$ & far, CA & 60.5 - 168 & 0.62 $\pm$ 0.14 & 543 & BGPS & Serend.\\ 
G048.656+00.228 & 19 19 52.3 & +14 03 51 & 48.6564 & 0.2278 & 12.7 $\pm$ 2.5 & 1.0 $\pm ^{0.3}_{0.2}$ & 10.3 $\pm ^{1.4}_{1.3}$ & far, CA & \nodata & 0.67 $\pm$ 0.16 & 564 & BGPS &Select. \\
G048.751-00.142 & 19 21 24.0 & +13 58 25 & 48.7506 & -0.1421 & 66.3 $\pm$ 1.0 & 5.3 $\pm 0.7$ \tablenotemark{\diamond} & 5.3 $\pm 0.7$ \tablenotemark{\diamond} & far, HISA & \nodata & 0.39 $\pm$ 0.12 & 87 & BGPS & Select.\\
G048.771-00.148  & 19 21 27.6 & +13 59 18 & 48.7706 & -0.1481 & 66.9 $\pm$ 1.2 & 5.3 $\pm 0.7$ \tablenotemark{\diamond} & 5.3 $\pm 0.7$ \tablenotemark{\diamond} & far, HISA & \nodata & 0.14 $\pm$ 0.08 & 31 & BGPS & Serend.\\
G049.830+00.370  & 19 21 37.9 & +15 10 03 & 49.8303 & 0.3703 & 5.2 $\pm$ 1.8 & 0.4  $\pm ^{0.2}_{0.1}$ & 10.6 $\pm {1.4}$ & far, CA & 127 & 1.09 $\pm$ 0.18 & 972 & BGPS & Serend.\\
G049.912+00.370 & 19 21 47.5 & +15 14 23 & 49.9123 & 0.3704 & 8.1 $\pm$ 1.4 & 0.6 $\pm ^{0.2}_{0.1}$ & 10.3 $\pm ^{1.4}_{1.3}$ & far, HISA & 7.61 - 21.2 & 0.50 $\pm$ 0.13 & 421 & BGPS &Select.\\
G050.271-00.442  & 19 25 27.5 & +15 10 18 & 50.2706 & -0.4415 & 14.8 $\pm$ 1.3 & 1.2 $\pm {0.2}$ & 9.7 $\pm {1.3}$ & far, HISA & 0.960 - 254 & 0.21 $\pm$ 0.10 & 157 & BGPS &Select. \\
G050.283-00.390 & 19 25 17.6 & +15 12 25 & 50.2826 & -0.3895 & 16.1 $\pm$ 1.5 & 1.2 $\pm ^{0.3}_{0.2}$ & 9.6 $\pm {1.3}$ & far, CA & 281 - 286 & 1.40 $\pm$ 0.21 & 1024 & BGPS & Select.\\

IRAS~18256-0742 Clump~1 & 18 28 18.9 & -07 40 06 & 23.4730 & 1.6041 & \nodata & 3.0 & \nodata & near, B06 & 10.5  & 0.59 & 52  & B06 &Select.\\

IRAS~18424-0329 Clump~2 & 18 45 00.5 & -03 27 04 & 29.1280 & -0.1449 & 47.4 $\pm$ 1.5 & 3.2 $\pm$ 0.5 & 11.6 $\pm$ 1.5 & far, HISA & 55 \tablenotemark{\dag} & 0.53 & 710 &  B06 &Select.\\
IRAS~18424-0329 Clump~4 & 18 45 01.6 & -03 27 20 & 29.1261 & -0.1510 & 47.6 $\pm$ 1.4 & 3.2 $\pm$ 0.5 & 11.6 $\pm$ 1.5 & far, HISA & 55 \tablenotemark{\dag} & 0.28 & 372  &  B06 & Serend.\\
IRAS~18424-0329 Clump~6 & 18 45 00.5 & -03 27 36 & 29.1201 & -0.1490 & 47.6 $\pm$ 1.7 & 3.2 $\pm$ 0.5 & 11.6 $\pm$ 1.5 & far, HISA & 55 \tablenotemark{\dag}  & 0.24 & 326 &  B06 &Serend.\\

IRAS~18571+0349 Clump~1 & 18 59 42.7 & +03 53 42 & 37.3409 & -0.0615 & 55.5 $\pm$ 1.0 & 3.7 $\pm ^{0.6}_{0.5}$ & 9.8 $\pm {1.3}$ & far, KB94 & 106 & 1.55 & 1509 &  B06 &Select.\\
IRAS~18571+0349 Clump~3 & 18 59 49.0 & +03 56 30 & 37.3944 & -0.0635 & 56.7 $\pm$ 2.6 & 3.8 $\pm ^{0.7}_{0.6}$ & 9.7 $\pm {1.3}$ & far, KB94 & \nodata & 0.31 & 291 & B06 &Select.\\
IRAS~18571+0349 Clump~4 & 18 59 51.2 & +03 55 18 & 37.3808 & -0.0808 & 57.1 $\pm$ 1.1 & 3.8 $\pm ^{0.7}_{0.6}$ & 9.7 $\pm {1.3}$  & far, KB94 & \nodata & 0.23 & 217 & B06 &Serend.\\

IRAS~18586+0106 Clump~1 & 19 01 15.8 & +01 12 28 & 35.1276 & -1.6345 & \nodata & 2.7 & \nodata & near, B06 & \nodata & 1.47 & 110 & B06 &Select.\\
IRAS~18586+0106 Clump~3 & 19 00 59.8 & +01 13 40 & 35.1150 & -1.5661 & \nodata & 2.7 & \nodata & near, B06 &  \nodata & 0.43 & 32 &  B06 &Serend.\\
IRAS~18586+0106 Clump~4 & 19 01 01.4 & +01 13 16 & 35.1121 & -1.5751 & \nodata & 2.7 & \nodata & near, B06 &  \nodata  & 0.52 & 39 &  B06 &Serend. \\
IRAS~18586+0106 Clump~5 & 19 01 12.1 & +01 10 44 & 35.0949 & -1.6340 & \nodata & 2.7 & \nodata & near, B06 &  4.4  & 0.49 & 36 &  B06 &Serend.\\
IRAS~18586+0106 Clump~6 & 19 01 27.0 & +01 10 28 & 35.1193 & -1.6912 & \nodata & 2.7 & \nodata & near, B06 &  \nodata  & 0.30 & 22 &  B06 &Serend.\\
IRAS~18586+0106 Clump~7 & 19 00 59.3 & +01 11 08 & 35.0765 & -1.5835 & \nodata & 2.7 & \nodata & near, B06 &  \nodata & 0.21 & 16 &  B06 &Serend.\\

\enddata
\tablecomments{Columns: 
1. Millimeter clump name.
2. \& 3. Equatorial J2000 coordinates of millimeter clump.
4. \& 5. Galactic coordinates of millimeter clump.
6. Mean velocity of associated GRS $^{13}$CO emission at position of mm source.
7. \& 8. Near and far distance to millimeter clump in kpc.
9. Assumed distance used to calculate the IRAS luminosity and clump mass. The method used to determine whether the source is at the near or far distance is also given: HISA: HI self-absorption, CA: 21\,cm continuum absorption, B06: taken from B06. KB94: \citet{kuchar94}.
10. Luminosity derived from associated IRAS source fluxes.
11. Millimeter flux measured at a wavelength of 1.1~mm for the sources taken from the BGPS, and 1.2~mm for those taken from B06.
12. The calculated dust mass of the millimeter clump.
13. References -- BGPS: Bolocam Galactic Plane Survey preliminary images, Aguirre et al. (2009), in preparation. B06: \citet{beltran06}
14. Denotes whether source was selected (Select.) or serendipitously fell within the VLA field (Serend.).
}
\tablenotetext{\dag}{It is not certain which of the clumps listed by B06 is associated with IRAS~18424-0329, however the general 1.2~mm emission in this field is coincident with the IRAS source.}
\tablenotetext{\diamond}{The velocity of this source is too high to be explained by the galactic rotation curve at this galactic longitude, therefore the highest possible velocity at this longitude was instead assumed. These sources were originally thought to be at a different velocity, placing them at $d_{\rm{near}} \sim$1\,kpc.}
\end{deluxetable}

\clearpage

\begin{deluxetable}{llccccc}
\tabletypesize{\scriptsize}
\tablecolumns{8}
\tablewidth{0pt}
\tablecaption{Summary of VLA Pointing Centers \label{obssum}}
\tablehead{\colhead{Source Name} & \colhead{Observation Date} & \multicolumn{2}{c}{Pointing Center}  \\ 
 & & R.A.~(2000) & Dec.~(2000)}
\startdata
G44.587  & 2007 April 6 & 19 11 36.2 & +10 31 47.0 \\
\nodata  & 2007 May 1 & 19 11 33.9 & +10 31 13.0  \\
G44.661  & 2007 April 6 & 19 11 48.5 & +10 35 10.0  \\
\nodata  & 2007 May 1 & 19 11 46.2 & +10 34 48.0 \\
G48.580  & 2007 April 6 & 19 20 22.0 & +13 54 52.0 \\
 \nodata  & 2007 May 1 & 19 20 25.4 & +13 52 22.3 \\
G48.598  & 2007 April 6 & 19 19 40.4 & +14 01 22.0\\
\nodata  & 2007 May 1 & 19 19 43.9 & +14 01 48.2 \\
G48.616 & 2007 April 6  & 19 20 18.8 & +13 57 44.0 \\
\nodata & 2007 May 1  & 19 20 22.8 & +13 56 18.6 \\
G48.656  & 2007 April 6 & 19 19 53.3 & +14 03 53.0 \\
\nodata  & 2007 May 1 & 19 19 53.9 & +14 03 53.2 \\
G48.751 & 2007 April 6 & 19 21 23.4 & +13 58 27.0 \\
\nodata & 2007 May 1 & 19 21 23.4 & +13 58 27.4 \\
G49.912 & 2007 April 6 & 19 21 48.3 & +15 14 32.0  \\
\nodata & 2007 May 1 & 19 21 47.7 & +15 14 09.0  \\
G50.271  & 2007 April 6 & 19 25 28.3 & +15 10 22.0 \\
\nodata  & 2007 May 1 & 19 25 27.7 & +15 10 22.0 \\
G50.283  & 2007 April 6 & 19 25 19.3 & +15 12 23.0 \\
\nodata  & 2007 May 1 & 19 25 22.6 & +15 13 03.1  \\
IRAS~18256-0742 & 2007 April 6 & 18 28 20.5 & -07 40 22.0 \\
IRAS~18424-0329 & 2007 April 6 & 18 45 03.2 & -03 26 48.0  \\
\nodata & 2007 May 1 & 18 45 01.2 & -03 26 58.0 \\
IRAS~18571+0349 & 2007 April 6 & 18 59 40.0 & +03 53 34.0 \\
\nodata & 2007 May 1 & 18 59 42.9 & +03 54 20.0 \\
IRAS~18586+0106 & 2007 April 6 & 19 01 10.5 & +01 11 16.0 \\ 
\nodata & 2007 May 1 & 19 01 10.5 & +01 11 16.0 \\
\enddata

\end{deluxetable}

\clearpage

\begin{deluxetable}{lllcclcccp{2in}}
\rotate
\tabletypesize{\tiny}
\tablecolumns{8}
\tablewidth{0pt}
\tablecaption{Observed Parameters of VLA 3.6~cm Sources \label{vlaresult}}
\tablehead{\colhead{Observed} & \colhead{VLA Source} & \colhead{Flux} & \multicolumn{2}{c}{Peak}  & \colhead{Peak} & \colhead{Angular} & \colhead{Position} & \colhead{Solid} & \colhead{Alternative Identifiers} \\ 
\colhead{Field} & \colhead{Name} &\colhead{Density} & \multicolumn{2}{c}{Position}  & \colhead{Flux} &   \colhead{Size} &   \colhead{Angle} & \colhead{Angle} & \colhead{for VLA Source} \\ 
& & (mJy)  & R.A.~(J2000) & Dec.~(J2000)& (mJy\,beam$^{-1}$) & (arcsec) & (degrees) & (steradians $\times$10$^{-8}$)\tablenotemark{\dagger} & }
\startdata
G44.587 \& G44.661 & VLA~1 & 231 $\pm$ 5 & 19 11 25.1 & +10 28 37 & 31.6 $\pm$ 0.6 & 52 $\times$ 40 & 20 & 4.08 & IRAS~19090+1023, G44.521+0.388 \\
\nodata  & VLA~2 & 1.34 $\pm$ 0.07  & 19 11 38.8 & +10 33 13 & 1.40 $\pm$ 0.08 & \nodata & \nodata & \nodata & New \\
\nodata  & VLA~2-P & 1.43 $\pm$ 0.07 & 19 11 38.8 & +10 33 14 & 1.52 $\pm$ 0.03 & $<$ 1.1 & unres. & \nodata & New \\
\\
G48.580 \& G48.616 & VLA~1 & 51.8 $\pm$ 1.5 & 19 20 16.6 & +13 54 14  & 10.6 $\pm$ 1.2 & 27 $\times$ 18 & -70 & 2.57 & New \\
\nodata & VLA~2 & 168 $\pm$ 4 & 19 20 17.6 & +13 56 32 & 10.8 $\pm$ 1.2 & 68 $\times$ 23 & -50 & 7.62 & \nodata \\
\nodata & VLA~3 & 42.0 $\pm$ 1.2 & 19 20 21.7 & +13 52 04 & 7.28 $\pm$ 1.21 & 24 $\times$ 23 & -40 & 2.73 & New \\
\nodata & VLA~4 &  434 $\pm$ 9 & 19 20 29.6 & +13 51 26 & 140 $\pm$ 3 & 79  $\times$ 62 & 90 & 11.0 & G48.54-0.00, IRAS~19181+1346 \\
\nodata & VLA~4A & 140 $\pm$ 3 \tablenotemark{\diamond} & 19 20 29.6 & +13 51 26 & 140 $\pm$ 3 & \nodata & \nodata & \nodata & \nodata \\
\nodata & VLA~4A-P & 140 $\pm$ 3 & 19 20 29.6 & +13 51 26 & 141 $\pm$ 3 & 1.8 $\times$ $<$1.3 & 59 & \nodata & \nodata \\
\nodata & VLA~4B & 294 $\pm$ 6 \tablenotemark{\diamond} & 19 20 31.2 & +13 51 56 & 15.4 $\pm$ 1.2 & 64 $\times$ 55 & 0 & 9.91 & \nodata \\
\nodata & VLA~5 &  3230 $\pm$ 65 & 19 20 30.7 & +13 55 42 & 127 $\pm$ 3 & 187 $\times$ 144 & -40 & 51.4 & NRAO~608, W51D, RRF~376, IRAS~19181+1349, G48.61+0.02\\
\nodata & VLA~5A & 2370  $\pm$ 48 \tablenotemark{\diamond}& 19 20 25.4 & +13 55 14 & 49.8 $\pm$ 1.6 & 180 $\times$ 130 & -80 & 43.4 & G48.59+0.04 \\
\nodata & VLA~5B & 360 $\pm$ 7 \tablenotemark{\diamond} & 19 20 30.2 & +13 55 16 & 63.0 $\pm$ 1.7 & 44 $\times$ 43 & 90 & 3.20 & 48.603+0.026, G48.60+0.03 \\
\nodata & VLA~5C &  254 $\pm$ 5 \tablenotemark{\diamond} & 19 20 30.7 & +13 55 42 & 127 $\pm$ 3 & 50 $\times$ 24 & -20 & 2.25 & IRAS~19181+1349, G48.61+0.02 \\
\nodata & VLA~5C-P & 321 $\pm$ 5 & 19 20 30.7 & +13 55 42 & 120 $\pm$ 2 & 9.8 $\times$ 7.2 & 160 & \nodata & IRAS~19181+1349, G48.61+0.02 \\
\nodata & VLA~5D & 252 $\pm$ 5 \tablenotemark{\diamond} & 19 20 31.2 & +13 55 24 & 109 $\pm$ 2 & 51 $\times$ 30 & 50 & 2.70 & IRAS~19181+1349, G48.61+0.02 \\
\nodata & VLA~5D-P & 284 $\pm$ 6 & 19 20 31.0 & +13 55 23 & 94.3 $\pm$ 1.9 & 14.2 $\times$ 11.0 & 48 & \nodata & IRAS~19181+1349, G48.61+0.02 \\
\nodata & VLA~6 &  293 $\pm$ 6 & 19 20 31.5 & +13 56 18 & 34.2 $\pm$ 1.4 & 51 $\times$ 30 & 50 & 5.03 & \nodata \\
\\
G48.598 \& G48.656 & VLA~1 & 72.8 $\pm$ 1.6 & 19 19 42.9 & +14 01 56 & 1.85 $\pm$ 0.21 & 94 $\times$ 59 & 80 & 12.3 & 1917+1356 (Taylor~et~al~1996) \\
\nodata & VLA~2 & 3.21 $\pm$ 0.14 & 19 19 43.6 & +14 01 22 & 1.52 $\pm$ 0.21 & \nodata & \nodata & \nodata & 1917+1356 (Taylor~et~al~1996) \\
\nodata & VLA~2-P & 2.83 $\pm$ 0.07 & 19 19 43.6 & +14 01 23 & 1.38 $\pm$ 0.03 & 8.7 $\times$ 6.5 & 134 & \nodata & 1917+1356 (Taylor~et~al~1996) \\
\nodata & VLA~3 & 7.69 $\pm$ 0.27 & 19 19 46.9 & +14 03 44 & 1.60 $\pm$ 0.21 & \nodata & 40 & \nodata & New  \\
\nodata & VLA~3-P & 6.65 $\pm$ 0.14 & 19 19 46.9 & +14 03 43 & 1.44 $\pm$ 0.03 & 17.2 $\times$ 11.3 & 135 & \nodata & New \\
\nodata & VLA~4 & 28.0 $\pm$ 0.6 & 19 19 49.3 & +14 02 52 & 3.38 $\pm$ 0.22 & 36 $\times$ 35 & 50 & 3.04 & WFS70, WFS71, IRAS~19175+1357 \\
\nodata & VLA~5 &  189 $\pm$ 4 & 19 19 51.5 & +14 03 06 & 11.0 $\pm$ 0.3 & 143 $\times$ 98 & 60 & 15.5 & New  \\
\nodata & VLA~5A & 19.1 $\pm$ 0.5 \tablenotemark{\diamond} & 19 19 49.9 & +14 04 28 & 2.97 $\pm$ 0.22 & 25 $\times$ 24 & -40 & 1.73 & New \\
\nodata & VLA~5B & 16.5 $\pm$ 0.4 \tablenotemark{\diamond} & 19 19 50.6 & +14 04 14 & 2.41 $\pm$ 0.22 & 31 $\times$ 26 & 50 & 1.79 & New \\
\nodata & VLA~5C & 58.1 $\pm$ 1.2 \tablenotemark{\diamond} & 19 19 51.5 & +14 03 06 & 11.0 $\pm$ 0.3 & 65 $\times$ 29 & 90 & 4.23 & New  \\
\nodata & VLA~5C-P & 25.5 $\pm$ 0.5 & 19 19 51.6 & +14 03 06 & 9.43 $\pm$ 0.19 & 10.4 $\times$ 9.0 & 0 & \nodata & New \\
\nodata & VLA~5D & 11.7 $\pm$ 0.3 \tablenotemark{\diamond} & 19 19 52.1 & +14 03 48 & 2.49 $\pm$ 0.22 & 22 $\times$ 22 & -30 & 1.38 & New \\
\nodata & VLA~5E & 42.9 $\pm$ 1.0 \tablenotemark{\diamond} & 19 19 53.5 & +14 03 06 & 4.48 $\pm$ 0.23 & 54 $\times$ 40 & -50 & 3.73 & New \\
\nodata & VLA~5F &  40.3 $\pm$ 0.9 \tablenotemark{\diamond} & 19 19 55.3 & +14 03 34 & 5.75 $\pm$ 0.24 & 38 $\times$ 32 & -60 & 2.60 & New  \\
\nodata & VLA~6 & 28.7 $\pm$ 0.6 & 19 19 55.4 & +14 04 58 & 25.5 $\pm$ 0.6 & \nodata & \nodata & \nodata & IRAS 19176+1359, New \\
\nodata & VLA~6-P & 29.8 $\pm$ 0.6 & 19 19 55.5 & +14 04 58 & 27.0 $\pm$ 0.5 & 2.4 $\times$ 2.3 & 129 & \nodata &  IRAS 19176+1359, New \\
\\
G48.751 & VLA~1 & 5.74 $\pm$ 0.14 & 19 21 18.5 & +13 58 19 & 0.431 $\pm$ 0.071 & 47 $\times$ 21 & 40  & 5.25 & New \\
\\
G49.912 & VLA~1 & 459 $\pm$ 9 & 19 21 39.0 & +15 09 45 & 52.0 $\pm$ 1.0 & 98 $\times$ 55 & -50 & 9.47 & G49.8+0.4, IRAS~19193+1504, RFS~849\\
\nodata  & VLA~1A & 313 $\pm$ 6 \tablenotemark{\diamond} & 19 21 39.0 & +15 09 45 & 52.0 $\pm$ 1.0 & 57 $\times$ 47 & -50 & 3.90 &  \nodata  \\
\nodata  & VLA~1B & 145 $\pm$ 3 \tablenotemark{\diamond} & 19 21 40.0 & +15 10 13 & 18.7 $\pm$ 0.4 & 44 $\times$ 33 & 30 & 5.57 & \nodata  \\
\\
G50.271 \& G50.283  & VLA~1 & 463 $\pm$ 9 & 19 25 18.5 & +15 12 27 & 55.9 $\pm$ 1.1 & 71 $\times$ 59 & 70 & 7.90 &  IRAS~19230+1506, G50.28-0.39 \\
\nodata  & VLA~2 & 17.8 $\pm$ 4.1 & 19 25 21.1 & +15 11 05 & 1.16 $\pm$ 0.15 & 55 $\times$ 38 & 80 & 7.31 & New \\
\nodata  & VLA~3 & 163 $\pm$ 3 & 19 25 27.9 & +15 13 29 & 45.4 $\pm$ 0.9 & 57 $\times$ 37 & -80 & 4.61 & 050.318-0.418, 050.308-0.418 \citep{zoonematkermani90} \\
\nodata  & VLA~3A & 69.1 $\pm$ 1.4 \tablenotemark{\diamond} & 19 25 26.8 & +15 12 57 & 35.6 $\pm$ 0.7 & \nodata &  \nodata  & \nodata & 050.308-0.418  \citep{zoonematkermani90}  \\
\nodata  & VLA~3A-P & 61.4 $\pm$ 1.2 & 19 25 26.8 & +15 12 57 & 34.2 $\pm$ 0.7 & 8.9 $\times$ 6.0 & 35 & \nodata & 050.308-0.418  \citep{zoonematkermani90}  \\
\nodata  & VLA~3B & 93.5 $\pm$ 1.9 \tablenotemark{\diamond} & 19 25 27.9 & +15 13 29 & 45.4 $\pm$ 0.9 & \nodata &  \nodata & \nodata & 050.318-0.418  \citep{zoonematkermani90}  \\
\nodata  & VLA~3B-P & 84.4 $\pm$ 1.7 & 19 25 28.0 & +15 13 28 & 44.2 $\pm$ 0.9 & 10.7 $\times$ 6.3 & 127 & \nodata & 050.318-0.418  \citep{zoonematkermani90}  \\
\nodata  & VLA~4 & 12.2 $\pm$ 0.3 & 19 25 32.7 & +15 08 09 & 11.2 $\pm$ 0.3 & \nodata &  \nodata &  \nodata & New \\
\nodata  & VLA~4-P & 12.2 $\pm$ 0.3 & 19 25 32.7 & +15 08 10 & 11.2 $\pm$ 0.2 & 3.7 $\times$ 1.1 & 150 &  \nodata &  New \\
\\
IRAS~18256-0742 & VLA~1 & 1.29 $\pm$ 0.03 & 18 28 16.2 & -07 39 36 & 1.32 $\pm$ 0.06 & \nodata &  \nodata & \nodata & New \\
\nodata & VLA~1-P & 1.35 $\pm$ 0.10 & 18 28 16.2 & -07 39 36 & 1.35 $\pm$ 0.07 & 1.5 $\times$ $<$ 1.8 & 153 & \nodata & New \\
\nodata  & VLA~2 & 2.15 $\pm$ 0.06  & 18 28 18.2 & -07 40 12 & 0.630 $\pm$ 0.052 & 30 $\times$ 24 & -50 & 2.24 & Mol~57, New \\
\nodata  & VLA~3 & 0.912 $\pm$ 0.030  & 18 28 19.7 & -07 38 54 & 0.890 $\pm$ 0.053 & \nodata &  \nodata & \nodata & New \\
\nodata  & VLA~3-P & 0.958 $\pm$ 0.100 & 18 28 19.7 & -07 38 54 & 0.895 $\pm$ 0.066 & 4.3 $\times$ $<$ 2.2 & 178 & \nodata &  New \\
\nodata  & VLA~4 & 9.96 $\pm$ 0.20  & 18 28 23.7 & -07 40 60 & 0.460 $\pm$ 0.051 & 66 $\times$ 61 & 10 & 8.88 & New \\
\nodata  & VLA~5 & 6.83 $\pm$ 0.10 & 18 28 28.4 & -07 42 46 & 6.27 $\pm$ 0.14 & \nodata & \nodata & \nodata &  Mol~57 \\
\nodata  & VLA~5-P & 6.74 $\pm$ 0.10 & 18 28 28.4  & -07 42 46 & 6.28 $\pm$ 0.14 & 2.5 $\times$ 2.0 & 59 & \nodata & Mol~57 \\
\\
IRAS~18424-0329 & VLA~1 & 14.7 $\pm$ 0.1  & 18 45 00.4 & -03 28 26 & 13.1 $\pm$ 0.3 & \nodata & \nodata & \nodata &  Mol~70, 029.107-0.155 \citep{becker94} \\
\nodata & VLA~1-P & 14.0 $\pm$ 0.3  & 18 45 00.3 & -03 28 26 & 13.5 $\pm$ 0.3 & 2.0 $\times$ 0.7 & 114 & \nodata & Mol~70, 029.107-0.155 \citep{becker94} \\
\nodata  & VLA~2 & 44.9 $\pm$ 0.9 & 18 45 02.1 & -03 26 50 & 4.35 $\pm$ 0.13 & 70 $\times$ 50 & 50 & 7.78 & 029.134-0.148  \citep{becker94} \\
\\
IRAS~18571+0349 & VLA~1 &  22.9 $\pm$ 0.6 & 18 59 35.1 & +03 52 46 & 2.58 $\pm$ 0.25 & 35 $\times$ 18 & 10 & 2.82 &  New \\
\nodata & VLA~2 & 15.9 $\pm$ 0.4 & 18 59 40.9 & +03 52 28 & 1.79 $\pm$ 0.24 & 27 $\times$ 22 & -80 & 2.65 &  New \\
\nodata & VLA~3 & 3.07 $\pm$ 0.14 & 18 59 41.7 & +03 53 28 & 1.47 $\pm$ 0.24 & \nodata &  \nodata & \nodata & New \\
\nodata & VLA~3-P & 3.37 $\pm$ 0.14 & 18 59 41.7 & +03 53 27 & 1.31 $\pm$ 0.05 & 10.3 $\times$ 9.1 & 6 & \nodata &  New \\
\nodata & VLA~4 & 13.9 $\pm$ 0.4 & 18 59 42.5 & +03 52 36 & 1.90 $\pm$ 0.24 & 28 $\times$ 20 & 60 & 2.39 &  New \\
\nodata & VLA~5 & 27.8 $\pm$ 0.6 & 18 59 43.1 & +03 53 38 & 4.11 $\pm$ 0.25 & 49  $\times$ 31 & -80 & 4.32 & Mol~86 \\
\nodata & VLA~5-P & 13.1 $\pm$ 0.3 & 18 59 43.2 & +03 53 40 & 3.78 $\pm$ 0.09 & 12.5  $\times$ 11.8 & 35 & \nodata & Mol~86 \\
\nodata & VLA~6 & 36.1 $\pm$ 0.8 & 18 59 45.9 & +03 55 20 & 2.04 $\pm$ 0.24 & 51 $\times$ 48 & 50 & 5.82 &  New \\
\nodata & VLA~6A & 15.0 $\pm$ 0.4 \tablenotemark{\diamond}& 18 59 45.5 & +03 55 00 & 1.84 $\pm$ 0.24 & 44 $\times$ 19 & 60 & 2.74 & New \\
\nodata & VLA~6B & 21.1 $\pm$ 0.5 \tablenotemark{\diamond}& 18 59 45.9 & +03 55 20 & 2.04 $\pm$ 0.24 & 30 $\times$ 26 & 10 & 3.08 & New \\
\nodata & VLA~7 & 48.0 $\pm$ 1.0 & 18 59 52.0 & +03 55 22 & 9.85 $\pm$ 0.31 & 30 $\times$ 24 & -80 & 1.98 & New \\
\nodata & VLA~8 & 83.3 $\pm$ 1.8 & 18 59 53.8 & +03 55 02 & 11.6 $\pm$ 0.3 & 35 $\times$ 24 & 80 & 2.29 & New \\
\\
IRAS~18586+0106 & \nodata  & \nodata & \nodata & \nodata & \nodata & \nodata &  \nodata & \nodata & \nodata \\
\enddata

\tablenotetext{\dagger}{The solid angle is given for irregular/extended sources only, found from the number of pixels above 1$\times \Delta S$ within the photometry aperture.}
\tablenotetext{\diamond}{An additional error of $\sim$10-20\% of the measured flux should be added to the flux errors of the components of multiply peaked sources. This is to account for the uncertainty in where the aperture is placed to divide the components of the source.}

\end{deluxetable}

\clearpage

\begin{deluxetable}{lccccccc}
\tabletypesize{\tiny}
\tablecolumns{8}
\tablewidth{0pt}
\tablecaption{Evolutionary Indicators Associated with The Observed Millimeter Clumps\label{assoc_2}}
\tablehead{\colhead{Millimeter} & \colhead{3.6\,cm Emission} & \colhead{Maser} & \colhead{Dense} & \colhead{Mid-IR} & \colhead{Outflow?}& \colhead{IRAS} & \colhead{L$_{\rm{IRAS}}$}\\
\colhead{Clump Name} & \colhead{Within 60"?} & \colhead{Emission} & \colhead{ Gas} & \colhead{Emission?}& \colhead{}& \colhead{Source} & \colhead{(10$^3$ L$_{\odot}$)}}
\startdata
G044.521+00.387           & 		Y & \nodata		& CS [1] 		& Y 	& \nodata	& IRAS~19090+1023	& 0.932 - 2.13 \\
G044.587+00.371           & 		N & \nodata		& \nodata		& Y	& \nodata	& \nodata				& \nodata		\\
G044.617+00.365           & 		Y & \nodata		& \nodata		& N	& \nodata	& \nodata				& \nodata		\\
G044.661+00.351           & 		N & \nodata		& \nodata		& Y	& \nodata	& IRAS~19094+1029	& 24.3 - 24.6 \\
G048.540+00.040           &  		Y & \nodata		& \nodata		& N	& \nodata	& \nodata				& \nodata		\\
G048.580+00.056           &  		N & \nodata		& \nodata		& N	& \nodata	& \nodata				& \nodata		\\
G048.598+00.252           & 		Y & \nodata		& \nodata		& Y	& \nodata	& \nodata				& \nodata		\\
G048.605+00.024           & 		Y & H$_2$O [2,3],  OH [3] & CO [4,5,6] CS [1,6,7,8] & Y	& Y\tablenotemark{\star}	& IRAS~19181+1349	& 927 - 932 \\
					&		&				& HCN [8] 	     &		&					&					&			\\
G048.610+00.220           &  		N & \nodata		& \nodata		& N	& \nodata	& \nodata				& \nodata		\\
G048.616+00.088           & 		N & \nodata		& \nodata		& Y	& \nodata	& \nodata				& \nodata		\\
G048.634+00.230           & 		Y & \nodata		& CO[9,10] CS [11] & Y	& \nodata	& IRAS~19175+1357	& 60.5 - 168  \\
					&		&				& 1.2\,mm [11], Others [12]       &		&					&					&			\\
G048.656+00.228           & 		Y & \nodata		& \nodata		& Y	& \nodata	& \nodata				& \nodata		\\
G048.751-00.142           & 		N & \nodata		& \nodata		& N	& \nodata	& \nodata				& \nodata		\\
G048.771-00.148           & 		N & \nodata		& \nodata		& Y	& \nodata	& \nodata				& \nodata		\\
G049.830+00.370           & 		Y & \nodata		& \nodata		& Y	& \nodata	& IRAS~19193+1504	& 127 		\\
G049.912+00.370           & 		N & \nodata		& \nodata		& Y	& \nodata	& IRAS~19195+1508	& 7.61 - 21.2 \\
G050.271-00.442           & 		N & \nodata		& \nodata		& Y	& \nodata	& IRAS~19232+1504	& 0.960 - 254 \\
G050.283-00.390           & 		Y & OH? [13,14]		& CS [1]		& Y	& Y\tablenotemark{\star}	& IRAS~19230+1506	& 281 - 286 \\
IRAS~18256-0742~Clump~1   & 	Y & \nodata		& CO [15] NH$_3$ [16] & Y	& N	& IRAS~18256-0742		& 10.5 		\\
IRAS~18424-0329~Clump~2   & 	Y & OH? [17]		& CO [15] NH$_3$ [16] & Y	& N	& IRAS~18424-0329		& 55 \tablenotemark{\dag}\\
IRAS~18424-0329~Clump~4   & 	Y & OH? [17]		& CO [15] NH$_3$ [16] & N	& N	& IRAS~18424-0329		& 55 \tablenotemark{\dag} \\
IRAS~18424-0329~Clump~6   & 	Y & OH? [17]		& CO [15] NH$_3$ [16] & N	& N	& IRAS~18424-0329		& 55  \tablenotemark{\dag}\\
IRAS~18571+0349~Clump~1   & 	Y & \nodata		& CO [15]                     & Y	& N	& IRAS~18571+0349 	& 106  		\\
IRAS~18571+0349~Clump~3   & 	N & \nodata		& CO [15] 			& N	& N	& \nodata				& \nodata		\\
IRAS~18571+0349~Clump~4   & 	Y & CH$_3$OH [18]	& CO [15] 			& Y	& N	& \nodata				& \nodata		\\
IRAS~18586+0106~Clump~1   & 	N & \nodata		& CO [15] 		& N	& N	& \nodata				& \nodata		\\
IRAS~18586+0106~Clump~3   & 	N & \nodata		& CO [15] 		& N	& N		& \nodata				& \nodata		\\
IRAS~18586+0106~Clump~4   & 	N & \nodata		& CO [15] 		& N	& N		& \nodata				& \nodata		\\
IRAS~18586+0106~Clump~5  & 	N & OH [17]		& CO [15] NH$_3$ [16]	& Y	& N		& IRAS~18586+0106	& 44			 \\
IRAS~18586+0106~Clump~6   & 	N & \nodata		& CO [15] 		& N	& N		& \nodata				& \nodata		\\
IRAS~18586+0106~Clump~7   & 	N & \nodata		& CO [15] 		& N	& N		& \nodata				& \nodata		\\
\enddata

\tablecomments{References:
[1] \citet{bronfman96}
[2] \citet{kurtz050}
[3] \citet{forster89}
[4] \citet{shepherd960}
[5] \citet{solomon87}
[6] \citet{plume92}
[7] \citet{shirley03}
[8] \citet{wu03}
[9] \citet{sridharan02}
[10] \citet{thomas08}
[11] \citet{beuther020}
[12] \citet{fuller05}
[13] \citet{te-lintel-hekkert96}
[14] \citet{baudry97}
[15] \citet{zhang05}
[16] \citet{molinari96}
[17] \citet{edris07}
[18] \citet{pandian07}
}

\tablenotetext{\dag}{It is not certain which of the clumps listed by B06 is associated with IRAS~18424-0329, however the general 1.2~mm emission in this field is coincident with the IRAS source.}
\tablenotetext{\star}{Diffuse 4.5$\mu$m emission towards the source suggests the presence of an outflow.}

\end{deluxetable}

\clearpage

\begin{deluxetable}{llcccrrrrrrrcr}
\rotate
\tabletypesize{\tiny}
\tablecolumns{12}
\tablewidth{0pt}
\tablecaption{Derived Physical Properties for VLA 3.6~cm Sources \label{vlaproperties}}
\tablehead{ \colhead{Observed} & \colhead{VLA Source} & \colhead{Irr./} &\colhead{$d$} & \colhead{$\Delta \theta$} & \colhead{$\Delta s$} & \colhead{$T_b$} & \colhead{$\tau_{\nu} \times 10^3$} & \colhead{$EM/10^6$} & \colhead{$n_e$/10$^3$} & \colhead{$U$} & \colhead{$\log{ N_{Ly}}$} & \colhead{Spectral} & \colhead{L$_{\rm{cm}}$/10$^3$} \\ 
\colhead{Field} & \colhead{Name} & \colhead{Unres.} &\colhead{(kpc)}  & \colhead{(")} & \colhead{(pc)} & \colhead{(K)} & \colhead{}& \colhead{(pc cm$^{-6}$)} & \colhead{(cm$^{-3}$)} &\colhead{(pc cm$^{-2}$)} & \colhead{(s$^{-1}$)} &\colhead{Type} & \colhead{(L$_{\odot}$)} }
\startdata

G44.587 \& G44.661   & VLA 1      & I &  3.8 &   46 & 0.84 &  2.57 &    0.314 &  0.066 &   0.28 & 17.99 & 47.47 & B0   &   25.12 \\
\nodata              & VLA 2      & U & 10.8 & \nodata & \nodata & \nodata & \nodata & \nodata & \nodata &  \nodata  & 46.14 & B0.5 &   10.96 \\
\nodata              & VLA 2-P    & U & 10.8 &   $<$2.1 & $<$0.11 &  $>$7.78 &  $>$0.949 &  $>$0.200 &   1.64 &  7.74 & 46.17 & B0.5 &   10.96 \\
G48.580 \& G48.616   & VLA 1      & I & 10.0 &   22 & 1.07 &  0.92 &    0.112 &  0.024 &   0.15 & 14.98 & 47.66 & O9.5 &   38.02 \\
\nodata              & VLA 2      & I & 10.0 &   40 & 1.92 &  1.00 &    0.122 &  0.026 &   0.12 & 22.77 & 48.17 & O8.5 &   53.70 \\
\nodata              & VLA 3      & I & 10.0 &   23 & 1.14 &  0.70 &    0.085 &  0.018 &   0.13 & 14.28 & 47.57 & B0   &   25.12 \\
\nodata              & VLA 4      & I & 10.0 &   70 & 3.39 &  1.79 &    0.219 &  0.046 &   0.12 & 40.46 & 48.59 & O7   &  100.00 \\
\nodata              & VLA 4A     & U & 10.0 & \nodata & \nodata & \nodata & \nodata & \nodata & \nodata &  \nodata  & 48.10 & O9   &   45.71 \\
\nodata              & VLA 4A-P   & U & 10.0 & $<$2.9 & $<$0.14 & $>$414.62 &  $>$51.886 & $>$10.916 &  10.83 & 33.88 & 48.09 & O9   &   45.71 \\
\nodata              & VLA 4B     & I & 10.0 &   59 & 2.88 &  1.35 &    0.165 &  0.035 &   0.11 & 32.95 & 48.42 & O8   &   64.57 \\
\nodata              & VLA 5      & I &  9.9 &  164 & 7.88 &  2.86 &    0.349 &  0.073 &   0.10 & 82.84 & 49.45 & O5.5 &  398.11 \\
\nodata              & VLA 5A     & I & 10.0 &  153 & 7.42 &  2.48 &    0.302 &  0.064 &   0.09 & 75.87 & 49.32 & O5.5 &  398.11 \\
\nodata              & VLA 5B     & I &  9.9 &   43 & 2.09 &  5.13 &    0.626 &  0.132 &   0.25 & 41.54 & 48.50 & O7.5 &   83.18 \\
\nodata              & VLA 5C     & I &  9.9 &   35 & 1.66 &  5.15 &    0.628 &  0.132 &   0.28 & 35.73 & 48.35 & O8   &   64.57 \\
\nodata              & VLA 5C-P   & U &  9.9 &  15.1 & 0.72 &  24.98 &    3.051 &  0.642 &   1.15 & 39.80 & 48.30 & O8   &   64.57 \\
\nodata              & VLA 5D     & I &  9.9 &   39 & 1.88 &  4.25 &    0.518 &  0.109 &   0.24 & 36.34 & 48.34 & O8   &   64.57 \\
\nodata              & VLA 5D-P   & U &  9.9 &  21.9 & 1.05 &  14.62 &    1.784 &  0.375 &   0.73 & 42.65 & 48.39 & O8   &   64.57 \\
\nodata              & VLA 6      & I &  9.9 &   39 & 1.88 &  2.65 &    0.324 &  0.068 &   0.19 & 31.06 & 48.41 & O8   &   64.57 \\
G48.598 \& G48.656   & VLA 1      & I & 10.6 &   74 & 3.83 &  0.27 &    0.033 &  0.007 &   0.04 & 23.28 & 47.86 & O9.5 &   38.02 \\
\nodata              & VLA 2      & U & 10.6 & \nodata & \nodata & \nodata & \nodata & \nodata & \nodata &  \nodata  & 46.51 & B0.5 &   10.96 \\
\nodata              & VLA 2-P    & U & 10.6 &  13.4 & 0.69 &   0.39 &    0.047 &  0.010 &   0.15 &  9.60 & 46.45 & B0.5 &   10.96 \\
\nodata              & VLA 3      & U & 10.5 & \nodata & \nodata & \nodata & \nodata & \nodata & \nodata &  \nodata  & 46.88 & B0   &   25.12 \\
\nodata              & VLA 3-P    & U & 10.5 &  23.9 & 1.22 &   0.29 &    0.035 &  0.007 &   0.10 & 12.69 & 46.82 & B0   &   25.12 \\
\nodata              & VLA 4      & I & 10.5 &   35 & 1.81 &  0.42 &    0.051 &  0.011 &   0.08 & 16.37 & 47.44 & B0   &   25.12 \\
\nodata              & VLA 5      & I & 10.5 &  118 & 6.03 &  0.55 &    0.068 &  0.014 &   0.05 & 40.11 & 48.27 & O8.5 &   53.70 \\
\nodata              & VLA 5A     & I & 10.3 &   24 & 1.22 &  0.50 &    0.061 &  0.013 &   0.10 & 13.41 & 47.26 & B0   &   25.12 \\
\nodata              & VLA 5B     & I & 10.3 &   28 & 1.42 &  0.42 &    0.051 &  0.011 &   0.09 & 13.93 & 47.19 & B0   &   25.12 \\
\nodata              & VLA 5C     & I & 10.5 &   43 & 2.21 &  0.62 &    0.076 &  0.016 &   0.09 & 21.39 & 47.76 & O9.5 &   38.02 \\
\nodata              & VLA 5C-P   & U & 10.5 &  17.0 & 0.86 &   2.18 &    0.266 &  0.056 &   0.31 & 19.87 & 47.40 & B0   &   25.12 \\
\nodata              & VLA 5D     & I & 10.3 &   22 & 1.10 &  0.38 &    0.047 &  0.010 &   0.09 & 11.41 & 47.04 & B0   &   25.12 \\
\nodata              & VLA 5E     & I & 10.3 &   46 & 2.32 &  0.52 &    0.064 &  0.013 &   0.08 & 20.81 & 47.61 & O9.5 &   38.02 \\
\nodata              & VLA 5F     & I & 10.3 &   35 & 1.74 &  0.70 &    0.086 &  0.018 &   0.10 & 18.98 & 47.58 & B0   &   25.12 \\
\nodata              & VLA 6      & U & 10.3 & \nodata & \nodata & \nodata & \nodata & \nodata & \nodata &  \nodata  & 47.43 & B0   &   25.12 \\
\nodata              & VLA 6-P    & U & 10.3 &   4.5 & 0.22 &  37.02 &    4.525 &  0.952 &   2.54 & 20.64 & 47.45 & B0   &   25.12 \\
G48.751              & VLA 1      & I &  5.3 &   31 & 0.81 &  0.05 &    0.006 &  0.001 &   0.04 &  4.70 & 46.16 & B0.5 &   10.96 \\
G49.912              & VLA 1      & I & 10.6 &   73 & 3.77 &  2.20 &    0.269 &  0.057 &   0.12 & 46.51 & 48.66 & O7   &  100.00 \\
\nodata              & VLA 1A     & I & 10.6 &   52 & 2.66 &  3.65 &    0.446 &  0.094 &   0.19 & 43.60 & 48.50 & O7.5 &   83.18 \\
\nodata              & VLA 1B     & I & 10.6 &   38 & 1.96 &  1.19 &    0.145 &  0.030 &   0.12 & 24.45 & 48.16 & O8.5 &   53.70 \\
G50.271 \& G50.283   & VLA 1      & I &  9.6 &   65 & 3.01 &  2.66 &    0.325 &  0.068 &   0.15 & 42.63 & 48.58 & O7   &  100.00 \\
\nodata              & VLA 2      & I &  9.6 &   46 & 2.13 &  0.11 &    0.014 &  0.003 &   0.04 & 11.73 & 47.17 & B0   &   25.12 \\
\nodata              & VLA 3      & I &  9.6 &   46 & 2.14 &  1.60 &    0.196 &  0.041 &   0.14 & 28.64 & 48.13 & O9   &   45.71 \\
\nodata              & VLA 3A     & U &  9.6 & \nodata & \nodata & \nodata & \nodata & \nodata & \nodata &  \nodata  & 47.75 & O9.5 &   38.02 \\
\nodata              & VLA 3A-P   & U &  9.6 &  13.2 & 0.62 &   8.66 &    1.056 &  0.222 &   0.74 & 25.07 & 47.70 & O9.5 &   38.02 \\
\nodata              & VLA 3B     & U &  9.6 & \nodata & \nodata & \nodata & \nodata & \nodata & \nodata &  \nodata  & 47.89 & O9.5 &   38.02 \\
\nodata              & VLA 3B-P   & U &  9.6 &  14.8 & 0.69 &   9.47 &    1.156 &  0.243 &   0.73 & 27.88 & 47.84 & O9.5 &   38.02 \\
\nodata              & VLA 4      & U &  9.7 & \nodata & \nodata & \nodata & \nodata & \nodata & \nodata &  \nodata  & 47.01 & B0   &   25.12 \\
\nodata              & VLA 4-P    & U &  9.7 &   3.9 & 0.18 &  20.03 &    2.446 &  0.515 &   2.06 & 14.74 & 47.01 & B0   &   25.12 \\
IRAS 18256-0742      & VLA 1      & U &  3.0 & \nodata & \nodata & \nodata & \nodata & \nodata & \nodata &  \nodata  & 45.01 & B1   &    5.25 \\
\nodata              & VLA 1-P    & U &  3.0 & $<$3.1 & $<$0.05 &  $>$3.39 &   $>$0.413 & $>$0.087 &   1.69 &  3.23 & 45.03 & B1   &    5.25 \\
\nodata              & VLA 2      & I &  3.0 &   27 & 0.39 &  0.04 &    0.005 &  0.001 &   0.05 &  2.77 & 45.24 & B1   &    5.25 \\
\nodata              & VLA 3      & U &  3.0 & \nodata & \nodata & \nodata & \nodata & \nodata & \nodata &  \nodata  & 44.86 & B2   &    2.88 \\
\nodata              & VLA 3-P    & U &  3.0 &  $<$5.8 & $<$0.08 & $>$0.70 & $>$0.085 & $>$0.018 &   0.56 &  2.89 & 44.89 & B2   &    2.88 \\
\nodata              & VLA 4      & I &  3.0 &   63 & 0.92 &  0.05 &    0.006 &  0.001 &   0.04 &  5.18 & 45.90 & B0.5 &   10.96 \\
\nodata              & VLA 5      & U &  3.0 & \nodata & \nodata & \nodata & \nodata & \nodata & \nodata &  \nodata  & 45.74 & B1   &    5.25 \\
\nodata              & VLA 5-P    & U &  3.0 &   4.3 & 0.06 &   9.02 &    1.101 &  0.232 &   2.36 &  5.53 & 45.73 & B1   &    5.25 \\
IRAS 18424-0329      & VLA 1      & U & 11.6 & \nodata & \nodata & \nodata & \nodata & \nodata & \nodata &  \nodata  & 47.25 & B0   &   25.12 \\
\nodata              & VLA 1-P    & U & 11.6 &  2.2 & 0.13 &  69.05 &    8.457 &  1.779 &   4.61 & 17.38 & 47.23 & B0   &   25.12 \\
\nodata              & VLA 2      & I & 11.6 &   59 & 3.33 &  0.26 &    0.032 &  0.007 &   0.04 & 21.04 & 47.73 & O9.5 &   38.02 \\
IRAS 18571+0349      & VLA 1      & I &  9.8 &   25 & 1.19 &  0.37 &    0.045 &  0.009 &   0.09 & 11.90 & 47.29 & B0   &   25.12 \\
\nodata              & VLA 2      & I &  9.8 &   24 & 1.16 &  0.27 &    0.033 &  0.007 &   0.08 & 10.54 & 47.13 & B0   &   25.12 \\
\nodata              & VLA 3      & U &  9.8 & \nodata & \nodata & \nodata & \nodata & \nodata & \nodata &  \nodata  & 46.42 & B0.5 &   10.96 \\
\nodata              & VLA 3-P    & U &  9.8 &  17.0 & 0.81 &   0.29 &    0.035 &  0.007 &   0.12 &  9.66 & 46.46 & B0.5 &   10.96 \\
\nodata              & VLA 4      & I &  9.8 &   24 & 1.12 &  0.26 &    0.032 &  0.007 &  0.08 & 10.23 & 47.08 & B0   &   25.12 \\
\nodata              & VLA 5      & I &  9.8 &   39 & 1.85 &  0.29 &    0.036 &  0.007 &   0.06 & 14.75 & 47.38 & B0   &   25.12 \\
\nodata              & VLA 5-P    & U &  9.8 &  21.1 & 1.00 &   0.73 &    0.089 &  0.019 &   0.17 & 15.21 & 47.05 & B0   &   25.12 \\
\nodata              & VLA 6      & I &  9.7 &   49 & 2.33 &  0.28 &    0.034 &  0.007 &   0.06 & 16.98 & 47.48 & B0   &   25.12 \\
\nodata              & VLA 6A     & I &  9.7 &   29 & 1.36 &  0.25 &    0.030 &  0.006 &   0.07 & 11.40 & 47.10 & B0   &   25.12 \\
\nodata              & VLA 6B     & I &  9.7 &   28 & 1.31 &  0.31 &    0.038 &  0.008 &   0.08 & 11.98 & 47.25 & B0   &   25.12 \\
\nodata              & VLA 7      & I &  9.7 &   27 & 1.26 &  1.10 &    0.134 &  0.028 &   0.15 & 17.78 & 47.61 & O9.5 &   38.02 \\
\nodata              & VLA 8      & I &  9.7 &   29 & 1.36 &  1.65 &    0.201 &  0.042 &   0.18 & 21.42 & 47.84 & O9.5 &   38.02 \\
\enddata

\end{deluxetable}

\clearpage

\begin{deluxetable}{llcccc}
\tabletypesize{\scriptsize}
\tablecolumns{6}
\tablewidth{0pt}
\tablecaption{Ionized Gas Associated with Observed Millimeter Clumps\label{assoc}}
\tablehead{\colhead{Millimeter} & \colhead{Nearest } & \colhead{3.6~cm Emission} & \multicolumn{2}{c}{Distance to Nearest 3.6~cm Emission} & \colhead{L$_{\rm{cm}}$/10$^{3}$}\\
\colhead{Clump Name} & \colhead{3.6~cm Source} & \colhead{Within 60"?} & \colhead{(arcsec)} & \colhead{(pc)} & \colhead{L$_{\odot}$}}
\startdata
G044.521+00.387           & G44.587 \& G44.661 VLA 1  & Y &   9 &   0.2 & 25.12 \\
G044.587+00.371           & G44.587 \& G44.661 VLA 2  & N &  98 &   5.2 & \nodata \\
G044.617+00.365           & G44.587 \& G44.661 VLA 2  & Y &  22 &   1.2 & 10.96\\
G044.661+00.351           & G44.587 \& G44.661 VLA 2  & N & 183 &   9.6 & \nodata \\
G048.540+00.040           & G48.580 \& G48.616 VLA 3  & Y &  34 &   1.6 & 25.12 \\
G048.580+00.056           & G48.580 \& G48.616 VLA 5A & N &  66 &   3.2 & \nodata \\
G048.598+00.252           & G48.598 \& G48.656 VLA 1  & Y &  48 &   2.5 & 38.02 \\
G048.605+00.024           & G48.580 \& G48.616 VLA 5D & Y &   6 &   0.3 & 64.57\\
G048.610+00.220           & G48.598 \& G48.656 VLA 2  & N &  75 &   3.8 & \nodata \\
G048.616+00.088           & G48.580 \& G48.616 VLA 2  & N &  76 &   3.7 & \nodata\\
G048.634+00.230           & G48.598 \& G48.656 VLA 4  & Y &   8 &   0.4 & 25.12\\
G048.656+00.228           & G48.598 \& G48.656 VLA 5D & Y &   3 &   0.2 & 25.12\\
G048.751-00.142           & G48.751 VLA 1             & N &  81 &   2.1 & \nodata\\
G048.771-00.148           & G48.751 VLA 1             & N & 146 &   3.8 & \nodata\\
G049.830+00.370           & G49.912 VLA 1A            & Y &  24 &   1.2 & 83.18 \\
G049.912+00.370           & G49.912 VLA 1B            & N & 273 &  13.6 & \nodata\\
G050.271-00.442           & G50.271 \& G50.283 VLA 2  & N & 105 &   4.9 & \nodata\\
G050.283-00.390           & G50.271 \& G50.283 VLA 1  & Y &  13 &   0.6 & 100.00\\
IRAS~18256-0742~Clump~1   & IRAS 18256-0742 VLA 2     & Y &  12 &   0.2 & 5.25\\
IRAS~18424-0329~Clump~2   & IRAS 18424-0329 VLA 2     & Y &  28 &   1.6 & 38.02 \\
IRAS~18424-0329~Clump~4   & IRAS 18424-0329 VLA 2     & Y &  31 &   1.7 & 38.02\\
IRAS~18424-0329~Clump~6   & IRAS 18424-0329 VLA 1     & Y &  50 &   2.8 & 25.12\\
IRAS~18571+0349~Clump~1   & IRAS 18571+0349 VLA 5     & Y &   7 &   0.3 & 25.12\\
IRAS~18571+0349~Clump~3   & IRAS 18571+0349 VLA 7     & N &  82 &   3.8 & \nodata\\
IRAS~18571+0349~Clump~4   & IRAS 18571+0349 VLA 7     & Y &  13 &   0.6 & 38.02\\
IRAS~18586+0106~Clump~1   & \nodata     & N &  \nodata &  \nodata &  \nodata\\
IRAS~18586+0106~Clump~3   & \nodata     & N &  \nodata &  \nodata &  \nodata\\
IRAS~18586+0106~Clump~4   & \nodata     & N &  \nodata &  \nodata &  \nodata\\
IRAS~18586+0106~Clump~5   & \nodata     & N &  \nodata &  \nodata &  \nodata\\
IRAS~18586+0106~Clump~6   & \nodata     & N &  \nodata &  \nodata &  \nodata\\
IRAS~18586+0106~Clump~7   & \nodata     & N &  \nodata &  \nodata &  \nodata\\
\\
\enddata

\end{deluxetable}

\clearpage

\begin{deluxetable}{llcccc}
\tabletypesize{\tiny}
\tablecolumns{6}
\tablewidth{0pt}
\tablecaption{Comparison of Cloud and Clump Masses Derived from $^{13}$CO and  1\,mm \label{m13co}}
\tablehead{\colhead{Observed} & \colhead{Associated} & \colhead{M$_{\rm{cloud}}(^{13}$CO)} & \colhead{M$_{\rm{cloud}}(1\,\rm{mm})$}  & \colhead{$\frac{\rm{M}_{\rm{cloud}}(1\,\rm{mm})}{\rm{M}_{cloud}(^{13}\rm{CO})}$} &  \colhead{$\frac{\Sigma \rm{M}_{\rm{clump}}}{\rm{M}_{\rm{cloud}}(^{13}\rm{CO})}$}\\
\colhead{Field} & \colhead{Millimeter Clumps} & \colhead{(M$_{\odot}$)} & \colhead{(M$_{\odot}$)} & \colhead{} & \colhead{} }
\startdata
G44.587 \& G44.661 & G44.521 & 1200 & 310 \tablenotemark{\dagger} & 0.26 & 0.05 \\
G44.587 \& G44.661 & G44.587, G44.617, G44.661 & 9200 & 2600 & 0.28 & 0.14  \\
G48.580 \& G48.616 & G48.580, G48.616, G48.540, G48.605 & 42000 & 15000 & 0.36 & 0.22 \\
G48.598 \& G48.656 & G48.598, G48.656, G48.634, G48.610 & 53000 & 5700 & 0.11 & 0.05 \\
G48.751 & G48.751, G48.771 & 2700 & 650 & 0.24 & 0.04  \\
G49.912 & G49.830 & 13000 & 2500 & 0.19 & 0.07 \\
G49.912 & G49.912 & 3200 & 1400 & 0.44 & 0.13 \\
G50.271 \& G50.283 & G50.271 & 2100 & 940 & 0.45 & 0.07 \\
G50.271 \& G50.283 & G50.283 & 2400 & 2400 & 1.00 & 0.43  \\
IRAS 18424-0329 & Clumps 2, 4 and 6 & 29000 & 2200 & 0.08 & 0.05 \\
IRAS 18571+0349 & Clumps 1, 3 and 4 & 12000 & 2500 & 0.21 & 0.17 \\
\enddata
\tablenotetext{\dagger}{The photometry apertures were different for this source to avoid unrelated emission in the 1\,mm image.}
\end{deluxetable}

\clearpage

\begin{deluxetable}{llccc}
\tabletypesize{\scriptsize}
\tablecolumns{5}
\tablewidth{0pt}
\tablecaption{Comparison of Clump and Stellar Masses \label{mclump_mstar}}
\tablehead{\colhead{Millimeter} & \colhead{Associated} & \colhead{M$_{\rm{clump}}$} & \colhead{M$_{\star}$} & \colhead{M$_{\star}$/(M$_{\rm{clump}}$+M$_{\star}$)} \\
\colhead{Clump Name} & \colhead{3.6~cm source} & \colhead{M$_{\odot}$} & \colhead{M$_{\odot}$} }
\startdata
G044.521+00.387 & G44.587 \& G44.661 VLA 1  & 56 & 17 & 0.23\\
G048.605+00.024 & G48.580 \& G48.616 VLA 5B, C, D \tablenotemark{\dagger} &  5725 & 67 & 0.01 \\
G048.634+00.230 & G48.598 \& G48.656 VLA 4  & 543 & 17 & 0.03 \\
G048.656+00.228 & G48.598 \& G48.656 VLA 5D & 564 & 17 & 0.03 \\
G049.830+00.370 & G49.912 VLA 1A and B  \tablenotemark{\dagger} & 972 & 44 & 0.04 \\
G050.283-00.390 & G50.271 \& G50.283 VLA 1& 1024 & 24 & 0.02 \\
IRAS 18256-0742 Clump 1 & IRAS 18256-0742 VLA 2 & 52 & 11 & 0.17 \\
IRAS 18571+0349 Clump 1 & IRAS 18571+0349 VLA 5 & 1509 & 17 & 0.01 \\
IRAS 18571+0349 Clump 4 & IRAS 18571+0349 VLA 7 & 217 & 19 & 0.08 \\
\enddata
\tablenotetext{\dagger}{The stellar masses derived for VLA 5B, C, and D were summed to find the stellar mass associated with G48.605. Similarly, the stellar masses derived for VLA 1A and VLA 1B were also combined for the millimeter source G48.830.}
\end{deluxetable}

\end{document}